\documentclass[11pt,letterpaper]{article}
\usepackage[margin=1in]{geometry}
\usepackage[T1]{fontenc}
\usepackage{lmodern,microtype,amsmath,amssymb,amsthm,mathtools,booktabs,tabularx}
\usepackage[hidelinks]{hyperref}
\hypersetup{hypertexnames=false,
  pdftitle={Space-Optimal Algorithms for Spanning Forests in Dynamic Streams at Every Confidence Level: With and Without Deletion Constraints},
  pdfauthor={Zemin Chao; Jianhe Cen; Hengzhao Ma; Hongzhi Wang}}
\newtheorem{theorem}{Theorem}[section]
\newtheorem{lemma}[theorem]{Lemma}
\newtheorem{proposition}[theorem]{Proposition}
\newtheorem{corollary}[theorem]{Corollary}
\numberwithin{equation}{section}
\newcommand{\E}{\mathbb E}
\newcommand{\poly}{\operatorname{poly}}
\DeclareMathOperator{\supp}{supp}
\title{Space-Optimal Algorithms for Spanning Forests in Dynamic Streams at Every Confidence Level: With and Without Deletion Constraints}
\author{%
  Zemin Chao\textsuperscript{1}\quad
  Jianhe Cen\textsuperscript{1}\quad
  Hengzhao Ma\textsuperscript{2}\quad
  Hongzhi Wang\textsuperscript{1}\thanks{Corresponding author.}\\[0.6em]
  \small\textsuperscript{1}Harbin Institute of Technology, Harbin, China\\
  \small\textsuperscript{2}Northeastern University, Shenyang, China}
\date{}
\begin{document}
\pagenumbering{roman}
\maketitle
\thispagestyle{empty}

\begin{abstract}
We study the space complexity of one-pass spanning-forest computation in dynamic graph streams on $v$-vertex simple graphs, with failure probability at most $\delta$.
Nelson and Yu prove a space lower bound of $\Omega(v\log^2 v\log(v/\delta))$ bits for every fixed
$0<\epsilon<1$ and $2^{-v^{1-\epsilon}}<\delta\le1/4$, thereby establishing the space optimality of the well-known AGM algorithm in this range.
However, these results do not determine the optimal space throughout the higher-confidence regime.
We prove space lower bounds in the previously unresolved high-confidence regime and design a matching space-optimal algorithm, thereby establishing that, for every $0<\delta\le1/4$, the optimal space is
\[
\Theta\!\left(
  \min\left\{
    v^2,\;
    v\log(v/\delta)
    \log^2\!\left(2+\frac{v}{\log(v/\delta)}\right)
  \right\}
\right)
\quad\text{bits}.
\]
Our results advance the state of the art by achieving space savings by up to a factor of $\Theta((\log v/\log\log v)^2)$ in the higher-confidence regime and establishing the optimal space complexity across all confidence levels.

We further study the problem under the promise that at most $Z$ previously inserted edges have been deleted and not yet reinserted when the report is requested, where the integer $Z\ge0$ is given in advance.
The total number of deletions and restorations, and intermediate absence counts, are unrestricted.
For every $0<\delta\le1/4$, we prove space lower bounds and design a matching space-optimal one-pass randomized algorithm for a single report, establishing an optimal space complexity of $\Theta(\mathcal L(v,Z,\log(1/\delta)))$ bits, where $x_Z(t)=\min\{Z+1,v,\sqrt{vt}\}$, and
\[
\mathcal L(v,Z,t)=v\log v\min\{\log^2v,\log^2(Z+2)\}
 +vt\log\!\left(1+\frac{x_Z(t)}t\right)
      \log\!\left(1+\frac v{x_Z(t)}\right).
\]
At zero error, we prove a lower bound for all zero-error algorithms and design a matching deterministic algorithm, establishing an optimal space complexity of $\Theta(\min\{v^2,v(Z+1)\log(1+v/(Z+1))\})$ bits.
Together, these results characterize how failure probability and report-time edge loss determine the memory required for dynamic spanning-forest computation, and provide algorithms attaining the optimal space bounds.
\end{abstract}
\clearpage
\begingroup
\pagestyle{empty}
\small
\setcounter{tocdepth}{2}
\tableofcontents
\thispagestyle{empty}
\endgroup
\clearpage
\pagenumbering{arabic}

\section{Introduction}
\label{sec:graph-introduction}

Computing connected components of graphs is an important task in data analysis~\cite{TenchEtAl2025Landscape}.
At a fixed distance threshold, single-linkage clusters are precisely the connected components of the graph joining pairs of objects within that threshold~\cite{GowerRoss1969}.
In record linkage, connected components can group records linked by candidate matches for subsequent refinement~\cite{MamunAseltineRajasekaran2016}.
When the relationship graph is too large to retain and its edges arrive as insertions and deletions, these tasks motivate recovering connectivity from a compact summary~\cite{AGM2012}.

A spanning forest contains one tree spanning each connected component, and is a single spanning tree when the graph is connected.
Two vertices are connected in the forest exactly when they are connected in the original graph, with paths made of currently present original edges.
For a graph with $v$ vertices and $c$ connected components, this spanning forest has $v-c$ edges and can be represented in $O(v\log v)$ bits, even though the graph itself may have $\Theta(v^2)$ edges.
Storing the full graph can therefore require far more memory than storing the requested answer.

In a one-pass stream, however, discarded edges cannot be revisited unless information about them has been retained.
With insertions alone, an algorithm can retain a forest and discard each edge whose endpoints are already connected.
If a forest edge is later deleted, a replacement may exist among those discarded edges, but the stored forest does not reveal whether such an edge exists or where to find it.
The central space question is how much information about these alternatives must be retained, and how this requirement grows as the tolerated failure probability decreases.

We study this question in the fully dynamic graph-streaming model.
An initially empty undirected simple graph has a fixed set of $v$ labeled vertices.
The algorithm processes a \emph{legal dynamic graph stream} in one pass: an insertion adds an absent edge, and a deletion removes a present edge.
At a requested report, it must output a spanning forest consisting of currently present original edges, with failure probability at most $\delta$.
We design algorithms and prove matching space lower bounds, determining the optimal space complexity in bits for one report, both with unrestricted deletions and under a report-time bound on missing historical edges.
We then determine the additional space required for joint correctness across multiple reports under the same promise.
All update and report sequences are fixed independently of the algorithm's random coins.

\subsection{Prior algorithms, space lower bounds, and the unresolved reliability threshold}

For any fixed constant $c>0$, the AGM algorithm recovers a spanning forest using $O(v\log^3v)$ bits with failure probability at most $\delta=v^{-c}$~\cite{AGM2012}.
Nelson and Yu prove that every algorithm for the same task requires $\Omega(v\log^3v)$ bits, even when constant failure probability is allowed~\cite{NelsonYu2019}.
Together, these results establish the optimal space at constant failure probability and at $\delta=v^{-c}$ for fixed $c>0$.
Furthermore, the parameterization of AGM analyzed by Nelson and Yu gives an algorithm using
\[
 O\!\left(v\log^2v\log(v/\delta)\right)
\]
bits at failure probability at most $\delta$~\cite[Appendix~A, Theorem~5]{NelsonYu2019}.
Separately, for every fixed $0<\epsilon<1$ and $2^{-v^{1-\epsilon}}<\delta\le1/4$, their lower-bound theorem states that every algorithm solving this task requires
\[
 \Omega\!\left(v\log^2v\log(v/\delta)\right)
\]
bits~\cite[Theorem~3]{NelsonYu2019}.
The two bounds match in this range, establishing the space optimality of the reparameterized AGM algorithm there.
An algorithm storing the full adjacency matrix uses $O(v^2)$ bits and has zero error.
Standard confidence amplification by independent repetition and majority voting does not directly apply: successful runs may return different forests~\cite[Appendix~A]{NelsonYu2019}.

Consider now the smaller failure probability $\delta=2^{-v/\log^2v}$.
At this parameter, the published AGM memory guarantee becomes $O(v^2)$ bits, the same order as storing the full adjacency matrix.
The earlier lower-bound theorem does not determine the optimal space at this parameter.
This leaves the following question:
\begin{quote}
Can a spanning-forest algorithm use subquadratic space at this failure probability, and how small can the failure probability become before every algorithm requires quadratic space?
\end{quote}

Deletion restrictions offer another way to reduce memory.
With insertions alone, retaining a spanning forest uses the optimal $O(v\log v)$ bits, whereas unrestricted insertions and deletions require $\Theta(v\log^3v)$ bits at constant failure probability~\cite{AGM2012,NelsonYu2019}.
The bounded-deletion graph-streaming model studies this gap by limiting the total number of deletion operations.
Recent work in this model addresses maximal matching~\cite{KKD2025} and hypergraph cut sparsification~\cite{KPS2025}.
For a given integer $Z\ge0$ and streams with at most $Z$ deletion operations, specializing the latter result to ordinary graphs with fixed multiplicative error below one and extracting a spanning forest gives a $\widetilde O(v\log(Z+2))$-bit algorithm with failure probability at most $v^{-c}$ for some fixed constant $c>0$.
The sparsifier is a reweighted subgraph with the same zero cuts, so its connected components coincide with those of the original graph.
The exact logarithmic factors and the dependence on the full range of failure probabilities remain unresolved for spanning forests.

A total-deletion budget counts every deletion, including deletions followed by restoration before the next report.
We instead consider a bound on the number of previously inserted edges that are still absent when an answer is requested.
Jayaram and Woodruff study a related query-time survival promise through their $L_0$ $\alpha$-property~\cite{JayaramWoodruff2018}, which compares current and historical support sizes by a ratio.
Our graph-stream promise bounds their difference and allows arbitrarily many deletions and restorations between reports.
The resulting question is how the optimal space depends jointly on this report-time bound and the allowed failure probability.
Section~\ref{sec:intro-historical-deficit} states our algorithms and matching space lower bounds for this setting; Appendix~\ref{sec:extended-related-work} gives further comparisons with related work.

\subsection{Space-optimal algorithms at every confidence level}

The following two theorems give a space lower bound and a space-optimal algorithm for every $0<\delta\le1/4$.
We recover the known optimal space orders in the previously resolved parameter range; the additional results determine the optimum in the remaining high-confidence range.
Throughout the paper, $v\ge2$.
The space measures are worst-case peak space and worst-input expected peak space, in bits.
Following standard asymptotic notation in theoretical computer science, an optimal space complexity of $\Theta(f)$ combines an algorithm using $O(f)$ bits with a matching $\Omega(f)$ space lower bound for all algorithms meeting the same task requirements, under the specified space measure.
Running-time guarantees are stated separately.
The two peak-space measures are specified in Section~\ref{sec:model-organization}.

\begin{samepage}
\begin{theorem}[Space lower bound at every confidence level]
\label{thm:unrestricted-space-lower}
For $v\ge2$ and $0<\delta\le1/4$, every algorithm that recovers a spanning forest with one report and failure probability at most $\delta$ on unrestricted legal dynamic streams of simple graphs requires, in bits,
\[
\Omega\!\left(\min\left\{v^2,\,
 v\log(v/\delta)\log^2\!\left(2+\frac{v}{\log(v/\delta)}\right)\right\}\right).
\]
The constants are absolute. The lower bound applies to arbitrary algorithms, including nonlinear algorithms.
\end{theorem}
\end{samepage}

\begin{theorem}[Space-optimal fully dynamic algorithm]
\label{thm:unrestricted-space}
For $v\ge2$ and $0<\delta\le1/4$, the one-pass algorithm recovers a spanning forest with one report and failure probability at most $\delta$ on unrestricted legal dynamic streams of simple graphs, using, in bits,
\begin{equation}
O\!\left(\min\left\{v^2,\,
 v\log(v/\delta)\log^2\!\left(2+\frac{v}{\log(v/\delta)}\right)\right\}\right).
\label{eq:graph-linear-main}
\end{equation}
The constants are absolute. Conditional on its stored setup, the algorithm maintains a binary linear sketch over $\mathbb F_2$. It matches Theorem~\ref{thm:unrestricted-space-lower} and is therefore space-optimal.
\end{theorem}

Sections~\ref{sec:unrestricted-cut-interface}--\ref{sec:compiler} present the algorithm; Sections~\ref{sec:pointwise-source}--\ref{sec:indexed-confidence} prove the space lower bound.

For example, at $\delta=2^{-v/\log^2v}$, these theorems give:
\[
\begin{aligned}
\text{Our algorithm uses}\quad
 &O\!\left(\frac{v^2(\log\log v)^2}{\log^2v}\right)\text{ bits},\\
\text{Our space lower bound}\quad
 &\Omega\!\left(\frac{v^2(\log\log v)^2}{\log^2v}\right)\text{ bits}.
\end{aligned}
\]
Comparing algorithmic memory guarantees, its space is a factor $\Theta((\log v/\log\log v)^2)$ smaller than the previous $O(v^2)$ guarantee.

More generally, for any sequence $v\to\infty$ and $0<\delta(v)\le1/4$, with unrestricted deletions and one report, an algorithm using $o(v^2)$ bits exists if and only if $\log(1/\delta)=o(v)$.
Every algorithm requires $\Omega(v^2)$ bits if and only if $\log(1/\delta)=\Omega(v)$; full adjacency storage then gives a matching $O(v^2)$-bit algorithm.
Both equivalences hold for both peak-space measures (Corollary~\ref{cor:quadratic-boundary}).

\subsection{Optimal space under a report-time deletion promise}
\label{sec:intro-historical-deficit}

For a stream prefix $w$, let $E^{\rm ever}(w)$ be the set of edge names inserted at least once and let $E(w)$ be the current edge set.
Define the \emph{historical deficit} by
\[
 d(w)=|E^{\rm ever}(w)\setminus E(w)|.
\]
We are given an integer $Z\ge0$ in advance and require $d(w)\le Z$ only at reports.
The stream may contain arbitrarily many deletions and restorations, and its intermediate historical deficit may exceed $Z$.
The algorithm must nevertheless respect its space bound throughout the stream.

For a real $t\ge1$, put $x_Z(t)=\min\{Z+1,v,\sqrt{vt}\}$, and define
\begin{equation}
\mathcal L(v,Z,t)=v\log v\min\{\log^2v,\log^2(Z+2)\}
 +vt\log\!\left(1+\frac{x_Z(t)}t\right)
      \log\!\left(1+\frac v{x_Z(t)}\right).
\label{eq:intro-space-profile}
\end{equation}

\begin{theorem}[Optimal one-report space under a historical-deficit promise]
\label{thm:historical-deficit-main}
\label{thm:deletion-budget-lower}
\label{thm:deletion-budget-space}
\label{cor:intro-historical-deficit}
Let $v\ge2$, let the integer $Z\ge0$ be given in advance, and let $0<\delta\le1/4$.
Consider one-pass spanning-forest computation on arbitrary legal update histories of $v$-vertex simple graphs, with one report, where the history and report location are fixed independently of the algorithm's randomness.
The historical deficit is required to be at most $Z$ only at the report; its intermediate values and the total number of deletions and restorations are unrestricted.
\begin{enumerate}
\item \emph{Space lower bound.} Every algorithm returning a spanning forest with failure probability at most $\delta$ requires
\[
\Omega\!\left(\mathcal L(v,Z,\log(1/\delta))\right)\text{ bits}.
\]
\item \emph{Space-optimal algorithm.} Our one-pass randomized algorithm returns a spanning forest with failure probability at most $\delta$ and uses
\begin{equation}
O\!\left(\mathcal L(v,Z,\log(1/\delta))\right)\text{ bits}
\label{eq:graph-confidence-main}
\end{equation}
throughout every legal update history, including intermediate prefixes with historical deficit larger than $Z$.
It therefore has optimal space complexity $\Theta(\mathcal L(v,Z,\log(1/\delta)))$ under the report-time promise and under the stronger total-deletion restriction.
\end{enumerate}
The constants are absolute, and both conclusions hold for both peak-space measures.
\end{theorem}

\noindent\emph{Note.} The same space lower bound holds even when the stream is restricted to at most $Z$ deletion operations in total.

The randomized algorithm is presented in Sections~\ref{sec:forest-budget-sensitive-banks} and~\ref{sec:budget-upper-bound}; the lower bounds are proved in Sections~\ref{sec:indexed-block} and~\ref{sec:nearlinear-entropy-lower}.

\paragraph{The zero-error endpoint.}
The same input promise also admits an optimal deterministic algorithm.

\begin{theorem}[Optimal zero-error space under a historical-deficit promise]
\label{thm:zero-error-lower}
\label{thm:zero-error-space}
Let $v\ge2$ and let the integer $Z\ge0$ be given in advance.
Consider legal dynamic simple-graph streams with historical deficit at most $Z$ whenever a spanning forest is requested.
\begin{enumerate}
\item \emph{Space lower bound.} Every zero-error algorithm requires
\[
\Omega\!\left(\min\left\{v^2,\,v(Z+1)\log\!\left(1+\frac{v}{Z+1}\right)\right\}\right)\text{ bits}.
\]
The lower bound needs only one report and already holds for streams with at most $Z$ deletion operations in total.
It includes randomized zero-error algorithms that return a correct answer in finite time almost surely.
\item \emph{Space-optimal algorithm.} Our deterministic one-pass algorithm always returns a correct spanning forest at every promised report and uses
\begin{equation}
O\!\left(\min\left\{v^2,\,v(Z+1)\log\!\left(1+\frac{v}{Z+1}\right)\right\}\right)\text{ bits}.
\label{eq:graph-simple-exact}
\end{equation}
Its space guarantee holds throughout the stream, and it supports every finite family of reports without an advance bound on their number.
\end{enumerate}
Consequently, under both the report-time promise and the stronger total-deletion restriction, the optimal space complexity among all zero-error algorithms is
\[
\Theta\!\left(\min\left\{v^2,\,v(Z+1)\log\!\left(1+\frac{v}{Z+1}\right)\right\}\right)\text{ bits}.
\]
This conclusion holds for both peak-space measures.
\end{theorem}

Section~\ref{sec:simple-exact} gives the deterministic algorithm and the zero-error lower bound.

To our knowledge, this is the first characterization up to constant factors in bits over the full budget--confidence range, including zero error and the classical bounded-deletion subclass.

At any fixed $\delta\in(0,1/4]$, the optimal space simplifies to
\begin{equation}
\Theta\!\left(v\log v\,\log^2\!\bigl(\min\{Z+2,v\}\bigr)\right).
\label{eq:graph-randomized-main}
\end{equation}
A constant historical-deficit budget retains the insertion-only $\Theta(v\log v)$ space order, whereas $Z=v^\alpha$ for any fixed $\alpha>0$ already has the classical fully dynamic order.
For $Z=\lceil\log(v+1)\rceil$ and $\delta=1/4$, the optimal space changes from
\[
 \underbrace{\Theta(v\log^3v)}_{\text{no deletion promise}}
 \quad\text{to}\quad
 \underbrace{\Theta\!\left(v\log v(\log\log v)^2\right)}_{\text{historical deficit at most }Z\text{ at the report}}.
\]
Thus the promise saves a factor $\Theta((\log v/\log\log v)^2)$.
At the same logarithmic budget, requiring failure probability at most $v^{-c}$ for any fixed $c>0$ raises the optimum to $\Theta(v\log^2v)$ bits (Lemma~\ref{lem:familiar-regimes}).

For the one-report task, when $1+\log(1/\delta)\ge\min\{Z+1,v\}$, the optimal space has the zero-error order, and the deterministic algorithm remains space-optimal.
Corollary~\ref{cor:budget-saturation} gives the exact asymptotic criteria for when positive error saves space and when it has the zero-error space order.

\paragraph{Limits of current-graph linear sketches.}
The algorithm of Theorem~\ref{thm:historical-deficit-main} attains the general optimum by retaining a compressed certificate of historically inserted edges.
An additional lower bound shows that algorithms retaining only graph-independent setup and a linear sketch of current adjacency require a polynomial factor more space in some parameter ranges.
Both classes solve the same one-report task with the same historical-deficit promise and failure probability.
A \emph{standard retained-linear algorithm}, as defined in Section~\ref{sec:retained-state-model}, chooses a finite field $F$ from its input parameters and samples graph-independent setup data $\rho$ before the stream.
At every event boundary, its entire persistent state, including its encoding length, is an effective canonical encoding of $(\rho,M_\rho x)$, where $x$ is the current adjacency vector and $M_\rho:F^{\binom v2}\to F^{s_\rho}$ is linear with columns computable from the stored description.
The report distribution depends only on this pair and fresh random bits; the decoder may be nonlinear.

\begin{theorem}[Space limitation of retained-linear algorithms under the same deficit promise]
\label{thm:intro-polynomial-linear-gap}
Fix constants $0<\alpha<\beta<1$.
For every sufficiently large integer $v$, set $Z=\lfloor v^\alpha\rfloor$ and $\delta=2^{-v^\beta}$.
Consider one-pass spanning-forest recovery with one report and failure probability at most $\delta$ on every fixed legal simple-graph stream whose historical deficit is at most $Z$ at that report.
Intermediate deficits and the total number of deletions and restorations are unrestricted.
\begin{enumerate}
\item Theorem~\ref{thm:historical-deficit-main} gives a space lower bound of $\Omega_{\alpha,\beta}(v^{1+\alpha}\log v)$ bits for every general algorithm, matched by its history-retaining algorithm using $O_{\alpha,\beta}(v^{1+\alpha}\log v)$ bits.
\item Every standard retained-linear algorithm requires $\Omega_{\alpha,\beta}(v^{1+\beta}\log^2v)$ bits. The algorithm in Theorem~\ref{thm:unrestricted-space} belongs to this class and uses $O_{\alpha,\beta}(v^{1+\beta}\log^2v)$ bits, so this is the optimal space order within the restricted class.
\end{enumerate}
Both lower bounds already hold for streams with at most $Z$ deletion operations in total, and both algorithms respect their space bounds throughout every legal update history.
Consequently, the optimal space within the restricted class is $\Theta_{\alpha,\beta}(v^{\beta-\alpha}\log v)$ times the optimal space among general algorithms.
The statements hold for both peak-space measures, and the hidden constants depend only on $\alpha$ and $\beta$.
The proof is given in Section~\ref{sec:retained-linear-limitations}, culminating in Corollary~\ref{cor:polynomial-linear-gap}.
\end{theorem}

For example, $\alpha=1/4$ and $\beta=1/2$ give $Z=\lfloor v^{1/4}\rfloor$ and $\delta=2^{-\sqrt v}$.
Our history-retaining algorithm uses $O(v^{5/4}\log v)$ bits, whereas every algorithm in the restricted class needs $\Omega(v^{3/2}\log^2v)$ bits under the same requirements.
Since both bounds are attainable in their respective classes, the separation factor is $\Theta(v^{1/4}\log v)$.
Algorithms that retain extra historical state alongside a linear sketch are general algorithms and are not subject to the retained-linear restriction.

\subsection{Extension to jointly correct reports}

We next strengthen the output requirement for the same historical-deficit model.
A report is an output request made while the same one-pass algorithm continues processing the same stream; it asks for a spanning forest of the graph that is current at that moment.
Let the integer $Q\ge1$ be an advance upper bound on the number of such requests.
Together with $v$, $Z$, and the target error $\delta$, this bound is available to the algorithm at setup.
The update sequence and all report locations are fixed independently of the algorithm's random coins.
At each report the historical deficit is at most $Z$; intermediate deficits and the cumulative deletion count remain unrestricted.
The joint correctness requirement is
\[
 \Pr[\text{every requested forest is returned correctly in finite time}]
 \ge1-\delta.
\]
This is stronger than requiring each report separately to have failure probability at most $\delta$.
For the latter, marginal guarantee, the same single-report state with error parameter $\delta$ can be reused at every fixed report location, without knowing the number of requests in advance.
Joint correctness is an optional stronger requirement, and the report-count parameter measures its additional space cost.

Simultaneous guarantees over many outputs are standard in streaming: strong tracking asks for accurate answers at every prefix~\cite{BlasiokDingNelson2017Monitoring}, and explicit query budgets also appear in the study of repeated streaming queries~\cite{HassidimEtAl2020Robust}.
For dynamic connectivity, Gibb et al. guarantee correctness of all connectivity queries with high probability over polynomial-length update sequences~\cite{GibbKapronKingThorn2015}.
Here we separate the number of requested forests from the number of updates, which can be arbitrarily large between reports.
Our input model remains oblivious: allowing an algorithm to retain earlier answers does not allow the update sequence or report locations to depend on those answers.

Reusing our read-only single-report algorithm with error parameter $\delta/Q$ and applying a union bound gives joint correctness directly.
The main question is whether a different algorithm can use less space.
Joint correctness does not require each report to have failure probability at most $\delta/Q$: errors across reports may be correlated, and the algorithm may retain information acquired during earlier updates and reports.
We prove a matching space lower bound that permits this information to be retained and used, establishing that the space cost of the standard $\delta/Q$ allocation is optimal up to constant factors.

\begin{theorem}[Optimal space for jointly correct reports]
\label{thm:joint-report-lower}
\label{thm:joint-report-space}
Let $v\ge2$, let integers $Z\ge0$ and $Q\ge1$ be given in advance, and let $0<\delta\le1/4$.
Consider fixed legal dynamic simple-graph streams with at most $Q$ report requests and historical deficit at most $Z$ at each report.
\begin{enumerate}
\item \emph{Space lower bound.} Every algorithm making all requested forests correct with probability at least $1-\delta$ requires
\[
\Omega\!\left(\mathcal L(v,Z,\log(Q/\delta))\right)\text{ bits},
\]
even when it retains and uses information from earlier reports.
This lower bound already holds on streams whose historical deficit is at most $Z$ at every prefix.
\item \emph{Space-optimal algorithm by reuse.} Running the one-pass randomized algorithm of Theorem~\ref{thm:historical-deficit-main} with error parameter $\delta/Q$ makes all requested forests correct with probability at least $1-\delta$ and uses
\[
O\!\left(\mathcal L(v,Z,\log(Q/\delta))\right)\text{ bits}
\]
throughout the stream, including intermediate prefixes that violate the report-time promise.
It therefore has optimal space complexity $\Theta(\mathcal L(v,Z,\log(Q/\delta)))$.
\end{enumerate}
The constants are absolute, and both conclusions hold for both peak-space measures.
\end{theorem}

At $Q=1$, this theorem recovers the one-report historical-deficit characterization rather than defining a separate input model.
For $Q>1$, restored queries can reuse the same historical edges without increasing the deficit budget; their cumulative deletion count can grow with $Q$.
The resulting dependence on $Q$ therefore requires a separate lower-bound argument that accounts for information learned across reports.

For example, fix $Z=\lceil\log(v+1)\rceil$ and joint failure probability $\delta=1/4$.
Increasing the report allowance changes the optimal space from
\[
 \underbrace{\Theta\!\left(v\log v(\log\log v)^2\right)}_{Q=1}
 \quad\text{to}\quad
 \underbrace{\Theta(v\log^2v)}_{Q=v}.
\]
The memory requirement grows by a factor $\Theta(\log v/(\log\log v)^2)$ although the joint reliability target is unchanged, and the larger bound remains necessary even when earlier answers are retained.
It already matches the zero-error space order.
More generally, when $1+\log(Q/\delta)\ge\min\{Z+1,v\}$, the optimal space has the zero-error order in Theorem~\ref{thm:zero-error-space}.
If a single algorithm must handle every finite number of promised reports without an advance report-count bound, this same space order is necessary even at any fixed $0<\delta\le1/4$; the deterministic algorithm attains it.
Section~\ref{sec:restorations-reports} proves the joint lower bound and this saturation statement.

\subsection{Model conventions and organization}
\label{sec:model-organization}

All space bounds measure total peak memory in bits, including setup data, stored randomness, retained state, decoding workspace, and output buffers.
The algorithms stay within their stated memory limits on every random tape.
The matching lower bounds hold both for worst-case peak space and for worst-input expected peak space, with expectation taken over the algorithm's private randomness without conditioning on success.
After reading the input parameters, the algorithms use polynomially many bit operations in $v$ for setup, each update, and each report.
The linearity claim in Theorem~\ref{thm:unrestricted-space} is conditional on the stored setup data.
The space lower bounds apply to arbitrary nonlinear algorithms with unrestricted computation; the separate linear-sketch comparison explicitly restricts the retained state.
Stream length is unrestricted, and zero error requires each requested correct answer to arrive in finite time almost surely.
Appendix~\ref{app:machine-conventions} gives the full conventions.

Section~\ref{sec:technical-overview} explains the obstacles and the ideas that yield the matching bounds.
Section~\ref{sec:unrestricted-proof} presents the unrestricted one-report algorithm, analyzes its space, and proves a matching lower bound against arbitrary nonlinear algorithms.
Section~\ref{sec:deletion-budget-proof} develops the one-report historical-deficit characterization, including the space-optimal deterministic algorithm and lower bounds already valid for the bounded-deletion subclass.
Section~\ref{sec:restorations-reports} determines the additional cost of joint correctness across reports and proves the matching lower bound allowing retained answers.
Section~\ref{sec:retained-linear-limitations} proves the additional lower bound for retained-linear algorithms, showing why this class cannot match the optimal historical-deficit algorithm in some parameter ranges.
The appendices supply the full model, auxiliary information and probability arguments, algebraic constructions, and implementation details.
Appendix~\ref{sec:extended-related-work} gives further comparisons with related work.

\section{Technical Overview: Algorithm Design and Matching Space Lower Bounds}
\label{sec:technical-overview}

The overview begins with the unrestricted one-report problem and develops its basic recovery mechanism and space lower bound.
Historical information then makes deletion promises useful, while repeated reports introduce a further difficulty: earlier answers can teach the algorithm about the same source.
Throughout this overview, $A=\lceil\log(v+1)\rceil$, $t=1+\log(1/\delta)$, and $\tau_0=\lceil\log(64/\delta)\rceil=\Theta(t)$.

\subsection{Reducing row width, levels, and banks together}
\label{sec:graph-upper-main}

Classical graph sketches recover an edge leaving a component by combining vertex summaries so that internal edges cancel~\cite{AGM2012}.
At a suitable sampling level, a nonempty cut has a small nonempty sample that can be recovered.
Repeated cut recovery merges the components into a spanning forest.
Each independent \emph{bank} stores one copy of the recovery data, with fresh sampling randomness and nested sampling levels.
The main stored array has one recovery row per vertex, sampling level, and bank.

Replacing the sparse-recovery code alone cannot reduce space to the stated level.
For $b\le v$, encoding $b$ edge names from a $\Theta(v^2)$-element universe still costs $\Theta(b\log(v^2/b))=\Theta(bA)$ bits, even with an entropy-optimal code.
A narrower row would also remain replicated over the old sampling and merging schedules.
The algorithm therefore changes the encoded universe and the schedule together: neighbor encoding reduces row width, a sampled-universe bitmap shortens the hierarchy, and an initial neighbor-recovery step reduces the number of components needing repair.
The two logarithms in the final bound count levels and banks after these costs have been balanced.

\subsection{Algorithm design: vertex-domain recovery and shorter schedules}

\paragraph{Recovering a cut through its vertices.}
The algorithm encodes a vertex's neighbors over the $v$-vertex universe, rather than naming edges from the $\binom{v}{2}$ possible pairs.
Let $H$ be a binary syndrome matrix that recovers vectors supported on sufficiently few vertices.
If $X$ is the current symmetric adjacency matrix, vertex $u$ retains $y_u=HXe_u$; write $h_u=He_u$.
Insertion and deletion both toggle the two endpoint syndromes.
For a vertex set $C$, let $I_C$ be its diagonal indicator.
The crossing-edge adjacency is $X_C=I_CX+XI_C$ over $\mathbb F_2$, and
\begin{equation}
 \sum_{u\in C}(h_u y_u^{\mathsf T}+y_u h_u^{\mathsf T})
       =HX_CH^{\mathsf T}.
 \label{eq:graph-bilinear-main}
\end{equation}
Thus internal edges cancel while the remaining matrix can be decoded through vertex coordinates on both sides of an edge.

Consider triangles on $\{1,2,3\}$ and $\{4,5,6\}$, joined by $\{1,4\},\{2,4\},\{3,5\}$, and take $C=\{1,2,3\}$.
The internal edge $\{1,2\}$ contributes $h_1h_2^{\mathsf T}+h_2h_1^{\mathsf T}$ twice and cancels.
The remaining sum is
\[
 (h_1+h_2)h_4^{\mathsf T}+h_4(h_1+h_2)^{\mathsf T}
       +h_3h_5^{\mathsf T}+h_5h_3^{\mathsf T}.
\]
Writing $B=HX_C$, this matrix is $BH^{\mathsf T}$ and is decoded in two passes.
First, decoding its rows after transposition recovers $B$, whose nonzero columns are $h_4,h_4,h_5,h_1+h_2,h_3$.
Second, decoding those columns returns $\{4\},\{4\},\{5\},\{1,2\},\{3\}$, respectively, recovering all three crossing edges, including the two incident to vertex $4$.

A cut with $O(b)$ edges has only $O(b)$ endpoints.
This permits $O(b\log(1+v/b))$ bits per neighbor syndrome, with one deterministic code shared by all levels and banks.
The shared matrix $H$ and serial decoding workspace fit $O(vb\log(1+v/b))$ bits.
Realizing this bound requires charging the code's construction and decoder, not just its output length.
Section~\ref{sec:unrestricted-cut-interface} gives the precise interface and cut decoder.
Construction takes polynomial time, and decoding does so even on syndromes with no sparse preimage.

\paragraph{Shortening the schedule.}
\label{sec:graph-bootstrap-branch}
At sampling rate $\Theta(b/v)$, a bank stores one live bit for each selected edge name in the entire universe.
The sampling seed fixes the bitmap's coordinate set independently of the graph, and selected positions can be recomputed without storing edge identifiers.
This top representation uses $O(vb)$ bits and reduces the number of retained sampling levels to $O(\log(2+v/b))$.

During the bootstrap phase, singleton queries recover $\min\{b,\deg(u)\}$ neighbors at each vertex with the required confidence.
A spanning forest of their union has at most $O(v/b)$ \emph{unfinished components}, meaning components that still have an outgoing edge in the current graph: any such component must contain more than $b$ vertices.
Fresh banks then process these unfinished components.
Section~\ref{sec:forest-bitmap-bootstrap} proves this invariant and gives the full algorithm, including the bitmap overflow cap and the error allocation.

\paragraph{Verifying recovered cuts.}
The per-vertex row combines a neighbor syndrome with an independent $O(A+\tau_0)$-bit verification fingerprint.
Verification must not bias the sampling analysis: first analyze an ideal execution with exact candidate checks, then bound the probability that independent fingerprints change any answer on that trace.
This also allows later cuts to depend on earlier recovered edges.

\paragraph{Balancing the threshold.}
For the sketching algorithm, let $r_*=\log(2+v/(A+\tau_0))$.
Choosing $b$ at scale $\max\{8,(A+\tau_0)/r_*\}$ makes the row width $O(b\log(1+v/b)+A+\tau_0)=O(A+\tau_0)$ and makes both the level count and total bank count $O(r_*)$.
Their product yields $O(v(A+\tau_0)r_*^2)$ space, while storing the adjacency matrix gives the quadratic cap.
With this choice, the algorithm uses the space stated in Theorem~\ref{thm:unrestricted-space}.
The space accounting and optimization are in Section~\ref{sec:compiler}.
Conditional on setup, neighbor rows, top bits, and binary-expanded fingerprints are linear in adjacency; report-time forests are temporary and discarded afterward.

\subsection{Matching space lower bounds for arbitrary algorithms}
\label{sec:graph-lower-main}

The lower-bound source is a bipartite graph with $\Theta(v)$ independent neighborhoods, each an $m$-subset of a $\Theta(v)$-vertex column universe.
Deleting a proper subset $T$ of one neighborhood $S$ leaves that row with a surviving edge.
Every valid spanning forest reveals such an edge, so the saved state lets a decoder find an element of $S\setminus T$.
This is the subset version of the universal relation problem~\cite{KNPWWY2017,NelsonYu2019}; the state may be entirely nonlinear, since only its information content matters.

The forest guarantee holds for every fixed source and query.
Lemma~\ref{lem:pointwise-source} converts this premise into a lossless source encoding with information cost $\Omega(L\log(2+m/L)\log(U/m))$, where $L=\ln(1/\beta)$ for error $\beta$ and $m\ge L$.
The column-universe size satisfies $U\ge Cm$ for an absolute constant $C$.
The factor $L\log(2+m/L)$ reflects the expected number of successful stages of this encoding, each saving approximately $\log(U/m)$ bits by recovering one source member.
These stages belong to the information-theoretic encoding and are distinct from the algorithm's recovery rounds.
By the information chain rule, the saved state must account for all source rows, although each query deletes edges from only one.

In the high-confidence range, neighborhoods of size about $\sqrt{vt}$ balance the two entropy factors, making each logarithm of order $\log(2+v/t)$ up to constant cutoffs.
The published constant-success lower bound covers smaller $t$~\cite{NelsonYu2019}; capping the confidence used by the reduction gives the quadratic endpoint.
Proposition~\ref{prop:direct-unrestricted} combines these ranges and accounts for the original expected peak space through a finite simulation of the saved state.

\subsection{Historical-deficit algorithms and matching space lower bounds}
\label{sec:graph-reserves-main}

For the historical-deficit analysis, we use the following profile for integers $Z\ge0$ and real $z\ge1$. It retains the integer parameter $A$ and is equivalent up to constant factors to $\mathcal L$ in \eqref{eq:intro-space-profile}:
\begin{equation}
\begin{gathered}
 B_Z=\min\{A^2,\log^2(Z+2)\},\qquad
 \mu=\min\{Z+1,v\},\qquad x_z=\min\{\mu,\sqrt{vz}\},\\
 \mathcal L'(v,Z,z)=vAB_Z+
 vz\log(1+x_z/z)\log(1+v/x_z).
\end{gathered}
\label{eq:analysis-frontier}
\end{equation}

Our algorithm exploits an advance bound $Z$ on the number of currently absent historical edges by maintaining compressed information about past connectivity.
The promise is imposed only at the report; intermediate deficits and the total number of deletions are unrestricted.
The algorithm adapts sparse connectivity certificates to retain historical edge names~\cite{NagamochiIbaraki1992,JinKapralovMahabadiVakilian2024}.
For threshold $q$, it maintains a monotone historical edge set $S$, accepting a new name exactly when its endpoints have minimum cut below $q$ in $S$.
Previously rejected names remain rejected on reinsertion because historical connectivity only increases.
At stream prefix $w$, one live bit per accepted edge identifies the currently present subset $W=S\cap E(w)$.

The historical edge set $S$ has at most $q(v-1)$ edges.
Each unfinished component of $W$ has at least $q$ stored names across its historical cut, all currently absent.
Since one absent name can be charged at most twice, the number of components of $W$ exceeds that of the current graph by at most $2d/q$, where $d=|E^{\rm ever}\setminus E|$.
The same cut argument shows that $d<q$ makes $W$ span the current graph.
A compressed representation and polynomial-time cut tests fit $O(vq\log(1+v/q))$ bits.
Choosing $q=Z+1$ for $Z<v$ yields a zero-error algorithm with the stated memory use; the adjacency bitmap covers larger $Z$.
The density and implementation bounds hold through arbitrary intermediate deficits and restoration cycles.
The matching space lower bound already holds for the subclass of streams with at most $Z$ deletion operations: its sources cannot share a prefix state because a common deletion query requires disjoint surviving edges.

For positive error, the algorithm combines historical state with independent randomized repair, an approach also used in bounded-deletion matching~\cite{KKD2025}.
Here, retaining a smaller historical edge set leaves only a bounded component excess to repair.
Maintaining historical edges and their live bits on sampled streams replaces the unrestricted algorithm's top bitmaps, and fresh banks complete the recovery using the same cut-decoding interface.
At a fixed report, the absent historical names form a fixed set of size at most $Z$, independently of the algorithm's coins.
The analysis of these sampled edge sets counts only the currently absent names; earlier deletions followed by restoration do not enter the count.
Historical sparsity holds throughout the stream, so intermediate violations of the report-time promise cannot increase the space cap.
This algorithm uses less space when $Z$ is small; the unrestricted algorithm remains available in the complementary regime.

To prove the stronger lower bound for the total-deletion subclass, partition source rows into blocks and query only one block, spending at most $k(m-1)\le Z$ deletions for $k$ rows of size $m$.
The information chain still charges every block.
To analyze a block, we reduce $k$ rows to one coordinate using the construction of Nelson and Yu~\cite[Lemma~3]{NelsonYu2019}.
Our information bound is measured under the original input law, although the auxiliary protocol is conditioned on joint success across the $k$ rows.
This yields the constant-error term $vAB_Z$.
A single-row query then captures the additional confidence cost without losing its error dependence.

The source-size optimization explains the scale $x_z$ in the main formula.
With $\mu=\min\{Z+1,v\}$, define
\begin{equation}
 \Gamma_{v,\mu}(z)=\max_{m\in[1,\mu]}
       z\log(1+m/z)\log(1+v/m),\qquad \Gamma_{v,\mu}(0)=0.
 \label{eq:graph-envelope-main}
\end{equation}
For $z\ge1$, its real maximizer is $m=\min\{\mu,\sqrt{vz}\}$: increasing the source size allows more encoding stages but reduces the information supplied by one recovered neighbor.
The classical square-root balance~\cite{KNPWWY2017,NelsonYu2019} persists until the total-deletion budget in the lower-bound subclass caps the source size.
Integer and graph-size restrictions change the bound only by absolute factors.
Section~\ref{sec:deletion-budget-proof} proves the invariants of the maintained edge sets, both lower bounds, and the matching algorithm choices.

\paragraph{Limits of current-graph linear sketches.}
An additional lower bound explains the advantage of this algorithm over the class retaining only graph-independent setup and a linear image of current adjacency.
Under this retained-state restriction, a communication decoder can subtract queried edges directly from the sketch; the resulting state has the same report law as an insertion-only presentation of the residual graph under the same advertised $Z$.
Section~\ref{sec:retained-linear-limitations} uses this virtual subtraction to prove that the restricted class needs a polynomial factor more space than the algorithm retaining historical edges in the regimes of Corollary~\ref{cor:polynomial-linear-gap}.

\subsection{Matching lower bounds for jointly correct reports}

For at most $Q$ fixed reports, each satisfying the same historical-deficit promise, running the single-report algorithm at error $\delta/Q$ makes the probability of any incorrect report at most $\delta$ by a union bound.
This standard extension reuses the same maintained state and its space cap throughout the stream.

The matching lower bound must allow earlier answers to teach the algorithm about later queries.
The reduction repeatedly queries one source graph and restores it after each report.
Deleting a proper subset of one row's neighborhood forces a forest to reveal a surviving neighbor; restoration returns the deficit to zero without introducing new historical names.
All queries are drawn in advance conditional on the source, independently of the algorithm's coins and answers.

Joint success identifies one fixed round with small first-failure probability, as in tracking lower bounds with independent inputs across rounds~\cite[Section~7 of the full version]{Blasiok2018}.
Here, globally conditioning on earlier success could change the source distribution and invalidate the information bound.
For each fixed source, the reduction conditions the boundary-state distribution on earlier success, then mixes these conditional distributions using the original source law.
The retained state still includes everything learned from earlier reports, and the next query remains fresh conditional on that source.
A uniform lower bound on sourcewise survival controls both the error and expected message length by constant factors.
The normalized one-query channel has error $O(\delta/Q)$, while its message length remains controlled by the original expected peak space.
Optimizing the source size yields the matching lower bound for joint correctness, including saturation at the zero-error space order.
Section~\ref{sec:restorations-reports} gives the finite query law, normalization, and saturation argument.

\section{An Optimal-Space Algorithm and Matching Lower Bound without Deletion Promises}
\label{sec:unrestricted-proof}

This section proves the unrestricted one-report lower bound and algorithm guarantees in Theorems~\ref{thm:unrestricted-space-lower} and~\ref{thm:unrestricted-space}.
The design of the spanning-forest algorithm combines small-cut recovery from vertex summaries with a reduction in the number of components before repair.
The analysis then selects its parameters and accounts for its space usage.
A matching space lower bound for arbitrary algorithms follows by encoding independent neighborhoods and querying one row at a time.
This section gives both arguments without assuming a deletion budget.

For the algorithm and its space analysis, $A=\lceil\log(v+1)\rceil$, $\tau_0=\lceil\log(64/\delta)\rceil$, and $U=A+\tau_0$.
We first analyze the sketching algorithm for a general recovery threshold $b\ge8$, then choose $b$ in Section~\ref{sec:compiler} after accounting for all memory.
This analysis assumes $\tau_0\le v$ and $b\le v/32768$; Section~\ref{sec:compiler} also justifies the exact-adjacency alternatives.
The cutoff corresponds to the codec's density threshold: its required recovery sparsity is $1024b$, which is then at most $v/32$.
All guarantees concern each fixed legal dynamic graph stream.
The space convention of Section~\ref{sec:model-organization} counts setup, retained state, and the peak update and report workspace.
Appendix~\ref{app:machine-conventions} gives the machine model and the finite-simulation lemma used by the lower bound.

\subsection{Cut recovery and verification in the algorithm}
\label{sec:unrestricted-cut-interface}

\paragraph{The binary-code interface.}
The first building block recovers every cut containing at most $K=512b$ edges from vertex-neighbor summaries.
For this value of $K$, Corollary~\ref{cor:binary-entropy-codec} supplies a deterministic binary matrix $H\in\mathbb F_2^{r\times v}$ with $r\le v$ and $r=O(bH_b)$, where $H_b=\log(1+v/b)$.
Every binary vector of weight at most $2K$ is recovered from its syndrome.
Construction, the stored matrix, and decoding use $O(vbH_b)$ bits of peak memory and polynomially many bit operations on every input.
On an arbitrary syndrome the decoder returns a vector passing its support and syndrome checks, or declares failure.
The same matrix serves all sampling levels and banks, because its guarantee is uniform over all sparse inputs, including inputs that depend on $H$.
The construction and its resource accounting appear in Appendix~\ref{sec:binary-codec-upgrade}, with the streamed constructor detailed in \S\ref{sec:binary-rr-constructor}.
Using the same matrix-existence guarantee, Appendix~\ref{sec:finite-search-alternatives} gives an exhaustive-search implementation with the same space order at the cost of exponential running time.

The following lemma constructs a two-sided cut measurement from stored vertex-neighbor syndromes and bounds the workspace for recovering its edges.
Its row-then-column decoder uses a two-stage algebraic structure also appearing in tensor-product recovery over the reals~\cite[Section~II-B]{HeJoseph2025Hierarchical}.

\begin{lemma}[Bilinear recovery of a sparse cut]
\label{lem:dense-bilinear-cut}
Let $1\le b\le v$, $K=512b$, and $H\in\mathbb F_2^{r\times v}$ with $r\le v$.
Suppose $H$ has a deterministic decoder that recovers every binary vector of weight at most $2K$ from its syndrome.
On every input, the decoder uses at most $S_{\rm dec}$ bits of workspace, takes polynomially many bit operations, and returns either a vector passing its support and syndrome checks or failure.
For a fixed simple graph $G$, retain at every vertex $u$ the $r$-bit vector
\[
 y_u=H\operatorname{adj}_G(u),
\]
where the adjacency column is a $v$-bit vector.
From this state, the stored matrix, and a given vertex cut $C$, a deterministic polynomial-time algorithm selects at most one candidate crossing-edge set of size at most $K$.
If the true cut has at most $K$ edges, the selected set is exactly that cut.
Beyond the retained state and matrix, the computation uses $O(vr+vA+S_{\rm dec})$ bits.
\end{lemma}
\begin{proof}
Let $X$ be the symmetric adjacency matrix of $G$, $I_C$ the diagonal indicator of $C$, and $h_u=He_u$ column $u$ of $H$.
Then $y_u=HXe_u$, and the crossing-edge adjacency matrix is $X_C=I_CX+XI_C$ over $\mathbb F_2$.
Consequently,
\begin{equation}
\begin{aligned}
M_C&=\sum_{u\in C}(h_uy_u^{\mathsf T}+y_uh_u^{\mathsf T})\\
   &=H(I_CX+XI_C)H^{\mathsf T}=HX_CH^{\mathsf T}.
\end{aligned}                                       \label{eq:forest-df4}
\end{equation}
An internal edge contributes twice and cancels, while a crossing edge contributes exactly its two symmetric outer products.

Use the sparse decoder on each of the $r$ rows of $M_C$ to obtain a row of a matrix $B\in\mathbb F_2^{r\times v}$; declare failure if a row has no sparse preimage.
Next decode each column of $B$ in turn.
Collect at most $2K$ ordered nonzero pairs from these column vectors, rejecting immediately on overflow.
Finally, check that the collected pairs contain no self-loops, occur in symmetric pairs, cross $C$, and represent at most $K$ unordered edges.
When the lemma is applied to a sampled graph, also require that each candidate name belongs to the sampled level.
If any check fails, select no candidate.
Otherwise, select this one edge set, without consulting a fingerprint.

When the true cut has at most $K$ edges, its endpoint support has size at most $2K$.
Every row of $HX_C$ is supported there, so decoding the rows of $M_C$ recovers $B=HX_C$.
Every column of $X_C$ has weight at most $K$, so decoding the columns then recovers the true cut, which passes all consistency checks.

The stored $M_C$ and $B$ use $r^2\le vr$ and $rv$ bits, respectively.
Only one decoded column is held at a time, and the ordered list costs $O(KA)=O(vA)$ bits because $K=512b$ and $b\le v$.
With serial decoder calls, the total additional workspace is $O(vr+vA+S_{\rm dec})$ bits.
There are at most $r+v$ calls, each on an input of polynomial length; all matrix operations and candidate checks are polynomial as well.
\end{proof}

The interface preceding the lemma meets these hypotheses with $r=O(bH_b)$ and $S_{\rm dec}=O(vbH_b)$.
The total cut-decoding space, including the stored matrix, vertex summaries, decoder scratch, candidates, and indices, is $O(vbH_b)$; the $O(vA)$ term fits this bound because $bH_b\ge\log(1+v)$.
The decoder's all-input time guarantee applies even when a dense cut has no sparse preimage, so such cuts cannot violate the time cap.

\paragraph{Independent verification.}
Structural checks alone cannot certify a candidate produced from a dense cut.
The algorithm therefore selects the entire candidate before consulting an independent fingerprint.
Write $N_E=\binom{v}{2}$ and identify edge names with $1,\ldots,N_E$.
Each bank chooses an independent uniform point $z$ in a binary extension field of degree
\begin{equation}
w=\tau_0+10A+20=O(A+\tau_0). \label{eq:forest-df5}
\end{equation}
At every vertex and retained level the algorithm stores $\sum_e z^e$ over incident present sampled names.
Internal edges cancel in the cut sum.
The selected candidate is tested once against this sum; a mismatch causes rejection without trying another candidate.
For distinct true and candidate crossing sets, their difference is a nonzero symmetric-difference polynomial of degree at most $N_E$.
Consequently, a false candidate passes with probability at most $N_E/2^w$.
The simple-graph promise keeps edge multiplicities in $\{0,1\}$, so no nonzero edge coefficient vanishes modulo two.

The shared field is constructed using the bounded random procedure of Lemma~\ref{lem:bounded-binary-field}, with degree and trial parameter
\begin{equation}
w=\tau_0+10A+20,\qquad a=\tau_0+10A+20.
                  \label{eq:forest-pu3}
\end{equation}
Corollary~\ref{cor:bounded-binary-cut-verification} bounds the construction scratch by $O(w+\log(a+1))$ bits and its running time by a polynomial on every tape.
Field-construction failure has probability at most $\delta/64$.
Conditional on successful construction, fresh independent $w$-bit strings are exactly uniform field points.
If construction fails, a retained flag makes the report procedure output an empty forest, with the same space and time caps.
The deterministic finite-search alternative is recorded in Appendix~\ref{sec:finite-core-alternatives}.

For the collision analysis, an ideal execution replaces fingerprint tests by exact equality tests.
Its cuts and candidates depend only on the fixed input, the deterministic code, and the sampling hashes; they are independent of the verification field and points.
If the ideal execution of this report makes at most $N_{\mathrm{test}}$ candidate tests, the polynomial root bound and a union bound give collision probability at most $N_{\mathrm{test}}N_E/2^w$, conditional on successful field construction.
In particular, $N_{\mathrm{test}}\le64v^3$ makes this probability at most $\delta/64$.
Section~\ref{sec:forest-bitmap-bootstrap} verifies this test bound after defining the sampling levels and query schedule.
On successful field construction, if no fingerprint collision occurs on the ideal trace, actual and ideal reports agree inductively, even when earlier returned edges determine later cuts.
This argument preserves the original law of the sampling hashes; it does not condition that law on successful fingerprint tests.

An insertion or deletion of $\{u,u'\}$ toggles $h_{u'}$ in $y_u$, $h_u$ in $y_{u'}$, and $z^{\operatorname{id}(\{u,u'\})}$ at both endpoints.
These updates maintain the summaries exactly through arbitrarily many legal deletions and reinsertions.

\subsection{A graph algorithm with fewer levels and rounds}
\label{sec:forest-bitmap-bootstrap}

The algorithm keeps an exact bitmap only for a sampled edge universe, while vertex syndromes handle shallower sampling levels.
At a report, it first recovers enough neighbors of each vertex to obtain large components and then repairs the remaining component boundaries.
Its parameters depend on $v$, $\delta$, and the chosen threshold $b$.

\par\medskip
\noindent\begin{minipage}{\linewidth}
\paragraph{Algorithm outline.}
The sketching branch has the following four steps.
\begin{enumerate}
\item \emph{Setup.} Construct the shared code and verification field, and initialize independent bootstrap and repair banks. Each bank stores its sampling seed, verification point, vertex rows, and top bitmap.
\item \emph{Stream update.} Toggle the two endpoint rows at every level sampling the edge and its top live bit when selected. Recompute selected-name ranks by universe scans, charging their indices and workspace.
\item \emph{Singleton bootstrap at a report.} Query each singleton in each bootstrap bank and insert returned edges into a temporary forest, discarding edges that would create cycles.
\item \emph{Component repair at the same report.} For each fresh repair bank, freeze the current components and try to recover an outgoing edge from each. Collect all proposals before merging; finally output the forest and discard report workspace.
\end{enumerate}
\end{minipage}
\par\medskip
The small hash field has degree $O(\log v)$, so deterministic irreducible-polynomial enumeration and hash evaluation are polynomial in $v$.

For every bank, Lemma~\ref{lem:sampling} supplies an independent $(2b+2)$-wise full-name hash into $h=\lceil\log(N_E+1)\rceil+8$ bits, using $O(bA)$ seed bits and polynomial construction and evaluation time.
A name belongs to level $j$ when its first $j$ hash bits are zero.
The bank uses depth
\begin{equation}
L=\max\{0,\lceil\log(v/(128b))\rceil\},\quad
 p=2^{-L},\quad
 C_{\rm top}=\min\{N_E,\lceil16N_Ep+2b\rceil\}.
                                                               \label{eq:forest-df13}
\end{equation}
The bank stores binary neighbor rows at levels $0,\ldots,L$.
During setup, enumerate the fixed universe of selected top-level names; if its size exceeds $C_{\rm top}$, permanently use an empty top graph.
Otherwise, retain one live bit per selected name, indexed by its position among the selected names in the fixed universe order.
To find this position, scan the original universe up to this name and count the names selected at level $L$; the list of edge names is never stored.
Thus the top graph always consists of present original edges and costs $O(vb)$ bits per bank.
Lemma~\ref{lem:sampling} bounds overflow by $16^{-2b}$, including the impossible-overflow case $C_{\rm top}=N_E$.

The algorithm's singleton bootstrap targets $\min\{b,\deg_G(u)\}$ distinct neighbors at each vertex to reduce the number of unfinished components before repair.
Related neighborhood-recovery and contraction steps appear in dynamic-stream spanner constructions~\cite{KapralovWoodruff2014}.
The bootstrap uses
\begin{equation}
\ell=\left\lceil\frac{\tau_0+A+2}{b-1}\right\rceil
                                                               \label{eq:forest-df14}
\end{equation}
independent banks.
In each bank every query remains the singleton $\{u\}$, even while returned edges are merged into a forest.
At each shallow level, apply Lemma~\ref{lem:dense-bilinear-cut} with $C=\{u\}$, followed by the independent verification procedure above, and interpret the accepted incident edges as candidate neighbors.
The top bitmap instead supplies present sampled neighbors directly.
Return $b$ distinct neighbors if its top bitmap or any accepted shallow candidate provides at least $b$; otherwise use its accepted level-zero list if that list has fewer than $b$ entries.
A bank may return fewer true neighbors on failure.
In the ideal execution, level zero recovers all neighbors whenever $\deg_G(u)\le K$.
For larger degree, a shallow sampling level or the top bitmap supplies at least $b$ neighbors except with probability at most $2^{1-b}$ for each fixed vertex and bank.
To obtain this bound, choose a shallow scale or top sample with mean at least $64b$.
The lower-tail calculation of Lemma~\ref{lem:sampling} gives
\begin{equation}
\Pr(X<b)\le2^{-b}.
                                                               \label{eq:forest-df15}
\end{equation}
Combining this lower-tail estimate with the shallow upper-tail bound, or with the bitmap overflow bound at the top level, gives the claimed per-vertex failure probability of at most $2^{1-b}$.

Independence of the $\ell$ banks and a union bound over the fixed singleton vertices bound the probability that some vertex has no successful bank by
\[
 v(2^{1-b})^\ell\le v2^{-\tau_0-A-2}\le2^{-\tau_0-2},
\]
since $v<2^A$.
The same ideal equality test covers all singleton queries, even when the union forest is updated between them.
When every vertex has a successful bank, every unfinished component of the union of returned edges has more than $b$ vertices.
Here a component is unfinished if it has an outgoing edge in the current graph $G$.
Indeed, if a component contains a vertex $u$ of degree at least $b$, it contains both $u$ and the $b$ distinct neighbors returned by a successful bank, so it has at least $b+1$ vertices.
If all its vertices have degree below $b$, a vertex with a true outgoing edge has a successful bank returning all its neighbors, also a contradiction.
Incremental cycle filtering stores just a forest of this union, with excess at most $v/(b+1)$, in $O(vA)$ bits; the union itself is never archived.

The repair phase uses fresh banks and freezes the current components at the beginning of each round.
Its parameters are
\[
 \lambda=\lceil\log(2+v/b)\rceil,\qquad
 R_2=8\lambda+2\lceil2\tau_0/b\rceil.
\]
If a cut's top-level sample has mean at least $64b$, its top bitmap misses the cut with probability at most $2^{-b}+16^{-2b}\le2^{1-b}$.
If the top-level mean is below $64b$, a nonempty cut has a retained shallow level whose sample is nonempty and $K$-sparse except with probability at most $2^{1-b}$.
Collect all proposals before making any merge.
The analysis of the fresh rounds uses the good/bad-round argument of Nelson and Yu~\cite[Appendix~A]{NelsonYu2019}, starting from the smaller excess established by the bootstrap.
If more than half of the unfinished components fail to propose an outgoing edge, call the round bad.
Markov's inequality bounds its conditional probability by $2^{2-b}$.
In a good round the proposal graph merges at least half as many components as there are successful proposals, so component excess decreases by a factor at most $3/4$.
Since the initial excess is at most $v/(b+1)<2^\lambda$ and $(3/4)^4<1/2$, $4\lambda$ good rounds suffice.
Failure requires at least $R_2/2$ bad rounds.
Repair sampling uses fresh randomness independent of the bootstrap, so conditioning on bootstrap success leaves its sampling law unchanged.
The conditional estimates and a union bound over the bad-round positions give
\[
 \Pr(\text{repair fails}\mid\text{bootstrap succeeds})
 \le 2^{R_2}(2^{2-b})^{R_2/2}
 \le2^{-bR_2/4}\le2^{-\tau_0}.
\]

To account for verification, put $J=\ell+R_2$.
Each bank handles at most $v$ singleton or frozen-component queries, and each queried cut selects at most one candidate per level, tested at most once.
The exact top bitmaps require no additional fingerprint tests.
Since $b\ge8$ and $\tau_0\le v$, the definitions give $\lambda\le A\le v$, $\ell\le v$, $R_2\le10v$, and $L+1\le A+4\le3v$.
Thus the ideal execution of a single report makes at most
\[
 N_{\mathrm{test}}\le J(L+1)v\le64v^3
\]
candidate tests, giving the $\delta/64$ collision allowance from Section~\ref{sec:unrestricted-cut-interface}.
Including verification and field construction, the total error bound is
\[
       2^{-\tau_0-2}+2^{-\tau_0}+\delta/64+\delta/64<\delta.
\]
All space and running-time caps hold on every tape, including the failure branches.

Before optimization, the algorithm uses peak space
\begin{equation}
O\left(v(L+1)\bigl[\lambda+(\tau_0+A)/b\bigr]
                 \bigl[bH_b+A+\tau_0\bigr]\right).                 \label{eq:forest-df16}
\end{equation}
This includes the stored dense matrix, every $O(vb)$ bitmap, the $O(JbA)$ hash seeds, all verification seeds, bilinear workspace, bootstrap and frozen-part arrays, universe-rank scans, and output.
Universe scans recompute ranks, so each top level retains only its $O(vb)$-bit live bitmap.
The bitmap remains exact after arbitrarily many simple updates because each selected name has one live bit.
\paragraph{Memory accounting.}
The following table records the retained arrays and the largest simultaneously needed workspace.
Temporary forests are maintained by cycle filtering, and report scratch is discarded before the next event.
Thus reports do not change the persistent graph state.
\begin{center}
\begin{tabularx}{\linewidth}{@{}>{\raggedright\arraybackslash}X l@{}}
\toprule
Component or workspace & Bits\\
\midrule
Neighbor rows and fingerprints, over all banks and levels
  & $O(Jv\lambda[bH_b+A+\tau_0])$\\
Top-universe bitmaps, over all banks
  & $O(Jvb)$\\
Sampling seeds and verification points, over all banks
  & $O(J[bA+A+\tau_0])$\\
Shared codec, its construction, and serial cut decoding
  & $O(vbH_b)$\\
Partitions, working forest and output; verification-field construction
  & $O(vA+A+\tau_0)$\\
\bottomrule
\end{tabularx}
\end{center}
The bilinear workspace fits the bound because $r^2\le vr$, and all candidate generation is serial.
Code setup and every decoding call are polynomial even on an unpromised input, and field construction has a fixed trial cap.
After saturating confidence at the exact-storage threshold, every retained parameter and loop count is polynomial in $v$.
Reading a longer parameter header adds its own input-reading time.

With the parameter choice justified next and the exact-adjacency alternative, the resulting algorithm uses peak space
\begin{equation}
O\left(\min\left\{v^2,
       v(A+\tau_0)\log^2\left(2+\frac v{A+\tau_0}\right)\right\}\right).
 \label{eq:forest-df17}
\end{equation}

\subsection{The optimized algorithm and its retained linear state}
\label{sec:compiler}

The following interface summarizes the preceding algorithm and allows its code implementation to be supplied independently.

\begin{lemma}[Vertex-syndrome compiler]
\label{lem:compiler}
Let $v\ge2$, $A=\lceil\log(v+1)\rceil$, and $\tau_0=\lceil\log(64/\delta)\rceil$ for $0<\delta\le1/4$, and suppose $\tau_0\le v$.
For $8\le b\le v/32768$, suppose a deterministic binary syndrome on $v$ coordinates recovers every $1024b$-sparse binary vector.
The syndrome has $r\le v$ rows with $r=O(bH_b)$, and uses $O(vbH_b)$ bits for construction, storage, and decoding, where $H_b=\log(1+v/b)$.
Its deterministic constructor must use polynomially many bit operations in $v$, with an absolute exponent.
Its decoder must have bounded polynomial work on \emph{every} input and either return a vector meeting its support and syndrome checks or fail.
Then one-report simple-graph spanning forests have a linear-sketch algorithm over $\mathbb F_2$ with peak space
\begin{equation}
O\!\left(v\lambda\left(\lambda+\frac{A+\tau_0}{b}\right)
                    \bigl[bH_b+A+\tau_0\bigr]\right),\qquad
 \lambda=\lceil\log(2+v/b)\rceil. \label{eq:vc1}
\end{equation}
The error is at most $\delta$ on every fixed legal dynamic graph stream, and running time is polynomial on every tape.
The algorithm uses $v$, $\delta$, and the chosen threshold $b$.
The range $\tau_0>v$ is handled by exact adjacency in the optimization below.
\end{lemma}
\begin{proof}
Sections~\ref{sec:unrestricted-cut-interface} and \ref{sec:forest-bitmap-bootstrap} give the algorithm for the stated free parameter $b$.
The two-pass decoder uses the hypothesized $1024b$-sparse code at cut threshold $512b$, and selects its candidate before verification.
The preceding error calculation is below $\delta$.
There are $O(\lambda)$ levels and $O(\lambda+(A+\tau_0)/b)$ banks; summing the resource table proves \eqref{eq:vc1}.
The code and field construction, universe scans, candidate checks, and output operations all have polynomial running time on every tape under the stated codec hypothesis.
Neighbor syndromes, top bits, and binary-expanded fingerprints are linear in current adjacency; setup and overflow flags are graph independent.
Proposition~\ref{prop:retained-binary-sketch} below records this map explicitly.
\end{proof}

\paragraph{Choosing the recovery threshold.}
The code of Corollary~\ref{cor:binary-entropy-codec} satisfies every hypothesis of Lemma~\ref{lem:compiler}, including setup space and all-input decoding time.
With $U=A+\tau_0$, choose
\begin{equation}
r_0=\lceil\log(2+v/U)\rceil,\qquad
 b=\max\{8,\lceil U/r_0\rceil\}. \label{eq:forest-df12}
\end{equation}
Then $L+1,\lambda,H_b=O(r_0)$, $U/b\le r_0$, $b\le8+U/r_0$, and $r_0=O(A)\le O(U)$.
Hence a row costs $bH_b+U=O(U)$ bits, and both the level count and bank count are $O(r_0)$.
Equations~\eqref{eq:forest-df16} and \eqref{eq:vc1} give $O(vUr_0^2)$ space.

The algorithm uses exact adjacency if $U>v$, if $b>v/32768$, if the fixed finite cutoff of the codec is not met, or if its full integer resource estimate favors exact storage.
The threshold $b>v/32768$ implies $vUr_0^2=\Omega(v^2)$ with an absolute constant.
Bounded small $v$ are absorbed by the same convention.
A code-table estimate reaching $v^2$ can therefore select adjacency before any code is constructed.
With these branches, the algorithm supports read-only reports and uses space
\begin{equation}
O\!\left(\min\left\{v^2,
     v(A+\tau_0)\log^2(2+v/(A+\tau_0))\right\}\right).
 \label{eq:vc2}
\end{equation}
The two savings are simultaneous: vertex-domain coding reduces a row to $O(U)$ bits at this threshold, and the bitmap and bootstrap reduce both scheduling counts to $O(r_0)$.
Keeping a $bA$-bit edge row or an $A$-scale repair depth would leave an additional $A/H_b$ factor when $H_b=o(A)$.

\paragraph{The persistent binary map.}
\label{sec:fully-dynamic-single-field-linear}
The following proposition makes explicit the retained-linear property of the algorithm attaining the space bound in Theorem~\ref{thm:unrestricted-space}: conditional on setup, its retained graph state is a binary linear image of current adjacency, as formalized in Section~\ref{sec:retained-state-model}.
For one report on an arbitrary fixed legal dynamic graph stream, put
\begin{equation}
A=\lceil\log(v+1)\rceil,\quad t=1+\log(1/\delta),\quad
 \mathcal S(v,t)=\min\left\{v^2,
 v(A+t)\log^2\left(2+\frac{v}{A+t}\right)\right\}.
 \label{eq:lf1}
\end{equation}
The setup depends on $v$ and $\delta$ alone, and the space bound includes its construction and decoding workspace.

\begin{proposition}[Optimal fixed-field retained sketch]
\label{prop:retained-binary-sketch}
The algorithm uses $O(\mathcal S(v,t))$ bits of peak space on every random tape.
Every algorithm meeting the same error requirement needs $\Omega(\mathcal S(v,t))$ bits, both for worst-case peak space and for worst-input expected peak space under the original coin law.
Thus the optimal space complexity is $\Theta(\mathcal S(v,t))$ in both measures, with absolute constants.
The algorithm has polynomial setup and per-event running time on every tape.
Conditional on its retained graph-independent setup $\rho$, every completed-event graph-dependent state is
\begin{equation}
M_\rho x\in\mathbb F_2^s,\qquad
 M_\rho:\mathbb F_2^{\binom{v}{2}}\longrightarrow\mathbb F_2^s
 \text{ linear},                                      \label{eq:lf2}
\end{equation}
where $x$ is current zero--one adjacency.
The dimension can be fixed by zero padding.
Both an insertion and a deletion toggle the corresponding column.
Reports are read only; their temporary forests and scratch are charged but not retained.
Every zero-error algorithm, including randomized algorithms, needs $\Omega(v^2)$ bits; a deterministic algorithm storing binary adjacency always returns a correct forest using $O(v^2)$ bits.
\end{proposition}

\begin{proof}
Lemma~\ref{lem:compiler} and its optimization establish the algorithm's stated memory use; Proposition~\ref{prop:direct-unrestricted} below proves the matching space lower bound.
The retained map, including failure branches, is specified below because Section~\ref{sec:retained-state-model} uses this exact state interface.
The parameters are
\begin{equation}
\tau_0=\lceil\log(64/\delta)\rceil,\quad U=A+\tau_0,\quad
 r_0=\lceil\log(2+v/U)\rceil,\quad
 b=\max\{8,\lceil U/r_0\rceil\},\quad H_b=\log(1+v/b).
 \label{eq:lf3}
\end{equation}
For the instantiated codec, $n=v$ and $k=2K=1024b$, where $K=512b$ is the cut threshold.
Corollary~\ref{cor:binary-entropy-codec} gives row width $O(b\log(ev/b))$ and a deterministic all-input checked decoder.
Both passes of cut recovery have binary sparse preimages, so no stronger field-input promise is required.
The deterministic code introduces no failure event, including for intermediate vectors that depend on its matrix.

Candidate-name lists fit the charged workspace.
If $4K\le v$, a nonzero kernel vector of weight at most $4K$ could be partitioned into two distinct binary vectors of weight at most $2K$ with the same syndrome, contradicting unique recovery.
Thus every set of at most $4K$ columns is linearly independent, and $r\ge4K$.
The $O(K\log v)$ list bits are absorbed by $O(vr+v\log v)$.
If $4K>v$, the identity or exact branch has $K=512b=O(v)$, so $O(v\log v)$ scratch suffices.
Exact adjacency may already be selected when $2K\ge v/32$.
The full code table and $O(vb\log(ev/b))$ decoder scratch are included in \eqref{eq:forest-df16}.

In the sketching branch, each bank's top depth and cap are
\begin{equation}
L=\max\{0,\lceil\log(v/(128b))\rceil\},\quad p=2^{-L},
 \qquad C_{\rm top}=\min\{N_E,\lceil16N_Ep+2b\rceil\}.
 \label{eq:lf4}
\end{equation}
The sampling seed determines overflow before any graph is processed.
An overflow flag selects a zero top block; otherwise every top bit is a fixed projection of current adjacency.
No selected-name list is retained.
For this retained-map implementation, the verification-field constants may be increased to
\begin{equation}
w_*=a_*=\tau_0+10A+40. \label{eq:lf5}
\end{equation}
Index the banks by $i$. Conditional on successful bounded field construction, each bank chooses an independent uniform $w_*$-bit point $z_i$.
At vertex $u$ and level $j$ its fingerprint is
\begin{equation}
Z_{u,i,j}=\sum_{e\ni u}\chi_{i,j}(e)x_e
                    z_i^{\operatorname{id}(e)}
             \in\mathbb F_{2^{w_*}}. \label{eq:lf6}
\end{equation}
In the fixed binary basis, these are $\mathbb F_2$-linear coordinates.
The ideal-trace argument above applies unchanged: field-search failure and fingerprint collisions are charged separately, and uniform points require no rejection sampler.
All failure flags are graph independent and select fixed zero blocks.

The neighbor syndromes, top bitmap, and expanded fingerprints concatenate to \eqref{eq:lf2}.
Reports discard partitions, proposals, and the working forest before the next event, leaving exactly $(\rho,M_\rho x)$ as persistent state.
The stored code tables and seeds generate columns of $M_\rho$ as needed.
Before optimization, the algorithm uses peak space
\begin{equation}
O\left(v(L+1)\bigl[\lambda+(\tau_0+A)/b\bigr]
                 \bigl[bH_b+A+\tau_0\bigr]\right),\qquad
 \lambda=\lceil\log(2+v/b)\rceil. \label{eq:lf7}
\end{equation}
It includes the full code tables, maximum decoder scratch, verification field and points, rank scans, hashes, partitions, and output.
The parameter relations above give $O(vUr_0^2)$; exact adjacency gives the cap in \eqref{eq:lf1}.
Confidence is read with saturation at $v$, so every field degree and loop count in the sketching branch is polynomial in $v$.
Reading the finite parameter headers is charged separately.
A zero-error algorithm also meets every positive-error guarantee.
The quadratic high-confidence lower bound of Proposition~\ref{prop:direct-unrestricted} therefore gives its $\Omega(v^2)$ lower bound, matched by exact adjacency.
\end{proof}

\subsection{Encoding a source under pointwise error}
\label{sec:pointwise-source}

For the lower bound, correctness on every fixed stream gives a strong premise: every fixed source and deletion query has small error.
The proof first converts this premise into an information bound, then embeds independent sources as graph neighborhoods.
Information is measured under the joint distribution of the source and the generated message, without conditioning on successful decoding.
Logarithms in information quantities are binary, while $\ln$ is natural.
The underlying one-way communication task is the subset version of universal relation: an encoder given a set $S$ sends one message to a decoder, which is given a proper subset $T\subsetneq S$ and must return an element of $S\setminus T$.
The argument adapts the universal-relation encoding of Kapralov et al.~\cite[Section~3.1]{KNPWWY2017} and its information formulation by Nelson and Yu~\cite[Appendix~B]{NelsonYu2019}, keeping the source cardinality explicit.

The lossless encoding reuses one message across a sequence of growing query subsets.
At each stage, it adds any successfully recovered element to the queried subset, then fills to the next prescribed size with unused members of $S$, ordered by an independent random permutation of the source universe.
The recovered elements influence later queries, whose distributions therefore need not remain uniform.
An adaptive-filling density estimate bounds this departure from uniformity.
The encoding argument below explains its role; Appendix~\ref{app:pointwise-source-proof} gives the full encoding proof and density calculation.

\begin{lemma}[Pointwise-error source encoding]
\label{lem:pointwise-source}
There are absolute constants $C,\beta_*>0$ with the following property.
Let $L=\ln(1/\beta)$, $0<\beta<\beta_*$, $m\ge L$, and $U\ge Cm$.
Let $S$ be uniform among the $m$-subsets of $[U]$.
A message $M$ is generated from $S$, independently of the query.
Suppose a decoder, given $M$ and any $T\subsetneq S$, outputs a member of $S\setminus T$ with error at most $\beta$ for every fixed pair $(S,T)$, where probability includes the message and fresh decoder coins.
Then
\begin{equation}
I(S;M)=\Omega\!\left(
 L\log(2+m/L)\log(U/m)\right).
 \label{eq:pe1}
\end{equation}
The decoder may be randomized, and message length is unrestricted.
The decoder is a total output channel: a failure symbol is permitted, but each simulated call terminates.
The common finite cutoff supplies this condition in the forest application.
\end{lemma}

\paragraph{Encoding argument.}
Reuse the message for $R=\Theta(L\log(2+m/L))$ queries whose sizes increase geometrically.
Each successful query reveals one new member of $S$; an independent random permutation fills the query to the next prescribed size.
Although the revealed members influence later queries, the adaptive-filling calculation bounds each query law, conditional on $S,M$, by $\beta^{-1/4}$ times the corresponding uniform-subset law.
Fresh decoder coins and the pointwise error premise then bound each stage's error by $\beta^{3/4}$, so a constant fraction of the $R$ stages succeed in expectation.

The message, one success flag per stage, the shared permutations and decoder tapes, and the residual set $S\setminus S_{\mathrm{rec}}$ reconstruct $S$ losslessly, where $S_{\mathrm{rec}}$ is the set of revealed members.
Reconstruction replays each successful output and fills from the remaining residual elements.
Elements revealed later could not have been chosen as earlier fillers, which makes the replay exact.
The flags cost $R$ bits, while each revealed member saves order $\log(U/m)$ bits in the residual description.
The resulting information bound is $I(S;M)\ge\E|S_{\mathrm{rec}}|\log((U-m)/m)-R$.
This proves the claimed order once $U/m$ is sufficiently large.
Appendix~\ref{app:pointwise-source-proof} gives the exact query sizes, rounding estimates, density bound, and lossless entropy calculation.

\subsection{The full-confidence spanning-forest space lower bound}
\label{sec:indexed-confidence}

The following graph reduction uses at most $Z$ deletion operations in its single query.
Consequently its hard streams satisfy the historical-deficit promise with budget $Z$.
The unrestricted application may take $Z$ large enough to impose no constraint; Section~\ref{sec:deletion-budget-proof} uses the same interface to prove lower bounds already under the stronger total-deletion restriction.
Only one row is queried, but the information chain rule charges every source row.

\begin{lemma}[Exact single-row interface]
\label{lem:exact-row-interface}
There are absolute $\beta_0,C_0>0$ such that, for every integer $Z\ge0$, every $0<\eta<\beta_0/2$, $L_\eta=\ln(1/(2\eta))$, and positive integers $N,U,m$ satisfying
\[
 m\ge L_\eta,\qquad U\ge C_0m,\qquad N+U\le v,
 \qquad m-1\le Z,
\]
every spanning-forest algorithm required to have error at most $\eta$ on all legal streams with at most $Z$ deletion operations requires space $s$ satisfying
\begin{equation}
s+1=\Omega\bigl(NL_\eta\log(2+m/L_\eta)\log(U/m)\bigr)
 \label{eq:ic1}
\end{equation}
for both peak-space measures, using one report and a simple graph.
\end{lemma}
\begin{proof}
Independently draw $N$ uniform $m$-subset source rows and insert their bipartite edges in a common prefix.
A query deletes only one row's subset; at most $m-1$ deletions suffice, and a spanning forest exposes an incident edge in that row's nonempty remaining set.
Use a common finite cutoff with extra error at most $\eta$ and let $M_0$ be the tagged original prefix state.
Its original-law entropy is $O(s+1)$.
For row $i$, view the simulation as a single-row encoder: given $S_i$, it privately samples the other independent rows and simulates the common insertion prefix to form $M_0$.
It sends $M_i=(S_{<i},M_0)$, revealing only the preceding rows alongside the saved state.
For every fixed pair $(S_i,T_i)$ its error, averaged over the other independent source rows, is at most $2\eta$.
This follows from correctness on each fixed stream and the common finite cutoff; the query remains independent of the prefix message given $S_i$.
Lemma~\ref{lem:pointwise-source} with $\beta=2\eta$ therefore bounds its external information $I(S_i;S_{<i},M_0)=I(S_i;M_0\mid S_{<i})$ from below by $\Omega(L_\eta\log(2+m/L_\eta)\log(U/m))$.
Summing the information lower bound by the chain rule bounds it by $I(S_1,\ldots,S_N;M_0)\le H(M_0)$, proving \eqref{eq:ic1}.
\end{proof}

\begin{proposition}[Matching space lower bound at every confidence level]
\label{prop:direct-unrestricted}
For spanning-forest output on fixed legal dynamic graph streams without a deletion or length promise, let $v\ge2$, $0<\delta\le1/4$, $A=\lceil\log(v+1)\rceil$, and $t=1+\log(1/\delta)$.
For any algorithm meeting these requirements, let $s$ denote either its worst-case peak space or its worst-input expected peak space.
Then
\begin{equation}
s+1=\Omega\!\left(\min\left\{v^2,
 v(A+t)\log^2\left(2+\frac v{A+t}\right)\right\}\right).
 \label{eq:du1}
\end{equation}
\end{proposition}
\begin{proof}
The constant-success lower bound of Nelson and Yu~\cite{NelsonYu2019} gives $s+1=\Omega(vA^3)$ for worst-case peak space.
It gives the same expected-peak lower bound: stop a simulation before its charged memory exceeds $16(s+1)$ bits.
For each fixed stream, Markov's inequality loses at most $1/16$ success probability.
The resulting hard-space algorithm has success at least $11/16>2/3$.
A memory-length guard uses $O(\log(s+2))$ additional bits, so the hard cap remains $O(s+1)$.
Applying the published theorem proves the claim with expectation taken over the original coin law.

The confidence term follows from a direct argument.
Let $C\ge4$ be a sufficiently large absolute constant for both Lemma~\ref{lem:pointwise-source} and Lemma~\ref{lem:exact-row-interface}.
Put
\begin{equation}
N=\lfloor v/2\rfloor,\qquad U=\lceil v/2\rceil,\qquad
 L=\min\left\{\ln\frac1{2\delta},\frac{U}{16C^2}\right\},
 \qquad \beta=e^{-L},\qquad
 m=\left\lceil\frac{\sqrt{UL}}{2C}\right\rceil.
 \label{eq:du2}
\end{equation}
First suppose $v$ and $t$ exceed sufficiently large absolute constants.
They may be chosen so that $U\ge16C^2$, $L\ge1$, and $\beta$ is below both lemmas' fixed thresholds.
Since $L\le U/(16C^2)$,
\begin{equation}
m\ge2L\ge L,\qquad
 m\le\frac{U}{8C^2}+1\le\frac{U}{4C^2}\le\frac UC,
 \qquad 2\delta\le\beta.
 \label{eq:du3}
\end{equation}
Thus all source-encoding hypotheses hold, even when the original confidence exponent is arbitrarily larger than $v$.

Apply Lemma~\ref{lem:exact-row-interface} with error $\eta=\beta/2\ge\delta$, so $L_\eta=L$, and the parameters $N,U,m$ in \eqref{eq:du2}.
Its source hypotheses follow from \eqref{eq:du3}, and $N+U=v$.
There is no deletion promise, so the required $m-1$ deletions are legal.
The lemma uses one row's query at a time and charges all $N$ independent source rows by the information chain rule, using a common finite cutoff and the tagged original state with entropy $O(s+1)$.
Hence, for both peak-space measures,
\begin{equation}
s+1=\Omega\!\left(
 vL\log(2+m/L)\log(U/m)\right).
 \label{eq:du4}
\end{equation}
Write $x=U/L\ge16C^2$.
Because the unrounded value in \eqref{eq:du2} is at least $2L\ge2$, rounding changes it by at most a constant factor.
Consequently,
\[
 \log(2+m/L)=\Theta(\log(2+x)),\qquad
 \log(U/m)=\Theta(\log(2+x)).
\]
The constants depend only on the fixed absolute $C$; the lower bound on $x$ keeps both logarithms positive.
Also $L=\Theta(\min\{t,v\})$.
Hence \eqref{eq:du4} gives
\begin{equation}
s+1=\Omega\!\left(
 v\min\{t,v\}
 \log^2\left(2+\frac{v}{\min\{t,v\}}\right)\right).
 \label{eq:du5}
\end{equation}
For bounded $t$, the required bound is absorbed by $vA^3$.
Bounded $v$ follows from the space-accounting convention.
When $t\le A$ the right-hand side of \eqref{eq:du1} is $O(vA^3)$; when $A\le t\le v$ it is $O(vt\log^2(2+v/t))$; and when $t\ge v$ it is at most $v^2$.
The baseline and \eqref{eq:du5} therefore prove \eqref{eq:du1} in all ranges.
\end{proof}

For the final simplification, $\log v\le A\le\log v+1$ and $\log(1/\delta)\ge2$ give
\[
\begin{aligned}
 \log(v/\delta)&\le A+t\le2\log(v/\delta),\\
 \log(v/\delta)&\le A+\tau_0\le4\log(v/\delta).
\end{aligned}
\]
For any $u>0$ and $c\ge1$, concavity and monotonicity of the logarithm imply
\[
 \frac1c\log(2+v/u)\le\log(2+v/(cu))\le\log(2+v/u).
\]
Thus replacing $u$ by $cu$ in $vu\log^2(2+v/u)$ changes its value by a factor between $1/c$ and $c$; taking the minimum with $v^2$ preserves this comparison.
The substitutions of $\log(v/\delta)$ for $A+t$ and $A+\tau_0$ therefore change the full space expressions only by absolute factors, absorbed by the $O$ and $\Omega$ notation.
The algorithm's memory use in \eqref{eq:vc2} and the space lower bound \eqref{eq:du1} prove Theorems~\ref{thm:unrestricted-space-lower} and~\ref{thm:unrestricted-space}, including both peak-space measures.
The lower bound uses legal edge deletions and applies to arbitrary nonlinear algorithms; the algorithm designed above uses the matching amount of memory with the retained binary map specified above.
The next consequence identifies exactly when its space advantage over explicit adjacency persists.

\begin{corollary}[The exact boundary of subquadratic space]
\label{cor:quadratic-boundary}
Consider any sequence $v\to\infty$ and errors $0<\delta(v)\le1/4$, with $t=1+\log(1/\delta)$.
The fully dynamic one-report optimal space complexity is $o(v^2)$ if and only if $t=o(v)$.
It is $\Theta(v^2)$ if and only if $t=\Omega(v)$.
These statements concern both peak-space measures in Theorem~\ref{thm:unrestricted-space}.
Oscillating sequences can satisfy neither alternative.
\end{corollary}
\begin{proof}
Put $u=(A+t)/v$ and $\phi(u)=\min\{1,u\log^2(2+1/u)\}$.
The normalized optimal space complexity is $\Theta(\phi(u))$ with absolute constants.
As $u\downarrow0$, $u\log^2(2+1/u)\to0$.
For every fixed $\epsilon>0$, $u\ge\epsilon$ gives $\phi(u)\ge\min\{1,\epsilon\}>0$.
Thus a vanishing ratio is equivalent to $u\to0$, whereas a uniform positive lower bound is equivalent to $\liminf u>0$: if $\liminf u=0$, select a subsequence tending to zero.
Since $A/v\to0$, these are respectively $t=o(v)$ and $t=\Omega(v)$.
Finally, $\log(1/\delta)\le t\le\tfrac32\log(1/\delta)$, so both criteria can equivalently use $\log(1/\delta)$ in place of $t$, as in the introduction.
\end{proof}

For example $t=v/f(v)$ with $f(v)\to\infty$ and $f(v)=o(v/A)$ has optimal space complexity $\Theta(v^2\log^2(2+f(v))/f(v))$.
Here $f$ may grow arbitrarily slowly, by Theorem~\ref{thm:unrestricted-space}.

The following section determines how far a report-time historical-deficit promise can reduce this space requirement.

\section{Optimal Spanning-Forest Space under a Historical-Deficit Bound}
\label{sec:deletion-budget-proof}
\label{sec:historical-deficit-proof}
\label{sec:snapshot-deficit}

At a fixed legal stream prefix $w$, write $E^{\rm ever}(w)$ for the edge names inserted at least once and $E(w)$ for the current edge set.
The historical deficit is
\begin{equation}
d(w)=|E^{\rm ever}(w)\setminus E(w)|.
\label{eq:sd1}
\end{equation}
The integer $Z\ge0$ is given in advance, and the single report is promised to satisfy $d(w)\le Z$.
The total number of deletions and restorations and all intermediate values of $d$ are unrestricted.
An edge deleted and then restored contributes zero to $d(w)$; the full set $E^{\rm ever}(w)$ is an analysis device and is never stored.
Update and report sequences are fixed independently of the algorithm's random coins.

We first design a deterministic algorithm that is correct at every promised report, and prove a matching lower bound for all zero-error algorithms.
For positive error, we combine the same compressed storage of historical edge names and live bits with the cut-recovery tools of Section~\ref{sec:unrestricted-proof}.
Both algorithms maintain their space bounds on every legal update history, even when the deficit exceeds $Z$ between reports.
The matching lower bounds use only the smaller class of streams with at most $Z$ deletion operations in total: an indexed-block argument gives the constant-error baseline, and a single-row argument retains the additional cost of confidence.
These algorithms for the full historical-deficit model and lower bounds for its bounded-deletion subclass establish Theorems~\ref{thm:deletion-budget-space} and~\ref{thm:zero-error-space}.
Throughout this section, $A=\lceil\log(v+1)\rceil$, $\mu=\min\{Z+1,v\}$, and $B_Z=\min\{A^2,\log^2(Z+2)\}$.
For positive failure probability, write $t=1+\log(1/\delta)$.
Write $c(G)$ for the number of connected components, including isolated vertices.
Use $N_E=\binom{v}{2}$ for the edge-universe size and $m$ for the size of one source neighborhood in the lower bounds.

\subsection{Maintaining historical connectivity in compressed space}

The number of currently absent historical edges limits how many stored historical edge names can be absent at a report.
The data structure adapts classical sparse connectivity certificates by maintaining a monotone set of historical edge names~\cite{NagamochiIbaraki1992,JinKapralovMahabadiVakilian2024}.
Its distinct edge names and live bits control the component excess caused by currently absent names; the implementation also accounts for compressed storage and update workspace.

\begin{lemma}[Maintaining historical edges and their live subset]
\label{lem:dense-compressed-certificate}
\label{lem:current-deficit-certificate}
For any integer $q\ge1$, a deterministic data structure processes any legal dynamic graph stream and maintains a set $S$ of historical edge names together with its live subset $W$, with the following properties.
Write $d_S=|S\setminus W|$ for the number of stored names currently absent.
\begin{equation}
\begin{gathered}
 |S[U]|\le q(|U|-1)
       \quad(\varnothing\ne U\subseteq[v]),\qquad
 W=S\cap E(G),\\
 c(W)-c(G)\le2d_S/q\le2d(w)/q,
 \end{gathered}                                                   \label{eq:forest-df6}
\end{equation}
These inequalities hold at every event boundary, since $S\setminus W\subseteq E^{\rm ever}(w)\setminus E(w)$.
If $d_S<q$, then $W$ spans $G$; in particular, this holds when $d(w)<q$.
The density invariant does not require any bound on $d(w)$ or on past deletions.
For $1\le q\le v+1$, its space, including its update and traversal scratch, is
\begin{equation}
O\bigl(vq\log(1+v/q)\bigr).                        \label{eq:forest-df7}
\end{equation}
The representation stores each accepted name once, together with its live bit.
For $1\le q\le v+1$, Lemma~\ref{lem:poly-historical-certificate} provides polynomial-time maintenance using this amount of memory.
\end{lemma}
\begin{proof}
The historical set $S$ is never reduced, and an insertion of a stored name only sets its live bit to one.
On insertion of a name $e=\{u,u'\}\notin S$, accept it if and only if the minimum $u$--$u'$ cut in the historical graph $([v],S)$ has size less than $q$; otherwise reject it.
Acceptance adds the name with a live bit equal to one, while a deletion clears a stored live bit and otherwise has no effect.
These operations preserve the live invariant because each edge has multiplicity zero or one throughout the stream.
Because all historical cuts only grow, rejection is permanent, and each accepted name is stored only once even after arbitrarily many reinsertion cycles.

Induct on accepted names for the density invariant.
If adding $e$ would first violate it, there is a set $U$ containing both endpoints with $|S[U]|=q(|U|-1)$ before insertion.
For any nontrivial $B\subset U$, the two smaller-set inequalities imply
\[
 |\partial_{S[U]} B|\ge
 q(|U|-1)-q(|B|-1)-q(|U\setminus B|-1)=q.
\]
Every global cut separating the endpoints consequently has at least $q$ historical edges, contradicting acceptance.
Thus $|S|\le q(v-1)$.

Let $C$ be a component of $W$ which is not a whole component of the current graph.
Some current edge crosses $C$.
Since $C$ is a component of $W=S\cap E(w)$, this edge cannot belong to $S$ and was therefore rejected when first inserted.
The historical set $S$ only grows, so the minimum cut between its endpoints in $([v],S)$ remains at least $q$.
It follows that $|\partial_S C|\ge q$.
Every such crossing historical name is absent from the current graph and belongs to $S\setminus W$.
Since $S\subseteq E^{\rm ever}(w)$ and $W=S\cap E(w)$, the number of absent stored names is at most $d(w)$.
Call a component of $W$ \emph{unfinished} when it has an outgoing edge in $G$.
If $N$ components are unfinished, summing their cuts gives $qN\le2d_S\le2d(w)$.
The component excess is at most $N$, proving \eqref{eq:forest-df6}.
If $d_S<q$, even one unfinished component would need at least $q$ distinct absent stored names on its boundary, which is impossible.

The set and its live bits are stored by combinatorial rank rather than by listing edge names.
With $C_q=\min\{N_E,q(v-1)\}$, the sparse branch uses
\[
 O\!\left(\log\binom{N_E}{C_q}+C_q+vA\right)
   =O\bigl(vq\log(1+v/q)\bigr)
\]
bits; a two-bit-per-name representation handles the dense branch.
Appendix~\ref{app:certificate-representation} verifies that rank updates, traversal, and all temporary copies fit within the same memory allowance.
The endpoint cut test fits the same cap by Lemma~\ref{lem:poly-historical-certificate}, whose implementation is given next.
\end{proof}

The graph invariant specifies which edge names the data structure accepts.
For efficient maintenance, integral flows directly on the compressed names implement this acceptance test.

\begin{lemma}[Polynomial-time operations on compressed historical edge storage]
\label{lem:poly-historical-certificate}
For every integer $1\le q\le v+1$, the data structure of \eqref{eq:forest-df6}, which stores each accepted edge name once, supports deterministic updates and traversals using polynomially many bit operations in $v$, on every legal dynamic graph stream, using space
\begin{equation}
O\bigl(vq\log(1+v/q)\bigr).
 \label{eq:forest-pu1}
\end{equation}
Its historical set $S$, current live subset $W$, density invariant, component-excess bound, and exact-spanning guarantee when $d_S<q$ are unchanged.
\end{lemma}
\begin{proof}[Implementation idea]
To test historical endpoint connectivity without constructing an explicit adjacency list, scan the compressed names to enumerate the residual arcs of a unit-capacity flow network, with two directed arcs per stored edge.
The transient flow uses two bits per edge, and the search uses $O(vA)$ bits for its queue, visited flags, and predecessor data.
These arrays, together with the compressed-name traversal workspace, fit the same space bound.
At most $v-1$ augmentations determine whether the endpoint cut is below $q$.
Appendix~\ref{app:certificate-polynomial} gives the residual-arc representation, exact acceptance test, and workspace accounting.
\end{proof}

\subsection{A space-optimal deterministic spanning-forest algorithm}
\label{sec:simple-exact}

Theorem~\ref{thm:zero-error-space} follows from a deterministic algorithm that always returns a correct forest at a promised report and a matching space lower bound that also applies to randomized zero-error algorithms.
The following statement also records an equivalent binomial expression for the space.

\begin{theorem}[Optimal zero-error space under a historical-deficit bound]
\label{thm:zero-error-implementation}
For every $v\ge2$ and advance integer $Z\ge0$, consider fixed legal dynamic graph streams whose report prefix $w$ satisfies $d(w)\le Z$.
The optimal space complexity among algorithms that return a correct spanning forest in finite time with probability one on every such stream is
\begin{equation}
\Theta\!\left(
  \min\left\{v^2,\,
    v(Z+1)\log\left(1+\frac v{Z+1}\right)\right\}\right)
 \label{eq:se1}
\end{equation}
bits, for both peak-space measures of Appendix~\ref{app:machine-conventions}.
Equivalently, this order is
\begin{equation}
\Theta\!\left(
  v\log\binom{v+\min\{Z,v\}+1}{v}\right).
 \label{eq:se2}
\end{equation}
The space lower bound applies to all zero-error algorithms, including randomized ones, and requires only one report after a common-length insertion prefix followed by at most $Z$ deletion operations in total.
The deterministic algorithm returns a correct spanning forest at every promised report, permits any finite number of such reports, and has polynomial time per event with all scratch and output charged.
Its space bound holds throughout every legal update history, without any intermediate deficit bound or limit on deletion and restoration operations.
In particular, $Z=0$ has order $v\log(v+1)$, and $Z\ge v$ has order $v^2$.
\end{theorem}

\begin{proof}
\emph{Algorithm and memory use.} For $Z\le v$, use the historical edge data structure of Lemma~\ref{lem:dense-compressed-certificate}, which stores each accepted name once, with $q=Z+1$.
At every promised report, $d_S\le d(w)\le Z<q$, so its live subset spans the current graph.
Scanning this subset and retaining a temporary forest gives a read-only report using space
\begin{equation}
 O\!\left(v(Z+1)\log(1+v/(Z+1))\right).
 \label{eq:forest-pu2}
\end{equation}
Lemma~\ref{lem:poly-historical-certificate} implements its acceptance test with bounded integral flow and polynomial work.
For $Z>v$, retain the $N_E$-bit adjacency bitmap and compute a spanning forest by a read-only scan at each report.
Its $O(v\log(v+1))$ traversal, component, and output scratch fits $O(v^2)$ bits.
The deficit header can be saturated at the branch threshold, so a large $Z$ does not require retaining its full description.
Both branches have polynomial time per event and a deterministic hard space cap.
For the branch that stores historical edges, the monotone historical set always has at most $q(v-1)$ names, regardless of the number of past restorations or the current deficit.
Report workspace is discarded after output, leaving the stored edge names and live bits, or the bitmap, unchanged.

\emph{Lower bound.} For $v\ge4$, put $N=\lfloor v/2\rfloor$, $U=\lceil v/2\rceil$, and $m=\min\{Z+1,\lfloor U/2\rfloor\}\ge1$.
Use $N$ row vertices and $U$ column vertices.
Independently choose an arbitrary $m$-subset $S_i\subseteq[U]$ at each row, and insert the corresponding bipartite edges in canonical order.
This gives
\[
 M=\binom{U}{m}^N
\]
sources, all of length $Nm$.

For any two different sources choose a row $i$ where the sets $S_i,S'_i$ differ.
Append the common continuation that deletes the edges of $S_i\cap S'_i$ at this row and requests a report.
Both streams are legal, and the continuation uses
\[
 |S_i\cap S'_i|\le m-1\le Z
\]
deletions in total, and therefore has historical deficit at most $Z$ at every prefix.
The two residual incident sets at row $i$ are nonempty and disjoint.
Every spanning forest must contain an incident edge at this nonisolated row, so no forest is valid for both continuations.
This remains true if a forest uses several edges incident to the row.

Thus the positive-probability supports of the original prefix-state distributions of distinct sources are disjoint.
Indeed, a shared state would induce the same future law on the common continuation, contradicting zero-error correctness and finite response with probability one.
Choose a shortest positive-probability state for each source.
These are $M$ distinct binary strings, whose average length is at least $\log M-2$.
For each fixed source, its original expected prefix-state length is at least its chosen minimum; that length is bounded by the expected peak space on a legal immediate-report completion.
Consequently, both space metrics are
\begin{equation}
\Omega\!\left(N\log\binom{U}{m}\right),
 \label{eq:se3}
\end{equation}
under the original state distribution.

If $Z+1\le U/2$, the standard binomial estimates and $N,U=\Theta(v)$ give
\[
 N\log\binom{U}{Z+1}
 =\Theta\!\left(v(Z+1)\log(1+v/(Z+1))\right).
\]
When the choice of $m$ is capped at $\lfloor U/2\rfloor$, $\log\binom{U}{m}=\Theta(U)$, so \eqref{eq:se3} is $\Omega(v^2)$.
In this case $Z+1=\Omega(v)$ and the expression in \eqref{eq:se1} is $\Theta(v^2)$ as well.
The space-accounting convention and exact bitmap cover $v<4$.
Finally, the binomial entropy estimate with $a=\min\{Z,v\}+1\le v+1$ gives
\[
 \log\binom{v+a}{v}
 =\Theta\bigl(a\log(1+v/a)\bigr),
\]
which proves the equivalent form \eqref{eq:se2} and completes the proof of Theorem~\ref{thm:zero-error-space}.
\end{proof}

The $v^2$ saturation relies on the simple-graph promise that edge multiplicities remain in $\{0,1\}$.
For example, an entire graph may be deleted and restored arbitrarily often.
Since complete restoration leaves no historical edge absent, the deterministic algorithm with budget $Z=0$ answers every such report exactly.
It uses $O(v\log v)$ space throughout the stream, regardless of how many reports are requested.

\subsection{A recovery algorithm using the report-time deficit}
\label{sec:forest-budget-sensitive-banks}

For positive error, the algorithm reuses the binary cut decoder and independent verification from Section~\ref{sec:unrestricted-proof}.
A spanning forest of the stored live edges supplies the initial partition, so no singleton bootstrap is needed.
Sampled historical-edge data structures replace the top bitmaps; at a report, each repair bank freezes the current components, collects outgoing-edge proposals, and then merges.
Write $\tau_0=\lceil\log(64/\delta)\rceil$ and first suppose $32\tau_0<Z\le v$.
Choose any integer $8\le b\le Z$ and set $K=512b$ as in the cut decoder.
Instantiate Corollary~\ref{cor:binary-entropy-codec} with $n=v$ and $k=\min\{2K,v\}=\min\{1024b,v\}$, using $H=I_v$ when $k\ge v/32$.
This meets the required recovery guarantee because every binary $v$-vector of weight at most $2K$ has weight at most $k$.
The row bound remains $r=O(b\log(1+v/b))$ and total codec space is $O(vb\log(1+v/b))$; in the identity branch these bounds follow from $v\le32768b$.
Put
\begin{equation}
\lambda=\lceil\log(2+2Z/b)\rceil,\quad
 L=\lceil\log(64Z/b)\rceil,\quad
 R=8\lambda+2\lceil2\tau_0/b\rceil.                        \label{eq:forest-df8}
\end{equation}
The algorithm maintains the compressed historical edge data structure with threshold $b$ on the entire raw stream.
At the fixed report prefix $w$, its live graph $W_0$ has component excess at most $2d(w)/b\le2Z/b$ by Lemma~\ref{lem:dense-compressed-certificate}.
Each independent bank has the standard $(2b+2)$-wise full-name hash into $\mathbb F_{2^h}$, where $h=\lceil\log(N_E+1)\rceil+8$; level $j$ retains names with $j$ leading zero bits.
Since $b\ge8$ and $Z\le v$, the inequality $64Z/b\le8v$ holds, hence $L\le\lceil\log v\rceil+3<h$.
Each bank maintains the same compressed data structure with threshold $b+1$ on its raw level-$L$ sample.
At all shallower levels it retains the binary vertex syndromes and independent verification fingerprints of Section~\ref{sec:unrestricted-proof}.
The live edge sets of all sampled data structures are always subgraphs of their current sampled graphs.

Lemma~\ref{lem:sampling} supplies the full-name hash and tails: a fixed cut with dyadic mean in $(64b,128b]$ is nonempty and $K$-sparse except with probability $2^{1-b}$; a top cut of mean at least $64b$ is nonempty except with probability $2^{-b}$.
At this report, put $E_{\rm miss}=E^{\rm ever}(w)\setminus E(w)$, so that $|E_{\rm miss}|\le Z$.
For the top-level selected edge universe $P$, let $Y=|E_{\rm miss}\cap P|$.
The choice of depth gives $p=\Pr(e\in P)\le b/(64Z)$, and hence $\E Y\le b/64$.
The set $E_{\rm miss}$ is fixed independently of the bank's coins, so $(b+1)$-wise independence gives
\begin{equation}
\begin{aligned}
\Pr(Y>b)&\le\E\binom{Y}{b+1}
 =\binom{|E_{\rm miss}|}{b+1}p^{b+1}\\
 &\le\left(\frac{eb}{64(b+1)}\right)^{b+1}
 \le16^{-(b+1)}.
\end{aligned}                                       \label{eq:forest-df9}
\end{equation}
The binomial coefficient is zero if $|E_{\rm miss}|<b+1$.
The sampled subhistory has ever-inserted and current edge sets whose difference is exactly $E_{\rm miss}\cap P$.
Thus its top-level data structure with threshold $b+1$ has at most $Y$ absent stored names and, on $Y\le b$, its live subset spans the current sample by Lemma~\ref{lem:dense-compressed-certificate}.
For a cut frozen before this bank is inspected and having top-level mean at least $64b$, the top-sample forest can miss the cut only if the current sample has no crossing edge or $Y>b$.
Under the fresh bank's sampling law, these events have probability at most $2^{-b}$ and $16^{-(b+1)}$, respectively.
A union bound therefore gives failure probability at most $2^{-b}+16^{-(b+1)}\le2^{1-b}$.
This argument uses only the deficit at the report and imposes no restriction on earlier deletions or intermediate deficits.

A report starts from a spanning forest of $W_0$.
In each bank the algorithm freezes its current connected components.
For each component, it selects an outgoing edge from the top-level forest if one exists, and otherwise queries the shallow vertex syndromes; all proposals are collected before any merge.
In the ideal execution every unfinished component fails to propose an outgoing edge with conditional probability at most $2^{1-b}$.
Call the round bad if more than half the unfinished components fail; by Markov's inequality, its conditional probability is at most $2^{2-b}$.
A good round reduces component excess by a factor at most $3/4$: at least half of the unfinished components propose an edge, and their proposal graph merges at least half that many components.
Since $(3/4)^4<1/2$ and $2Z/b<2^\lambda$ by \eqref{eq:forest-df8}, $4\lambda$ good rounds reduce the initial excess of at most $2Z/b$ below one, hence to zero.
Failure therefore requires more than $R-4\lambda\ge R/2$ bad rounds.
The conditional bounds imply that any specified $R/2$ rounds are all bad with probability at most $(2^{2-b})^{R/2}$.
Taking a union bound over their possible positions gives
\begin{equation}
2^R(2^{2-b})^{R/2}\le2^{-bR/4}\le2^{-\tau_0}.             \label{eq:forest-df10}
\end{equation}
Here the first inequality uses $b/2-2\ge b/4$ for $b\ge8$, and the second follows from the choice of $R$ in \eqref{eq:forest-df8}.
There are at most $v$ frozen components per bank and level, so the ideal trace has at most $N_{\mathrm{test}}\le R(L+1)v\le64v^3$ tests; here $\lambda\le A\le v$, $R\le10v$, and $L\le A+3$.
The independent verification and bounded field construction use the separate error allowances of \eqref{eq:forest-pu3}.
Together with \eqref{eq:forest-df10}, their total report error is below $\delta$.

With $H_b=\log(1+v/b)$, the binary matrix and bilinear scratch cost $O(vr+vA)$.
Each of the $vRL$ retained vertex rows has $O(bH_b+A+\tau_0)$ bits, and all historical edge data structures together cost $O(vbH_b(R+1))$.
Full-domain hash seeds use $O(RbA)$ bits and verification seeds use $O(R(A+\tau_0))$ bits.
Since $A\le v$, the hash seeds fit even when $H_b$ is small.
Including all component, enumeration, field, and output scratch, the algorithm uses peak space
\begin{equation}
O\left(v\lambda(\lambda+\tau_0/b)
             \bigl[bH_b+A+\tau_0\bigr]\right).                      \label{eq:forest-df11}
\end{equation}
Reports only read persistent syndromes, stored edge sets and live bits, matrices, and seeds; all working partitions and proposal arrays are reset.
The space bound also holds at unpromised intermediate prefixes: each copy of the data structure always has at most $q(v-1)$ stored names, its live bits and all source syndromes remain exact through arbitrary restoration cycles, and every other dimension is fixed from $v,Z,\delta$ at setup.
The algorithm neither stores the full historical edge set nor tests whether $d(w)\le Z$.
Since reports leave the persistent state unchanged, the same failure bound applies separately to each promised report in any fixed finite reporting schedule.
The joint correctness requirement for the entire schedule is treated in Section~\ref{sec:restorations-reports}.

\paragraph{Selecting an algorithm for each parameter range.}
For every $Z\ge0$ and $0<\delta\le1/4$, choose an algorithm using $O(\mathcal L'(v,Z,t))$ bits, as specified by \eqref{eq:analysis-frontier}.
The choice depends only on the input parameters, using the following cases in order:
\begin{enumerate}
\item If $Z\le32\tau_0$ or $t\ge v$, use the deterministic algorithm.
\item Otherwise, if $\mu\le\sqrt{vt}$, use the deficit-sensitive algorithm.
\item Otherwise, use the unrestricted algorithm of Section~\ref{sec:unrestricted-proof}.
\end{enumerate}
The analysis below verifies the space bound in each case.

The exact simple-graph algorithm uses space
\begin{equation}
E=v\mu\log(1+v/\mu)                                \label{eq:forest-df18}
\end{equation}
up to an absolute factor.
Whenever $t\ge\mu/97$, the second term of \eqref{eq:analysis-frontier} is $\Omega(E)$: one has $x_t\ge\mu/\sqrt{97}$, $x_t/t\le97$, and $\log(1+z)\ge z/[(1+97)\ln2]$ for $0\le z\le97$.
Since $\tau_0=\lceil\log(64/\delta)\rceil\le3t$, the condition $Z\le32\tau_0$ implies $\mu\le Z+1\le96t+1\le97t$.
Thus the deterministic algorithm meets the required space bound when $Z\le32\tau_0$, as well as when $t\ge v$.
It includes $Z=0$.
In \eqref{eq:vc2}, the additive $2$ keeps the logarithmic factor bounded below before the quadratic cap applies.
In \eqref{eq:analysis-frontier}, $x_t=\mu$ and $t\log(1+x_t/t)=\Theta(\mu)$ when $t\ge v$, so the confidence term already has the zero-error space order $E$.

It remains to consider $Z>32\tau_0$ and $t<v$.
First suppose $\mu\le\sqrt{vt}$.
Then $Z<v$, and the deficit-sensitive algorithm applies.
Choose
\[
 a_0=\lceil\log(2+Z/\tau_0)\rceil,\qquad
 b=\max\{8,\lceil \tau_0/a_0\rceil\}.
\]
Here $b\le Z$, $\lambda=O(a_0)$, $\tau_0/b\le a_0$, and $b\le8+\tau_0/a_0$.
Substituting in \eqref{eq:forest-df11} gives
\begin{equation}
O\left(v\left[Aa_0^2+
                 \tau_0 a_0(H_b+a_0)\right]\right).        \label{eq:forest-df19}
\end{equation}
Let $J_\mu=\log(1+v/\mu)$.
The branch condition $\mu^2\le vt$ and $\tau_0\ge t$ imply $\mu/\tau_0\le v/\mu$.
Therefore, $a_0=O(J_\mu)$, and, as $b\ge \tau_0/a_0$,
\[
 \begin{split}
 H_b&\le\log(1+va_0/\tau_0)\\
    &\le\log a_0+\log(1+v/\mu)+\log(1+\mu/\tau_0)
      =O(J_\mu).
 \end{split}
\]
Also $a_0=O(\log(1+\mu/t))$ and $\tau_0=\Theta(t)$.
Consequently, the second term of \eqref{eq:forest-df19} is at most an absolute constant times $vt\log(1+\mu/t)\log(1+v/\mu)$, exactly the confidence term of \eqref{eq:analysis-frontier}, because now $x_t=\mu$.
Its first term is $O(vAB_Z)$, using $Z<v$ and $a_0=O(\log(Z+2))$.
The algorithm therefore uses the stated amount of memory in this case.

Finally, suppose $\mu>\sqrt{vt}$.
Now $x_t=\sqrt{vt}$, and $B_Z=\Theta(A^2)$, since $t\ge3$ implies $Z+1>\sqrt{3v}$.
Use \eqref{eq:forest-df17}.
If $t\le A$, the sketching term is $O(vA^3)$ and is covered by the baseline.
If $t>A$, then $A+\tau_0=O(t)$ and
\[
 \log\left(2+\frac v{A+\tau_0}\right)
   =O\left(\log(1+\sqrt{v/t})\right).
\]
For this inequality use $t<v$ and $2+v/t\le(1+\sqrt{v/t})^2$.
The algorithm uses $O(vt\log^2(1+\sqrt{v/t}))$ space, matching the second term of \eqref{eq:analysis-frontier} in this case.

\subsection{Constant-error lower bounds under a total deletion bound}
\label{sec:indexed-block}

We prove the lower bound using only streams with at most $Z$ deletion operations in total.
These streams satisfy $d(w)\le Z$ at every prefix, so the bound also applies to algorithms whose only input promise is the deficit bound at the report.
The graph reduction starts with $N=hk$ row vertices and a shared universe of $U$ column vertices, with $N+U\le v$.
Each row has an independent $m$-element neighborhood, and the rows are partitioned into $h$ blocks of $k$ rows.
After all $Nm$ source edges have been inserted, a query selects one block and deletes a proper subset of the neighborhood of each of its rows.
This uses at most $k(m-1)\le Z$ deletions.
Every queried row remains nonisolated, so a correct spanning forest reveals a surviving neighbor for every one of the $k$ rows and solves their universal-relation instances jointly.
The information argument below charges the saved prefix state for all $h$ independent blocks, although each stream queries only one block.

The coordinate construction of Nelson and Yu~\cite[Lemma~3]{NelsonYu2019} admits an external-information accounting bound in place of its message-length bound.
The following argument gives this accounting directly and specifies its conditioning variables.
Combined with the sparse perturbed universal-relation bound stated below in Lemma~\ref{lem:sparse-perturbed-ur}, whose proof is in Appendix~\ref{sec:graph-frontier}, it gives the information bound for this reduction.

In this section, $W$ denotes the success event and $\mathbf{1}_W$ its binary indicator.
The notation $I_\pi(\cdot;\cdot\mid \mathsf{C},W)$ means conditional mutual information under the \emph{success-conditioned law} $\pi(\cdot\mid W)$, averaged over $\mathsf{C}$ under that same law.
We write $I_\pi(\cdot;\cdot\mid \mathsf{C},\mathbf{1}_W)$ when conditioning on the indicator as a random variable, which averages over its values.

\begin{lemma}[Two elementary success-conditioning bounds]
\label{lem:success-conditioning}
Let $\pi$ be any joint law and let $W$ be any event with probability $p>0$, including an event depending on private decoder coins.
For any collection of variables $V$,
\begin{equation}
D_{\rm KL}\bigl(\pi[V\mid W]\Vert\pi[V]\bigr)\le\log(1/p).
 \label{eq:bi1}
\end{equation}
If $\mathsf{A}$ and $\mathsf{B}$ are conditionally independent given $\mathsf{C}$ under $\pi$, then
\begin{equation}
I_\pi(\mathsf{A};\mathsf{B}\mid \mathsf{C},W)\le\log(1/p).
 \label{eq:bi2}
\end{equation}
For arbitrary $\mathsf{A},M$ one also has
\begin{equation}
p I_\pi(\mathsf{A};M\mid W)\le I_\pi(\mathsf{A};M)+H_\pi(\mathbf{1}_W)
 \le I_\pi(\mathsf{A};M)+1.
 \label{eq:bi3}
\end{equation}
\end{lemma}
\begin{proof}
The likelihood ratio in \eqref{eq:bi1} is $\Pr(W\mid V)/p\le1/p$.
To prove \eqref{eq:bi2}, expand the relative entropy in \eqref{eq:bi1} for $V=(\mathsf{A},\mathsf{B},\mathsf{C})$ against the reference factorization $\pi[\mathsf{C}]\pi[\mathsf{A}\mid \mathsf{C}]\pi[\mathsf{B}\mid \mathsf{C}]$.
The expansion is $I_\pi(\mathsf{A};\mathsf{B}\mid \mathsf{C},W)$ plus three nonnegative relative-entropy terms.
Finally,
\[
 pI_\pi(\mathsf{A};M\mid W)
 \le I_\pi(\mathsf{A};M\mid\mathbf{1}_W)
 \le I_\pi(\mathsf{A};M,\mathbf{1}_W)
 \le I_\pi(\mathsf{A};M)+H_\pi(\mathbf{1}_W),
\]
which proves \eqref{eq:bi3}.
None of these arguments requires $W$ to be a deterministic function of $\mathsf{A}$ and $M$.
\end{proof}

\begin{lemma}[Coordinate reduction with external information]
\label{lem:external-coordinate}
Let $k$ be even and let $(X_j,Y_j)_{j=1}^k$ be independent pairs with common law $\mathcal D$.
Consider a one-way protocol with randomized message $M$ and possibly randomized output $O=(O_1,\ldots,O_k)$ satisfying the Markov conditions
\[
 M\leftrightarrow X^k\leftrightarrow Y^k,\qquad
 O\leftrightarrow(M,Y^k)\leftrightarrow X^k.
\]
Any shared random seed used by the decoder is included in $M$.
Let $\pi$ be the joint law of $(X^k,Y^k,M,O)$.
Suppose it solves all $k$ relations with probability $p\ge1/2$, and put $\mathcal J=I_\pi(X^k;M)$.
There is one input law $\mathcal D'$ and one generalized one-coordinate protocol with message $\widehat M$ such that, simultaneously,
\begin{equation}
\begin{aligned}
D_{\rm KL}(\mathcal D'[X]\Vert\mathcal D[X]),\quad
 D_{\rm KL}(\mathcal D'[Y]\Vert\mathcal D[Y])&\le K,\\
 \E_xD_{\rm KL}(\mathcal D'[Y\mid x]\Vert\mathcal D[Y\mid x]),
 \quad
 \E_yD_{\rm KL}(\mathcal D'[X\mid y]\Vert\mathcal D[X\mid y])
 &\le K/k,\\
 I_{\mathcal D'}(\widehat M;Y\mid X)&\le K/k,\\
 \Pr_{\mathcal D'}(\text{error})&\le K/\sqrt k,\\
 I_{\mathcal D'}(X;\widehat M\mid Y)&\le K(\mathcal J+1)/k,
\end{aligned}
\label{eq:bi4}
\end{equation}
where $K$ is an absolute constant and the conditional expectations use the displayed $\mathcal D'$ marginals.
\end{lemma}
\begin{proof}[Coordinate reduction and original-law accounting]
Let $W$ be the event that all $k$ answers are correct and set $\mathsf{A}=(X^k,Y^k)$.
The original one-way law gives $I_\pi(\mathsf{A};M)=I_\pi(X^k;M)=\mathcal J$.
The coordinate construction of Nelson and Yu chooses a random coordinate together with partially exposed source and query coordinates, under the success-conditioned law.
Its information chain rule yields
\[
 \E_\lambda I_{\mathcal D'_\lambda}(X;\widehat M\mid Y)
 \le \frac{2}{k}I_\pi(\mathsf{A};M\mid W)
 \le \frac{2(\mathcal J+1)}{kp}
 \le \frac{4(\mathcal J+1)}{k}.
\]
The middle inequality is \eqref{eq:bi3}.
It charges the message under its original law, even though the auxiliary coordinate distribution is conditioned on success.
The construction has mean conditional KL and one-way dependence $O(1/k)$, mean error $O(k^{-1/2})$, and constant marginal KL.
Markov's inequality selects one coordinate view for which all seven required bounds hold simultaneously.
Appendix~\ref{app:coordinate-reduction-proof} specifies the two equivalent coordinate views and proves each of these estimates.
In particular, the accounting does not reveal future source rows or charge a success-conditioned message length.
\end{proof}

The coordinate protocol can have a perturbed input law.
The following source-size-sensitive interface specifies exactly which deviations the encoding tolerates.
Its proof is deferred to Appendix~\ref{sec:graph-frontier}; the statement makes the composition with Lemma~\ref{lem:external-coordinate} explicit.

\begin{lemma}[Sparse perturbed universal relation]
\label{lem:sparse-perturbed-ur}
Fix a constant $K_0$.
There are absolute constants $c_0,C_0>0$ and a sufficiently small $\beta_0>0$, with their admissible values allowed to depend on $K_0$, as follows.
Put $L=\ln(1/\beta)$ and suppose
\[
 0<\beta<\beta_0,\qquad m\ge L,\qquad U\ge C_0m.
\]
There is a distribution $\mathcal D=\mathcal D(U,m,\beta)$ on $T\subsetneq S\subseteq[U]$ with $|S|=m$ for which the following holds.
Suppose a distribution $\mathcal D'$ and a generalized one-way protocol with message $M$ and output $O$ satisfying the Markov condition $O\leftrightarrow(M,T)\leftrightarrow S$ have error at most $\beta$ and
\begin{align*}
 D_{\rm KL}(\mathcal D'[S]\Vert\mathcal D[S])&\le K_0,\\
 \E_{s\sim\mathcal D'[S]}
 D_{\rm KL}(\mathcal D'[T\mid S=s]\Vert\mathcal D[T\mid S=s])
 &\le\eta,\\
 \E_{t\sim\mathcal D'[T]}
 D_{\rm KL}(\mathcal D'[S\mid T=t]\Vert\mathcal D[S\mid T=t])
 &\le\eta,\\
 I(M;T\mid S)&\le\eta,\qquad \eta\le c_0\beta^2.
\end{align*}
Then
\[
 I(S;M\mid T)
 =\Omega\!\left(L\log(2+m/L)\log(U/m)\right).
\]
The marginal KL bound on $T$ supplied by Nelson and Yu is permissible but is not needed for this particular formulation.
\end{lemma}

\begin{proposition}[Sparse product relation: external information]
\label{prop:block-information}
For all sufficiently large even $k$, and integers $m\ge C_1\log(k+1)$ and $U\ge C_2m$, there is a product hard law for $k$ universal-relation instances, whose independent source sets are uniform $m$-subsets of $[U]$, with the following property.
Every randomized one-way protocol obeying both Markov conditions of Lemma~\ref{lem:external-coordinate}, with any shared random seed included in its message, and solving all $k$ instances with probability at least $1/2$ obeys
\begin{equation}
I(S^k;M)=\Omega\!\left(
  k\log(k+1)\log\left(2+\frac m{\log(k+1)}\right)\log\frac Um
 \right).
 \label{eq:bi10}
\end{equation}
\end{proposition}
\begin{proof}
Choose the hard law of Lemma~\ref{lem:sparse-perturbed-ur} with $\beta=C_\beta/\sqrt k$ before applying the coordinate lemma.
Choose the absolute constant $C_\beta$ large enough that the $K/k$ conditional KL and approximate-one-way bounds are at most $c_0\beta^2$, and the coordinate error is at most $\beta$.
The marginal KL terms are bounded by the fixed constant $K$.
For sufficiently large $k$, the sparse lemma applies with $L=\ln(1/\beta)=\Theta(\log(k+1))$ and yields coordinate information
\[
 \Omega\!\left(
 \log(k+1)\log\left(2+\frac m{\log(k+1)}\right)\log\frac Um
 \right).
\]
Compare this with $K(I(S^k;M)+1)/k$ from \eqref{eq:bi4}.
For sufficiently large $k$ the additive one is absorbed, proving \eqref{eq:bi10}.
\end{proof}

\begin{proposition}[Spanning-forest lower bound from indexed graph blocks]
\label{prop:indexed-block-forest}
Suppose $N=hk$, where $k,m,U$ satisfy the preceding proposition, and
\[
 N+U\le v,\qquad k(m-1)\le Z.
\]
A spanning-forest algorithm with error at most $1/4$ on every fixed legal stream with at most $Z$ deletion operations requires
\begin{equation}
s+1=\Omega\!\left(
  N\log(k+1)\log\left(2+\frac m{\log(k+1)}\right)\log\frac Um
 \right)
 \label{eq:bi11}
\end{equation}
for either space measure $s$ of Appendix~\ref{app:machine-conventions}.
The hard streams are legal, have exactly $Nm$ insertions followed by at most $k(m-1)\le Z$ deletions, and request one report.
In particular, their historical deficit never exceeds $Z$.
\end{proposition}
\begin{proof}
Use $N$ row vertices and $U$ column vertices.
Independently draw each row source $S_i$ uniformly among the $m$-subsets and insert all its incident edges in a common canonical order.
Partition the rows into $h$ consecutive blocks of size $k$; write $B_j$ for the entire source tuple in block $j$.
The query chooses one block $j$, supplies proper subsets $T_i$ of its $k$ source sets, deletes only those row edges, and reports.
Every queried row stays nonisolated.
A correct spanning forest provides a valid incident edge for every one of those $k$ rows, and hence solves all its $k$ relations.
At most $k(m-1)$ deletions are used, independently of the number $h$ of possible blocks.

Apply Lemma~\ref{lem:finite-simulation} with one common cutoff to this finite family, with added error at most $1/8$.
Let $M_0$ be the tagged original prefix state or the constant-size timeout symbol.
Each prefix has a legal immediate-report completion.
Under its original randomness $\E|M_0|\le s+O(1)$, uniformly over the fixed source prefix.
The state-entropy bound gives
\begin{equation}
H(M_0)\le \E|M_0|+
   \log\bigl(e(\E|M_0|+1)\bigr)=O(s+1).
 \label{eq:bi12}
\end{equation}
The simulation truncates running time while preserving the original state-length charge.
The protocol for every queried stream succeeds with probability at least $5/8>1/2$.

For block $j$, reveal only $Q_j=(B_1,\ldots,B_{j-1})$, independent of $B_j$.
Alice samples future blocks privately and generates $M_0$.
The message $M_j=(Q_j,M_0)$ is an ordinary randomized one-way channel from $B_j$; the queried subset tuple is generated independently from its product hard conditional law given $B_j$.
It satisfies the exact Markov condition required in \eqref{eq:bi10}.
The decoder continues from $M_0$ on this one block and obtains all $k$ requested elements with success at least $1/2$.
Thus \eqref{eq:bi10} bounds
\[
 I(B_j;M_j)=I(B_j;M_0\mid B_1,\ldots,B_{j-1})
\]
from below.
Sum over $j$ and apply the chain rule:
\[
 \sum_{j=1}^h I(B_j;M_0\mid B_1,\ldots,B_{j-1})
 =I(B_1,\ldots,B_h;M_0)\le H(M_0).
\]
Since $hk=N$, \eqref{eq:bi12} gives \eqref{eq:bi11}.
The argument uses only the guarantee for the single block selected by the query.
\end{proof}

\paragraph{Relation to prior lower bounds.}
The coordinate sampling, success conditioning, and approximate protocol construction are those of Nelson and Yu~\cite[Lemma~3]{NelsonYu2019}.
The constant-error external-information consequence also follows from exact channel simulation with shared randomness~\cite[Result~1]{HJMR2010}, constant-error truncation, and that coordinate lemma.
The simulation bound controls communication averaged over the source distribution and protocol randomness.
Equation \eqref{eq:bi6} in Appendix~\ref{app:coordinate-reduction-proof} proves the external-information bound directly under the original message law.
Lemma~\ref{lem:sparse-perturbed-ur} extends the encoding in their Appendix~B to track source cardinality.
Indexed direct-sum reductions spending deletions on a single graph block are also used by Khanna et al.~\cite[Sections~1.2 and~4]{KKD2025}.
To lower-bound spanning-forest reporting on fixed input sequences, the reduction combines this deletion-budget allocation with the source-cardinality-sensitive extension of the relation of Nelson and Yu.

The block construction covers growing budgets.
For bounded budgets, the insertion-only lower bound of Lemma~\ref{lem:randomized-forest-baseline} applies: a uniformly chosen labeled tree must be recovered exactly, and Fano's inequality charges $\Omega(v\log(v+1))$ bits under the original expected-state law.
Combining the two cases gives a bound uniform in $Z$.

\begin{corollary}[Matching constant-error lower bound on the bounded-deletion subclass]
\label{cor:all-budget-block-lower}
For all $v\ge2$ and integers $Z\ge0$, an algorithm returning a spanning forest with error at most $1/4$ on every fixed legal stream with at most $Z$ deletion operations requires
\begin{equation}
\Omega\!\left(
 v\log(v+1)\,
 \min\{\log^2(Z+2),\log^2(v+1)\}\right)
 \label{eq:ab1}
\end{equation}
bits, for both peak-space measures.
The hard streams are legal, use at most $Z$ deletion operations in total, and need only one report.
The algorithm in Theorem~\ref{thm:budget-space-implementation} uses space matching this lower bound on simple graphs.
\end{corollary}
\begin{proof}
Put
\[
 x=\min\{\sqrt{Z+1},v^{1/4}\}.
\]
First suppose $x$ and $v$ exceed sufficiently large absolute constants, and define
\[
 k=m=2\lfloor x/2\rfloor,\qquad
 N=k\left\lfloor\frac v{2k}\right\rfloor,\qquad
 U=\lceil v/2\rceil.
\]
Then $k$ is even, $k\ge x/2$, $N=\Theta(v)$, $N+U\le v$, $m\ge C_1\log(k+1)$, and $U\ge C_2m$.
For $k\ge2$ the number of deletions is at most
\[
 k(m-1)=k(k-1)\le k^2-2\le Z-1\le Z.
\]
Moreover,
\[
 \log(U/m)=\Theta(\log(v+1)),\qquad
 \log\left(2+\frac m{\log(k+1)}\right)=\Theta(\log(k+1)),
\]
and
\[
 \log(k+1)=\Theta\bigl(
   \min\{\log(Z+2),\log(v+1)\}\bigr).
\]
Proposition~\ref{prop:indexed-block-forest} therefore gives \eqref{eq:ab1}.

If the required absolute threshold on $x$ or $v$ fails, either $Z$ or $v$ is bounded by an absolute constant.
The minimum of the two squared logarithms in \eqref{eq:ab1} is then bounded by an absolute constant.
Lemma~\ref{lem:randomized-forest-baseline} supplies the claimed lower bound in all those cases, including $Z=0$.
\end{proof}

\paragraph{Tight dependence on the deficit bound.}
The bound grows as $v\log v\log^2(Z+2)$ until it reaches the $v\log^3v$ order.
In particular, for every fixed $\varepsilon>0$, $Z\ge v^\varepsilon$ already attains that order, with a lower-bound constant allowed to depend on $\varepsilon$.
The deficit-sensitive algorithm in Theorem~\ref{thm:budget-space-implementation} uses space matching this lower bound also for $Z$ smaller than every positive power of $v$, at constant error.
The external-information indexed-block proposition makes this possible by concentrating deletions on the queried block.

\subsection{Confidence lower bounds under a total deletion bound}
\label{sec:nearlinear-entropy-lower}

The proof retains the source-entropy factor in the local single-row information bound rather than replacing it by $\Theta(\log v)$.
The resulting bound has absolute constants for every positive error, including confidence exponents arbitrarily close to $v$.
This gives the matching confidence lower bound in Theorem~\ref{thm:budget-space-implementation}.
Again, the hard inputs use at most $Z$ deletion operations in total, which is stronger than the report-time historical-deficit promise.
Throughout, put
\[
 \begin{gathered}
 v\ge2,\qquad Z\in\mathbb Z_{\ge0},\qquad
 0<\delta\le1/4,\\
 A=\lceil\log(v+1)\rceil,\qquad t=1+\log(1/\delta),
 \qquad B_Z=\min\{A^2,\log^2(Z+2)\}.
 \end{gathered}
\]
The parameter $s$ denotes either worst-case peak space or worst-input expected peak space for the one-report problem, under the original coin law.
The lower bounds already hold when correctness is required only on the bounded-deletion subclass just described.
The argument uses the single-row information bound \eqref{eq:ic1} from Section~\ref{sec:unrestricted-proof}, the high-confidence exact-space regime \eqref{eq:ic2}--\eqref{eq:ic3} in Appendix~\ref{app:positive-error-exact}, and the constant-error lower bound \eqref{eq:ab1} proved above.

\begin{proposition}[Source-size-sensitive confidence lower bound]
\label{prop:nearlinear-source-bound}
There are absolute constants $c\in(0,1)$ and $v_0$ such that, for $v\ge v_0$, let
\begin{equation}
M=\min\{Z+1,\lfloor cv\rfloor\},\qquad
 \Psi=\max_{\substack{m\in\mathbb Z\\1\le m\le M}}
       t\log(1+m/t)\log(1+v/m).
 \label{eq:forest-ne1}
\end{equation}
Then both peak-space measures satisfy
\begin{equation}
s+1=\Omega\bigl(vAB_Z+v\Psi\bigr).
 \label{eq:forest-ne2}
\end{equation}
All constants are absolute.
The hard streams are legal, use at most $Z$ deletion operations in total, and require only one report.
\end{proposition}
\begin{proof}[How the source regimes combine]
For a fixed admissible neighborhood size $m$, the exact single-row interface \eqref{eq:ic1} gives the desired two entropy factors when $m$ is large relative to $t$ and the error is sufficiently small.
When $m=O(t)$, the same source argument with a relaxed positive error gives the exact-space order $vm\log(1+v/m)$, which dominates $vt\log(1+m/t)\log(1+v/m)$; the precise statement is Theorem~\ref{thm:small-budget-confidence}.
At bounded confidence the constant-error baseline absorbs this term.
The maximum over $m$ can therefore be taken while keeping the same advertised budget and the original coin law.
The maximum of the baseline and confidence lower bounds is at least half their sum.
Appendix~\ref{app:source-regimes-proof} gives the uniform constants, finite cutoffs, and endpoint checks.
\end{proof}

The source parameter has a direct interpretation: it is the number of possible neighbors retained in each hard input row.
In the source-encoding regime, increasing it adds encoding stages but reduces the information supplied by a recovered neighbor.
Their product, rather than either factor separately, determines the hardest source scale.

\begin{lemma}[Exact continuous optimization]
\label{lem:nearlinear-continuous-optimizer}
For positive real $v,t$, the function
\[
 f(m)=\log(1+m/t)\log(1+v/m),\qquad m>0,
\]
is strictly increasing up to $m=\sqrt{vt}$ and strictly decreasing thereafter.
If $M\ge1$ is an integer and $\sqrt{vt}\ge1$, its maximum over integers $1\le m\le M$ lies between $\tfrac12 f(y)$ and $f(y)$, where $y=\min\{M,\sqrt{vt}\}$.
\end{lemma}
\begin{proof}
Set $m=te^z$, $a=\ln(v/t)$, and $G(z)=\ln\ln(1+e^z)$.
Since $\ln(1+e^z)>0$ for every real $z$, the function $G$ is smooth on all of $\mathbb R$.
Direct differentiation gives
\[
 G''(z)=
 \frac{e^z\bigl(\ln(1+e^z)-e^z\bigr)}
 {(1+e^z)^2\ln^2(1+e^z)}<0.
\]
The inequality follows from $\ln(1+u)<u$ for $u>0$, so $G$ is strictly concave and $G'$ is strictly decreasing.
Using $\log x=\ln x/\ln 2$, we have $\ln f(te^z)=G(z)+G(a-z)-2\ln(\ln 2)$.
Its derivative $G'(z)-G'(a-z)$ is therefore positive for $z<a/2$ and negative for $z>a/2$, proving the continuous assertion.
For the integer maximum take $m=\lfloor y\rfloor$, so $y/2\le m\le y$.
Concavity gives $\log(1+m/t)\ge\tfrac12\log(1+y/t)$, and the second factor increases when $m$ decreases.
The continuous maximum is also an upper bound on the integer maximum.
\end{proof}

\begin{corollary}[A closed-form confidence lower bound with absolute constants]
\label{thm:nearlinear-entropy-lower}
Put
\begin{equation}
x_t = \min\{Z+1,v,\sqrt{vt}\}.
 \label{eq:forest-ne3}
\end{equation}
For all parameters above, both peak-space measures satisfy
\begin{equation}
s+1=\Omega\!\left(
 vA\min\{A^2,\log^2(Z+2)\}
 +vt\log(1+x_t/t)\log(1+v/x_t)\right).
 \label{eq:forest-ne4}
\end{equation}
The constant is absolute, independent of $Z$ and the error.
Legal streams with at most $Z$ total deletion operations and one report suffice.
For $v\ge v_0$, the second term is $\Theta(v\Psi)$, evaluating the full variational expression in \eqref{eq:forest-ne1}.
\end{corollary}
\begin{proof}
For sufficiently large $v$, let $M$ be as in \eqref{eq:forest-ne1}.
Since $\mu=\min\{Z+1,v\}$, it follows that $(c/2)\mu\le M\le\mu$.
For $y=\min\{M,\sqrt{vt}\}$ this implies $(c/2)x_t\le y\le x_t$.
Concavity of the first logarithm and monotonicity of the second give $f(y)\ge(c/2)f(x_t)$; continuous optimality gives the reverse inequality $f(y)\le f(x_t)$.
Lemma~\ref{lem:nearlinear-continuous-optimizer} therefore shows $\Psi=\Theta(t f(x_t))$.
Substitute this in \eqref{eq:forest-ne2}.

For bounded $v$, the first term of \eqref{eq:forest-ne4} is uniformly bounded in $Z$, while $t f(x_t)\le v/(\ln2)^2$ uniformly in $\delta$.
The space-accounting convention therefore applies; the guarantees for both peak-space measures and source legality are inherited from \eqref{eq:forest-ne2}.
\end{proof}

\subsection{Space-optimal algorithms at every deficit bound and confidence}
\label{sec:budget-upper-bound}

Combining the algorithm and space analysis from Section~\ref{sec:forest-budget-sensitive-banks} with the space lower bound of Corollary~\ref{thm:nearlinear-entropy-lower} establishes the following optimality and implementation guarantees.

\begin{theorem}[Optimal one-report space under a historical-deficit bound]
\label{thm:budget-space-implementation}
\label{thm:historical-deficit-space}
Let $v\ge2$, let $Z\ge0$ be an integer given in advance, and let $0<\delta\le1/4$.
Consider fixed legal dynamic graph streams whose single report prefix $w$ satisfies $d(w)\le Z$, without any restriction on previous deficits or the total number of deletion and restoration operations.
Put
\begin{equation}
\begin{gathered}
 A=\lceil\log(v+1)\rceil,\quad t=1+\log(1/\delta),
 \quad \mu=\min\{Z+1,v\},\quad x_t=\min\{\mu,\sqrt{vt}\},\\
 \mathcal L'(v,Z,t)=
 vA\min\{A^2,\log^2(Z+2)\}
 +vt\log(1+x_t/t)\log(1+v/x_t).
 \end{gathered}                                      \label{eq:forest-pu4}
\end{equation}
The algorithm uses $O(\mathcal L'(v,Z,t))$ bits, while every algorithm for this task requires $\Omega(\mathcal L'(v,Z,t))$ bits under either peak-space measure.
Thus the optimal space complexity in both measures is
\begin{equation}
\Theta\bigl(\mathcal L'(v,Z,1+\log(1/\delta))\bigr),
\label{eq:sd3}
\end{equation}
with absolute constants.
The lower bound already holds on the subclass of streams with at most $Z$ deletion operations in total.
The algorithm uses polynomially many bit operations in $v+\min\{t,v\}$ for setup, each update, and each read-only report, in addition to reading the parameter headers.
Its hard space cap holds on every random tape and every legal update history, including intermediate prefixes with $d(w)>Z$; it does not store or test $d(w)$.
Parameter saturation and the resulting finite running-time bounds are given in Appendix~\ref{app:budget-finite-work}.
At zero error, the deterministic algorithm and the lower bound for all zero-error algorithms give the optimal space in \eqref{eq:se1}; the algorithm is correct at every promised report, without a bound on their number, and retains its hard space cap throughout the stream (Theorem~\ref{thm:zero-error-implementation}).
\end{theorem}
\begin{proof}
The lower bound follows from Corollary~\ref{thm:nearlinear-entropy-lower} in both peak-space measures, because its hard streams with at most $Z$ total deletions satisfy the report-time deficit promise.
To implement the algorithm within this space, use the historical edge data structure of Lemma~\ref{lem:poly-historical-certificate}, the binary codec of Corollary~\ref{cor:binary-entropy-codec}, and the independent verification procedure of \eqref{eq:forest-pu3} in the algorithms of Section~\ref{sec:forest-budget-sensitive-banks}.
The data-structure lemma preserves its acceptance predicate and space bound; the uniform binary codec preserves sparse-cut recovery and row width; and the bounded field preserves the fingerprint collision bound along the ideal trace with the same setup-error allowance.
Thus \eqref{eq:forest-df11} and \eqref{eq:forest-df17} still hold, with the polynomial exact alternative \eqref{eq:forest-pu2}.
The parameter selection in Section~\ref{sec:forest-budget-sensitive-banks}, using \eqref{eq:forest-df18}--\eqref{eq:forest-df19}, proves $O(\mathcal L')$ over every parameter range, including all stored randomness and the setup, decoding, and output workspace.

After saturation, every sketching branch has $Z\le v$ when it stores historical edges, $\tau_0\le v+6$, $b\le v$, and polynomially many banks and retained levels.
All code setup, small-field arithmetic, cut recovery, unit-capacity flows, binomial ranks, bitmap rank scans, hash operations and bounded verification-field trials have fixed polynomial time in bit operations.
Every report has bounded running time, including on setup failure, and resets its working partitions without changing persistent tables, seeds or live bits.
The density invariant of the stored historical edge sets holds on all histories, and the other arrays have dimensions fixed at setup, so neither the hard cap nor the per-event time bound depends on intermediate deficits or on the number of restoration cycles.
The hard cap also bounds the expected memory usage under the original coin law.

To obtain the simplified expression in the introduction, note that $\log v\le A\le2\log v$, so replacing $A$ by $\log v$ changes the baseline by at most a factor of eight.
For the envelope in \eqref{eq:graph-envelope-main},
\[
 \Gamma_{v,\mu}(z)\le\Gamma_{v,\mu}(cz)
 \le c\Gamma_{v,\mu}(z)\qquad(z\ge1,\ c\ge1),
\]
because, for each fixed $m$, the function $z\log(1+m/z)$ is increasing and grows by at most a factor of $c$ when $z$ is multiplied by $c$.
Lemma~\ref{lem:nearlinear-continuous-optimizer} identifies the confidence term with $v\Gamma_{v,\mu}(t)$.
Since $\log(1/\delta)\ge2$ and $\log(1/\delta)\le t\le\tfrac32\log(1/\delta)$, these comparisons give
\[
 \mathcal L(v,Z,\log(1/\delta))
 \le \mathcal L'(v,Z,t)
 \le 8\mathcal L(v,Z,\log(1/\delta)).
\]
The comparison is uniform over all allowed parameters.
Absorbing these factors in the algorithmic $O$ bound and the matching $\Omega$ lower bound gives the simplified characterization in Theorem~\ref{thm:deletion-budget-space}.
\end{proof}

\subsection{The optimal space in different regimes and its saturation threshold}
\label{sec:parameter-consequences}

The closed form describes both the familiar confidence regimes and the point at which randomization ceases to save an asymptotic factor over exact recovery.
The following consequences express the space tradeoff directly in the parameters of the problem.

\begin{lemma}[Optimal space at constant and polynomially small error]
\label{lem:familiar-regimes}
At any fixed $\delta\in(0,1/4]$, the one-report optimal space under the historical-deficit promise has order $vA\min\{A^2,\log^2(Z+2)\}$.
If $t=\Theta(A)$ with fixed positive comparison constants, it has order
\begin{equation}
vA^2\min\{\log(1+(Z+1)/A),A\}. \label{eq:ps0}
\end{equation}
The hidden constants may depend on the stated fixed comparisons.
\end{lemma}
\begin{proof}
At constant $t$, the confidence envelope is at most $O(A\log(1+\mu))$ and is absorbed by $AB_Z$.
For $t=\Theta(A)$, first suppose $\mu\ge v^{1/4}$.
The constant-error baseline is $\Theta(vA^3)$ and the fully dynamic algorithm uses space of that order; the expression \eqref{eq:ps0} has the same order.
If $\mu<v^{1/4}$, then, for sufficiently large $v$, $x_t=\mu$ and $\log(1+v/\mu)=\Theta(A)$.
The confidence term is therefore $\Theta(vA^2\log(1+\mu/A))$.
It absorbs the baseline: for $\mu\le A$, use $\log^2(\mu+1)=O(\mu)=O(A\log(1+\mu/A))$; for $\mu>A$, expand $\log\mu=\log A+\log(\mu/A)$, and use $\log^2 A=O(A)$ and $\log(\mu/A)\le A$.
The minimum with $A$ does not bind in this smaller-deficit range.
Absolute factors cover bounded $v$.
At fixed error, $A=\Theta(\log v)$ and $\min\{\log^2v,\log^2(Z+2)\}=\log^2(\min\{v,Z+2\})$, giving the simplified form in \eqref{eq:graph-randomized-main}.
\end{proof}

\begin{corollary}[When positive error reaches the zero-error space order]
\label{cor:budget-saturation}
Suppose $v\to\infty$ and $\mu=\min\{Z+1,v\}\to\infty$.
Let $E=v\mu\log(1+v/\mu)$ be the zero-error space scale.
For positive error, the one-report optimal space complexity is $o(E)$ if and only if $t=o(\mu)$, and it is $\Theta(E)$ if and only if $t=\Omega(\mu)$.
Here too an oscillating sequence need not belong to either class.
If $\mu$ stays bounded, positive-error and zero-error optimal space complexities already have the same order $v\log v$; the growth hypothesis matters.
\end{corollary}
\begin{proof}
Write $J=\log(1+v/\mu)$.
The baseline divided by $E$ is $AB_Z/(\mu J)$.
If $\mu\le\sqrt v$, then $J=\Omega(A)$ and $B_Z\le\log^2(\mu+1)$, so the ratio is $O(\log^2(\mu+1)/\mu)=o(1)$.
If $\mu>\sqrt v$, it is at most $A^3/\sqrt v=o(1)$.
These estimates are uniform along the assumed growing sequence.

For $1\le t\le\mu$, put $r=\mu/t\ge1$.
Evaluating the envelope at $m=\mu$ gives the lower bound below.
If its maximizer is $\mu$, the normalized confidence term equals the lower bound.
Otherwise, $x=\sqrt{vt}<\mu$, hence $a=v/\mu<r$, and $\log(1+\sqrt{v/t})\le\log(1+r)$, while $J\ge1$.
Thus in both cases
\begin{equation}
\frac{\log(1+r)}r
 \le\frac{\Gamma_{v,\mu}(t)}{\mu J}
 \le\frac{\log^2(1+r)}r. \label{eq:ps1}
\end{equation}
Both bounds tend to zero as $r\to\infty$.
For $t\ge\mu$, the maximizer is $\mu$ and the same ratio lies between $1$ and $1/\ln2$.
More generally, if $t/\mu\ge\epsilon>0$, evaluation at $\mu$ gives a positive lower bound $\epsilon\log(1+1/\epsilon)$.
Combine these statements with the vanishing baseline and the absolute-constant characterization.
The converse directions follow by taking a subsequence on which $t/\mu$ is respectively bounded away from zero or tends to zero.
The deterministic zero-error algorithm always uses $O(E)$ bits.
\end{proof}

\paragraph{Tight space at near-linear confidence exponents.}
\label{app:nearlinear-specialization}
Fix $p,\kappa>0$, let $Z\ge\kappa v$, and set $t=v/A^p$.
For sufficiently large $v$, depending only on $p,\kappa$, $x_t=\sqrt{vt}=v/A^{p/2}$.
Consequently, \eqref{eq:forest-ne4} gives
\begin{equation}
s+1=\Omega_{p,\kappa}\!\left(
 \frac{v^2(\log A)^2}{A^p}\right),
 \label{eq:forest-ne5}
\end{equation}
because $\log(1+x_t/t)=\log(1+v/x_t)=\log(1+A^{p/2}) =\Theta_p(\log A)$.
The $vA^3$ baseline is smaller for sufficiently large $v$.
This conclusion retains both entropy factors without requiring a fixed polynomial gap between $t$ and $v$.
The algorithm of Theorem~\ref{thm:budget-space-implementation} matches the space lower bound and uses polynomially many bit operations per event.

\section{Joint Correctness over Multiple Reports}
\label{sec:restorations-reports}

We now extend the one-report result of Theorem~\ref{thm:historical-deficit-space} to jointly correct reports in the same historical-deficit model.
Let $Z\ge0$ and $Q\ge1$ be integers given in advance, where $Q$ bounds the number of report requests.
Fix a legal sequence of updates and report requests independently of the algorithm's random coins and answers.
If its report prefixes are $w_1,\ldots,w_r$, with $r\le Q$, the promise is $d(w_j)\le Z$ for every $j$.
At report $j$, the algorithm must output a spanning forest $F_j$ of the current graph $G(w_j)$, using currently present edges.
For $0<\delta\le1/4$, joint correctness requires
\[
 \Pr\!\left[\text{$F_j$ is a spanning forest of $G(w_j)$ for every $1\le j\le r$}\right]
 \ge1-\delta.
\]
The probability is over the algorithm's coins for the whole execution; a missing or nonterminating response counts as failure.
The number of deletions and restorations, and the historical deficit between reports, remain unrestricted; the algorithm's space guarantee must hold throughout the stream.

A failure probability of at most $\delta$ for each report separately does not by itself imply this joint guarantee.
Reusing our read-only one-report algorithm with error parameter $\delta/Q$ and applying a union bound gives an algorithm for the joint task directly.
The space question is whether this sufficient error allocation is also optimal.
Joint correctness does not require each report to have error at most $\delta/Q$, and errors across reports can be correlated.
Consequently, substituting $\delta/Q$ into the single-report space lower bound does not establish a lower bound for the joint task.

The main result of this section is a matching space lower bound for every algorithm satisfying joint correctness, including algorithms that retain and use information acquired during earlier updates and reports.
The proof must account for this accumulated information when the same source graph is restored and queried repeatedly.
Together with the direct algorithmic extension, this lower bound establishes Theorem~\ref{thm:joint-report-space}.

\subsection{Optimal joint recovery and a fixed query distribution}
\label{sec:snapshot-finite-joint}

The following theorem states the algorithm's space guarantee and a matching lower bound for every algorithm meeting the joint correctness requirement.
It uses the same function $\mathcal L'$ as the one-report result, with confidence parameter $1+\log(Q/\delta)$.

\begin{theorem}[Optimal space for jointly correct spanning-forest reports]
\label{thm:joint-report-implementation}
Let $v\ge2$, let the integers $Z\ge0$ and $Q\ge1$ be given in advance, and let $0<\delta\le1/4$.
Consider every fixed legal event sequence with at most $Q$ report requests and $d\le Z$ at each report.
Each request asks for a spanning forest of the current graph, and all answers must be correct together with probability at least $1-\delta$.
Put
\[
 A=\lceil\log(v+1)\rceil,\quad
 \tau=1+\log(Q/\delta),\quad
 \mu=\min\{Z+1,v\},\quad x_\tau=\min\{\mu,\sqrt{v\tau}\}.
\]
The optimal space complexity for this task is
\begin{equation}
\Theta\!\left(
 vA\min\{A^2,\log^2(Z+2)\}
 +v\tau\log(1+x_\tau/\tau)\log(1+v/x_\tau)
 \right)
 =\Theta\bigl(\mathcal L'(v,Z,\tau)\bigr).          \label{eq:sj9}
\end{equation}
The expression is in bits, the constants are absolute, and the characterization holds for both peak-space measures.
The read-only single-report algorithm, run with error parameter $\delta/Q$, uses $O(\mathcal L'(v,Z,\tau))$ bits with a hard cap on every legal update sequence and polynomial setup, update, and report time in bit operations after reading the parameter headers.
Every algorithm meeting the joint error requirement needs $\Omega(\mathcal L'(v,Z,\tau))$ bits, even when $Q$ is known in advance and information from earlier reports is retained.
The event sequence is fixed independently of the algorithm's coins and answers.
The algorithm's hard cap holds even at intermediate prefixes where $d>Z$.
The lower-bound streams can satisfy $d\le Z$ at every prefix.
\end{theorem}

To prove the matching dependence on $Q$, we repeatedly delete and restore edges of one fixed source graph, keeping the historical deficit at most $Z$ throughout.
An algorithm may retain information acquired during these updates and reports, so a later round cannot be treated as a new independent source instance.

We use the low-error-round approach for tracking lower bounds of B{\l}asiok~\cite[Section~7, Lemma~29 of the full version]{Blasiok2018}.
That lemma samples independent input pairs across rounds; here all rounds concern the same source graph.
Conditioning globally on earlier success can therefore bias the source distribution and invalidate the information lower bound.
Lemma~\ref{lem:restored-source-confidence} instead normalizes the surviving state channel separately for each fixed source, while drawing the source from its original distribution.
The next query remains independent of the retained state conditional on the source.
A uniform lower bound on the probability of earlier success controls both the auxiliary error and the expected message length, preserving the space charge under the algorithm's original random coins.

There is a second issue: the selected round supplies a small average error under a fixed input distribution, rather than a pointwise guarantee for every source and query.
Lemma~\ref{lem:finite-query-law} supplies a query distribution independent of the message channel for which this average guarantee implies the required information lower bound.
This also keeps the queries fixed independently of the algorithm's answers.

\begin{lemma}[A finite distributional query law]
\label{lem:finite-query-law}
There are absolute constants $C,\beta_*>0$ such that, if $0<\beta<\beta_*$, $L=\ln(1/\beta)$, $m\ge L$, and $U\ge Cm$, there is a finite law $\mathcal D_\beta$ on $T\subsetneq S\subseteq[U]$ with $S$ uniform among the $m$-subsets.
This law has the following property.
For every randomized message satisfying the Markov condition $M\leftrightarrow S\leftrightarrow T$, a total decoder that finds a member of $S\setminus T$ with distributional error at most $\beta$ satisfies
\begin{equation}
I(S;M)=\Omega\!\left(
 L\log(2+m/L)\log(U/m)\right).                   \label{eq:sj1}
\end{equation}
The law depends only on $U,m,\beta$, not on the message channel.
Private decoder randomness and an explicit failure output are allowed.
\end{lemma}

The law chooses one of the source-encoding query sizes uniformly, then a uniform subset of that size.
The adaptive-density estimate used in Lemma~\ref{lem:pointwise-source} bounds the loss when the same decoder is reused during the encoding.
Consequently an average error bound under this fixed law yields the required information cost.
Appendix~\ref{app:finite-query-law-proof} gives the complete proof.

\subsection{A lower bound allowing learning between reports}

The communication message below is the algorithm's actual retained state immediately before one selected round.
It therefore includes all information learned from earlier reports.
Sourcewise normalization changes only the auxiliary communication experiment; the streaming algorithm is always charged under its original random coins.

\begin{lemma}[Repeated queries to one restored source]
\label{lem:restored-source-confidence}
Let $q,N,U,m$ be positive integers, $N+U\le v$, $m-1\le Z$, and $0<\delta\le1/4$.
Suppose
\begin{equation}
16\delta/q\le\beta<\beta_*,\qquad
 L=\ln(1/\beta),\qquad m\ge L,
 \qquad U\ge Cm.                                  \label{eq:sj3}
\end{equation}
A spanning-forest algorithm with joint error at most $\delta$ on all fixed streams with at most $q+1$ reports and historical deficit at most $Z$ requires space $s$ satisfying
\begin{equation}
s+1=\Omega\!\left(
 NL\log(2+m/L)\log(U/m)\right).                    \label{eq:sj4}
\end{equation}
This holds for both peak-space measures of Appendix~\ref{app:machine-conventions}.
The source streams have $d\le Z$ at every prefix, and their graphs are connected at every report.
The original advertised headers are unchanged.
\end{lemma}
\begin{proof}
\emph{One source and legal restored cycles.} Independently choose $N$ uniform $m$-subsets $X=(S_1,\ldots,S_N)$ of a right universe of size $U$.
Include a fixed clique on the right vertices so that every report graph is connected.
Its edges remain present at every report and therefore contribute nothing to the report-time historical deficit.
Left vertex $i$ has exactly the neighbors $S_i$.
If $N+U<v$, attach each unused vertex to one fixed right vertex by a fixed edge.
Insert these fixed edges and the source edges in a canonical order.
The insertion prefix has the same length for every source.

At each of $q$ rounds, independently conditional on $X$, choose a uniform row $I$ and draw $T\mid S_I$ from the law of Lemma~\ref{lem:finite-query-law}.
Delete $T$ in that row, request a forest, and restore $T$.
After all $q$ cycles, we request one final report on the restored base graph.
Under the transducer model of Appendix~\ref{app:machine-conventions}, a jointly successful run must produce every requested answer in finite time.
This final request therefore ensures that such a run completes all preceding restorations and report cleanup; the analysis stops at the final response.

Each cycle returns to the same graph and never introduces a new historical name.
Its maximum deficit is $|T|\le m-1\le Z$.
At its report every left vertex retains a neighbor, and the right clique is intact, so the graph is connected.
If synchronized cycle boundaries are desired, after restoration perform $m-1-|T|$ deletion--reinsertion pairs on one fixed clique edge.
Each cycle then has length $2(m-1)+1$, independently of $I,T$.
The padding has deficit at most one; when $m=1$ it is empty.
It does not change any report graph or the ever-inserted set.
The argument below uses these padded cycles.
All input streams are fixed independently of the algorithm's private coins and responses.

\emph{Finite simulation and one round of low conditional loss.} This source and query law has finite support.
By Lemma~\ref{lem:finite-simulation}, choose one finite analytic cutoff $\ell_{\rm cut}$ for the source prefix and for each entire cycle, losing at most an additional $\delta$ on every complete stream.
The closing report justifies completion of each preceding segment on the original jointly successful event.
For $1\le j\le q$, let $E_j$ be the event that the source prefix and the first $j$ cycles complete within these cutoffs and all their reports are correct.
Let $E_0$ denote the sure event, so that any timeout during the source prefix belongs to $E_0\setminus E_1$.
Thus
\begin{equation}
\Pr(E_q\mid X=x)\ge1-2\delta\ge\tfrac12
       \quad\text{for every source }x.             \label{eq:sj5}
\end{equation}
Indeed, the same lower bound holds for every fixed complete query sequence, and can then be averaged conditional on $x$.
Let $\lambda_j=\Pr(E_{j-1}\setminus E_j)$.
These discarded masses are disjoint, so
\begin{equation}
\sum_{j=1}^q\lambda_j=1-\Pr(E_q)\le2\delta.       \label{eq:sj6}
\end{equation}
Choose a fixed index $j$ with $\lambda_j\le2\delta/q$.
Its choice depends only on the algorithm and the finite input law.

\emph{Normalize the channel separately at each source.} Set $a(x)=\Pr(E_{j-1}\mid X=x)$, so $a(x)\ge1/2$ by \eqref{eq:sj5}.
For each fixed source $x$, let the message channel sample the original state at the boundary immediately before round $j$, conditional on $E_{j-1}$ for this particular $x$.
At $j=1$, there is no past event; simulate the source prefix with its cutoff and tag a prefix timeout as failure.
The source marginal for this communication experiment is again the original uniform product law.
For $j\ge2$, an equivalent implementation lets an encoder knowing $x$ repeatedly sample its old query history and private bits until its simulated past passes the cutoffs and is correct.
The auxiliary channel transmits a tagged original boundary state, or the constant-size failure symbol for a prefix timeout at $j=1$.
This message $M$ includes any information the algorithm retained from earlier reports.
Query histories and sampling-trial counters remain private to the encoder; the streaming space premise continues to use the original coin law.

The next row and query are fresh, conditional on $X$.
Let $p_{\mathrm{err}}$ be the average error probability of extracting an edge in that row from its report, with the current simulation cutoff and any failure tag counted as errors.
A current query error entails loss of $E_j$.
Therefore, under the newly normalized channel,
\begin{equation}
\begin{split}
 p_{\mathrm{err}}
 &\le\E_X
       \frac{\Pr(E_{j-1}\cap E_j^c\mid X)}{a(X)}\\
 &\le2\Pr(E_{j-1}\cap E_j^c)
  =2\lambda_j\le4\delta/q.
 \end{split}
 \label{eq:sj7}
\end{equation}
The sourcewise normalization, rather than global conditioning on $E_{j-1}$, preserves the original source law.

For the expected-space assertion, let $P$ be the peak memory of the original run on the sampled complete stream, including its closing report.
The worst-input expected-peak premise gives $\E[P\mid X=x]\le s$ for every $x$, by averaging over the independent conditional query sequences.
The message is a tagged original boundary state, dominated by $P+O(1)$ before normalization.
Consequently,
\begin{equation}
\E[|M|\mid X=x]
 \le\frac{\E[P\mid X=x]+O(1)}{a(x)}
 \le2s+O(1),\qquad H(M)=O(s+1).                    \label{eq:sj8}
\end{equation}
At $j=1$ the same bound holds without division.
Hard space gives the same entropy bound by counting original states and the failure symbol.
Thus \eqref{eq:sj8} follows from the peak-space premise under the original coin law.

\emph{The indexed information sum.} Applying Markov's inequality to the uniform row choice in \eqref{eq:sj7}, at least $N/2$ rows have query error at most $8\delta/q\le\beta/2$.
For each such row $i$, expose only the preceding source rows and use the message $(S_{<i},M)$.
The original product source law and freshness of the current query given $S_i$ imply that $(S_{<i},M)\leftrightarrow S_i\leftrightarrow T_i$ forms the required Markov chain.
Future rows remain private to the encoder.
Applying \eqref{eq:sj1} gives
\[
 I(S_i;M\mid S_{<i})
 =I(S_i;S_{<i},M)
 \ge cL\log(2+m/L)\log(U/m).
\]
Summing these nonnegative row terms and using \eqref{eq:sj8} proves
\[
 \frac{cN}{2}L\log(2+m/L)\log(U/m)
 \le I(X;M)\le H(M)=O(s+1).
\]
The segment cutoffs belong to the finite simulation of Lemma~\ref{lem:finite-simulation}.
Its decoder receives only $T_i$, $(S_{<i},M)$, and fresh private coins.
This proves \eqref{eq:sj4}.
\end{proof}

\subsection{Joint space optimality and its saturation point}

The preceding lemma isolates the cost of repeated queries.
Optimizing the source size now gives the same entropy envelope as in the one-report theorem, with confidence parameter $\tau=1+\log(Q/\delta)$.

\begin{proof}[Proof of Theorem~\ref{thm:joint-report-implementation}]
Write $B_Z=\min\{A^2,\log^2(Z+2)\}$.
The one-report lower bound in Theorem~\ref{thm:historical-deficit-space} already gives $\Omega(vAB_Z)$.
It also proves \eqref{eq:sj9} when $Q=1$.
Suppose $Q\ge2$, put $q=Q-1$, and take $N=\lfloor v/2\rfloor$, $U=\lceil v/2\rceil$.
Choose an absolute $c>0$ sufficiently small that $m\le cv$ implies $U\ge Cm$.
For any integer
\[
 1\le m\le\min\{Z+1,\lfloor cv\rfloor\},
 \qquad
 \Phi_m=\tau\log(1+m/\tau)\log(1+v/m),
\]
the argument below proves the lower bound $\Omega(v\Phi_m)$.

If $\tau$ or $m$ is below a sufficiently large absolute constant, then $\Phi_m=O(AB_Z)$: bounded $m$ gives $\Phi_m=O(A)$, and bounded $\tau$ gives $\Phi_m=O(A\log(m+1))=O(AB_Z)$.
Otherwise, choose
\begin{equation}
L=\min\left\{\ln\frac{q}{16\delta},\frac m2\right\},
 \qquad \beta=e^{-L}.                             \label{eq:sj10}
\end{equation}
The absolute thresholds ensure $L>\ln(1/\beta_*)$.
Since $q\ge Q/2$, one has $L=\Theta(\min\{\tau,m\})$, as well as $\beta\ge16\delta/q$ and $m\ge L$.
Lemma~\ref{lem:restored-source-confidence} applies and gives
\begin{equation}
s+1=\Omega\!\left(
 vL\log(2+m/L)\log(U/m)\right)
     =\Omega(v\Phi_m).                            \label{eq:sj11}
\end{equation}
The last comparison holds in both regimes $\tau\le m$ and $\tau>m$, with absolute constants.
The cap $m\le cv$ makes $\log(U/m)$ comparable to $\log(1+v/m)$.

Maximize over the displayed integers $m$ and use Lemma~\ref{lem:nearlinear-continuous-optimizer}.
The capped endpoint $\min\{Z+1,\lfloor cv\rfloor\}$ is within an absolute factor of $\mu$ for sufficiently large $v$.
The maximum is therefore
\[
 \Theta\!\left(
 \tau\log(1+x_\tau/\tau)\log(1+v/x_\tau)\right).
\]
The space-accounting convention covers bounded $v$.
Combining this with the one-report baseline proves the lower bound.
Its repeated-source confidence family has connected report graphs; no such strengthening of the single-report baseline is needed here.

For joint recovery, we instantiate the read-only algorithm of Theorem~\ref{thm:historical-deficit-space} at error $\delta/Q$.
On a fixed input stream its at most $Q$ report prefixes are fixed independently of the coins; a union bound gives joint error at most $\delta$.
Its stored state is unchanged by reports, so the same per-prefix guarantee applies despite all earlier reports.
The algorithm uses $O(\mathcal L'(v,Z,\tau))$ bits, matching the optimal space order in \eqref{eq:sj9}.
Its hard space cap and polynomial per-event running time follow from Theorem~\ref{thm:historical-deficit-space}.

The finite parameter implementation in Appendix~\ref{app:joint-header-implementation} shows that this choice also has the claimed bounds when $Q$, $Z$, and the confidence exponent arrive as binary headers.
It preserves the error allowance under rounding and uses the deterministic algorithm when the parameters saturate.
The same hard cap bounds the expected peak memory under the original random coins.
This verifies the joint correctness and space guarantees of the reused algorithm.
Finally, $\log(Q/\delta)\le\tau\le\tfrac32\log(Q/\delta)$.
Applying the constant-factor comparisons at the end of the proof of Theorem~\ref{thm:historical-deficit-space} with error $\delta/Q$ gives
\[
 \mathcal L(v,Z,\log(Q/\delta))
 \le \mathcal L'(v,Z,\tau)
 \le 8\mathcal L(v,Z,\log(Q/\delta)).
\]
Thus the algorithm and the matching space lower bound above establish the simplified finite-report, positive-error statements of Theorem~\ref{thm:joint-report-space}.
\end{proof}

\paragraph{Saturation and repeated queries.}
For $\tau\ge\mu$ the joint optimal space complexity is already the zero-error space order $\Theta(v\mu\log(1+v/\mu))$.
In particular, with fixed positive error, a report allowance exponential in $\mu$ already requires this much space.
If no report-count bound is supplied in advance, any algorithm required to meet the same fixed joint error bound on every finite promised event sequence therefore needs the zero-error space order, even when positive error is allowed.
The deterministic algorithm is space-optimal for zero-error recovery already with one report.

The report factor is $Q$, not $\min\{Q,Z\}$.
Each cycle above restores the same historical source, so the number of queries does not increase the absent-name budget.
The construction therefore allows arbitrarily many tests of the same source while respecting the same historical-deficit promise throughout.

\section{Limitations of Retained Linear Sketches under a Historical-Deficit Bound}
\label{sec:retained-linear-limitations}

Kallaugher and Price~\cite{KP20} establish space separations between turnstile streaming and linear sketching for streams with restricted length or intermediate values.
Section~\ref{sec:deletion-budget-proof} already determines the optimal space among all algorithms under the report-time historical-deficit promise and gives an algorithm attaining it.
This section proves an additional lower bound for algorithms retaining only graph-independent setup and a linear sketch of the current graph.
In some parameter ranges, this class requires a polynomial factor more space than the already established optimum, attained by our algorithm retaining historical edges.
The comparison uses the same input promise and error requirement for both classes, and both lower bounds already hold on streams with at most $Z$ total deletion operations.

The distinction is about the entire persistent state.
A decoder may be nonlinear and use fresh randomness; the restriction is that no additional information about the update history survives between events.
This allows a communication decoder to subtract a query virtually, without spending any of the stream's deletion budget.

\subsection{The retained-state model}
\label{sec:retained-state-model}
A \emph{standard retained linear sketch} chooses a finite field $F$ from its parameters and samples graph-independent data $\rho$ before the stream.
Its entire persistent state at each event boundary, including its encoding length, is an effective canonical encoding of $(\rho,M_\rho x)$.
Here $x$ is the current adjacency vector, and $M_\rho:F^{\binom{v}{2}}\to F^{s_\rho}$ is linear with columns computable from the stored description.
Thus the pair determines both the stored bits and their length, and its canonical encoding can be computed from the pair.
Setup and updates finish almost surely.
The report law depends only on this pair and fresh coins; failure and nonresponse count as errors.
All descriptions and temporary workspace count toward peak space.
This standard linear interface is stricter than a controller retaining linear sketches together with nonlinear state.

\subsection{A matching class-specific lower bound and the resulting space gap}

\begin{corollary}[Polynomial space separation at positive error]
\label{cor:polynomial-linear-gap}
Fix constants $0<\alpha<\beta<1$, let $Z=\lfloor v^\alpha\rfloor$, and suppose $t=1+\log(1/\delta)=\Theta(v^\beta)$.
For one report under the historical-deficit bound $Z$, the optimal space complexity among general algorithms is $\Theta_{\alpha,\beta}(v^{1+\alpha}\log v)$, whereas among standard retained-linear sketches it is $\Theta_{\alpha,\beta}(v^{1+\beta}\log^2v)$.
Both space lower bounds already hold for the subclass with at most $Z$ total deletion operations.
Consequently, their ratio is $\Theta_{\alpha,\beta}(v^{\beta-\alpha}\log v)$.
The comparison uses exactly the retained-state class of Section~\ref{sec:retained-state-model} and holds for both space metrics.
The constants also depend on the fixed comparisons in the hypothesis on $t$.
\end{corollary}
\begin{proof}
For general algorithms, $\mu=\min\{Z+1,v\}=\Theta(v^\alpha)$, $t/\mu\to\infty$, and $x_t=\mu$ for sufficiently large $v$.
Thus the confidence term in \eqref{eq:analysis-frontier} is
\[
 vt\log(1+\mu/t)\log(1+v/\mu)
 =\Theta_\alpha(v^{1+\alpha}\log v),
\]
using $\log(1+u)=\Theta(u)$ for $0<u\le1$.
The baseline $O(vA^3)$ is smaller, so Theorem~\ref{thm:budget-space-implementation} gives the claimed optimal space complexity: its algorithm works under the report-time historical-deficit promise, and its lower bound already holds in the smaller subclass with at most $Z$ total deletion operations.

Proposition~\ref{prop:retained-binary-sketch} supplies a retained-linear algorithm: since $A=o(t)$ and $t=\Theta(v^\beta)$ with $\beta<1$, its memory use is $O_\beta(vt\log^2v)=o(v^2)$, independently of $Z$.
For the matching lower bound, it suffices to assume correctness only on the total-deletion subclass.
Replay the proof of the single-row interface \eqref{eq:ic1}, with virtual subtraction instead of streaming deletions.
Set
\[
 N=\lfloor v/2\rfloor,\qquad U=\lceil v/2\rceil,\qquad
 m=\lfloor\sqrt{vt}\rfloor,\qquad L=\ln(1/(2\delta)).
\]
For sufficiently large $v$, $L=\Theta(t)$, $m\ge L$, $U\ge Cm$, and $2\delta<\beta_*$, where $C,\beta_*$ are from Lemma~\ref{lem:pointwise-source}.
Draw $N$ independent uniform $m$-subset rows $S_i\subseteq[U]$ and insert their bipartite edges in one common prefix.
This prefix has no deletions and belongs to that subclass for the same advertised value of $Z$.

Alice sends her original boundary state, including $\rho$.
For a query $T\subsetneq S_i$, let $y$ be the adjacency of the edges from row $i$ to $T$.
Bob computes $M_\rho x-M_\rho y=M_\rho(x-y)$, canonically re-encodes the state, and invokes the report decoder with fresh coins.
The residual graph is simple.
By the exact retained-state definition, its state and report law equal those of its insertion-only presentation under the \emph{same} advertised $Z$.
That presentation uses no deletions and has historical deficit zero, so it is legal with this parameter $Z$.
Therefore, the error is at most $\delta$ for every fixed source and query.
A correct spanning forest contains an incident edge in row $i$ and supplies a member of $S_i\setminus T$.
Bob performs the subtraction and re-encoding locally in the communication protocol, so this step consumes none of the stream's deletion budget.

Write $s$ for either space charge.
Each source prefix has a legal immediate-report completion, so its original expected state length is at most $s+O(1)$.
Since the source/query family is finite, almost-sure setup and updates and finite correct report paths permit a common finite cutoff with total additional error at most $\delta$.
Let $M_0$ be the tagged original prefix state when reached, and a constant-size timeout symbol otherwise.
Then
\[
 H(M_0)\le\E|M_0|+
 \log\bigl(e(\E|M_0|+1)\bigr)=O(s+1).
\]
The cutoff produces total decoder channels and does not condition the original coin law.
For row $i$ expose only preceding rows: the message is $(S_{<i},M_0)$ and future rows are sampled privately.
Its error for every fixed $(S_i,T)$, averaged over the other independent rows, is at most $2\delta$; the query is independent of the message given $S_i$.
Apply \eqref{eq:pe1} with error $2\delta$ and sum by the information chain rule to obtain
\[
 O(s+1)\ge H(M_0)\ge
 \sum_{i=1}^N I(S_i;M_0\mid S_{<i})
 =\Omega\!\left(NL\log(2+m/L)\log(U/m)\right).
\]
Both logarithmic factors have order $\Theta_\beta(\log v)$, since $m/L$ and $U/m$ have order $\Theta(v^{(1-\beta)/2})$.
Hence $s+1=\Omega_\beta(vt\log^2v)$ for both metrics.
Combining the two optimal space complexities and dividing proves the ratio.
\end{proof}

A linear image of current adjacency attains the unrestricted optimum.
Under the historical-deficit promise, our algorithm retaining historical edges attains the smaller optimal space, while standard retained-linear sketches cannot match it in the stated parameter range.
The lower bound concerns the retained-state class defined above; algorithms that combine linear sketches with a separate historical edge set and its live bits belong to the general class.

\clearpage
\appendix

\section{Graph-Stream Conventions and Finite Simulation}
\label{app:machine-conventions}
The vertex universe is $[v]$, $v\ge2$, and the edge names are the $\binom{v}{2}$ loopless undirected pairs.
An insertion adds an absent name and a deletion removes a present name; every prefix is a simple graph.
The historical-deficit task has an advance integer parameter $Z\ge0$ and requires at most $Z$ previously inserted edge names to be absent at each report.
Intermediate deficits and the total number of deletion operations are unrestricted.
The total-deletion subclass permits at most $Z$ deletion operations over the entire stream; every such stream satisfies the historical-deficit promise.
The fully dynamic task has no deletion header or promise.
A finite legal dynamic graph stream contains at most $Q$ REPORT symbols, or asks for one answer at EOF.
Its length is not announced and the stream is fixed independently of the algorithm's private randomness.
An answer is any spanning forest consisting of current original edge names.
Failure, malformed output, and nonresponse count as errors.
A zero-error algorithm may use randomness, but must return every requested correct answer in finite time with probability one on each promised input.

An algorithm is an event-driven private-randomness transducer.
Its future law is determined by its complete state and the next input symbol.
Fresh bits are independent on request.
Retained seeds, readable indices, head positions, control registers, input/output buffers, and all update and decoding scratch count toward space.
There is no free clock, reusable random tape, input-position register, or archive of previous outputs.
A complete state has a charged binary encoding, so hard peak $s$ allows fewer than $2^{s+1}$ states.
An update ends before the next symbol is requested; a report is assessed when its answer is emitted.
The algorithms presented here are uniform with their stated all-tape work bounds.
Lower bounds permit arbitrary nonlinear transitions and unbounded computation.

Expected space is the supremum, over fixed legal dynamic graph streams, of the expected peak space over the algorithm's private randomness, without conditioning on success.
For each peak-space measure, the optimal space complexity is the infimum over algorithms satisfying the specified output and error requirements on every input allowed by the promise.
Space is measured as peak space plus one to include constant endpoints.
For the final graph-space bounds, which are uniformly $O(v^2)$, this convention absorbs bounded vertex counts into the absolute constants.
Every lower-bound source uses synchronized prefixes and arrival boundaries.
Finite cutoffs in communication proofs are analytic devices, not supplied resources.
All logarithms are binary unless written $\ln$.

Setup reads the structural parameters and then the confidence parameter as delimited canonical binary fields; confidence may be supplied by an integer exponent together with a separate zero-error flag.
Only the current bit is available as input, and every accumulator retained while parsing counts toward space.
Because the headers can be arbitrarily long, a field that has reached its saturation threshold is still read through its delimiter without retaining a position counter.
In particular, the deficit budget $Z$ can be saturated at its proved vertex threshold, so longer descriptions add reading time without increasing the retained value.
Reading the parameter headers adds work linear in their actual length; subsequent per-event running time has the dependence stated in each theorem.
Randomized forest claims concern streams fixed independently of the private coins.
Deterministic invariants hold pathwise and therefore also permit legal answer-adaptive input selection.

\subsection{Finite-family simulation}
\begin{lemma}[Finite-family simulation]
\label{lem:finite-simulation}
Consider a finite family of fixed legal update/request streams and any algorithm that produces all requested correct responses in finite time with probability at least $1-\delta$ on each stream.
For every $\eta>0$, their communication simulations may use one finite computation cutoff, with error at most $\delta+\eta$ per stream.
At a common boundary the message is a tagged original complete state or a constant-size failure symbol.
If every source prefix has a promised completion, its message entropy is $O(s+1)$ for hard peak space $s$, and $O(S+1)$ for worst-input expected peak space $S$.
The cutoff need not be stored by the original streaming algorithm.
\end{lemma}
\begin{proof}
For a fixed finite stream, the probability of producing all requested correct responses only after time $T_{\mathrm{sim}}$ tends to zero as $T_{\mathrm{sim}}\to\infty$.
Take a common $T_{\mathrm{sim}}$ making this probability at most $\eta$ on every stream in the finite family.
Simulate the prefix and each tested continuation for at most $T_{\mathrm{sim}}$ computation steps, and return failure on timeout.
Any run producing its required correct responses within $T_{\mathrm{sim}}$ steps also passes both parts; it need not halt after its final response.
A valid early terminal output is represented as an absorbing complete state; report completion is determined by the simulated execution.
Additional simulations beyond a true legal endpoint may be cut off or assigned failure without a correctness premise.

Without conditioning on successful reach, use a constant-size tag on the reached complete state and a constant-size timeout message.
For each prefix choose any one promised completion to bound its expected state length by $S+O(1)$.
A binary string of expected length $L$ has entropy at most $L+\log_2(e(L+1))=O(L+1)$.
The hard bound follows by counting state strings.
The communication observer's cutoff is fixed independently of the hidden input and private coins.
If $J$ legal simulations are unioned, choose the cutoff loss for each stream to be at most $\eta/J$; no uniform algorithmic running-time assertion is needed.
\end{proof}

\section{Auxiliary Probability and Information Arguments}
\label{app:probability-information}

\subsection{Sampling estimates for cut recovery}
\label{sec:sampling}

This section isolates the elementary probability facts from the graph and coding implementations.
Its random variables are sampled edge \emph{names}, not update occurrences.

\begin{lemma}[Explicit hashes and bounded-independence tails]
\label{lem:sampling}
Let $N_E=\binom{v}{2}$, $A=\lceil\log(v+1)\rceil$, and $b\ge8$.
A bank has a $(2b+2)$-wise independent map from the $N_E$ names to $h=\lceil\log(N_E+1)\rceil+8$ bits, with $O(bA)$ seed bits and polynomial setup and evaluation work in $v+b$.
Each level $j\le h$ retains names whose first $j$ hash bits are zero.

For any fixed set of names, its sampled size $X$ at rate $p=2^{-j}$ has mean $\mu$ and satisfies
\begin{equation}
\E|X-\mu|^{2b}\le4(4b\mu+16b^2)^b.
 \label{eq:sm1}
\end{equation}
Consequently,
\begin{equation}
\begin{aligned}
 \mu\ge64b&\ \Longrightarrow\ \Pr(X<b)\le2^{-b},\\
 \mu\le128b&\ \Longrightarrow\ \Pr(X>512b)\le2^{-b}.
 \end{aligned} \label{eq:sm2}
\end{equation}
If a fixed set of currently absent historical names has sampled mean at most $b/64$, its sampled size exceeds $b$ with probability at most $16^{-(b+1)}$.
If the \emph{whole universe} is sampled, a cap of $C=\min\{N_E,\lceil16N_E p+2b\rceil\}$ overflows with probability at most $16^{-2b}$.
All these probabilities are under the bank's original sampling law.
\end{lemma}
\begin{proof}
Construct $\mathbb F_{2^h}$ by enumerating monic binary polynomials and testing irreducibility.
There are $2^h=O(v^2)$ candidates; repeated squaring and gcd use $O(h)$ bits and polynomial work.
Evaluate a random polynomial of degree at most $2b+1$ on distinct field encodings of the names.
Vandermonde independence gives the claimed joint law, including when fewer than $2b+2$ names exist.
The full field word is evaluated before its first $j$ bits are tested.

The case $\mu=0$ is deterministic, so assume $\mu>0$.
Every term in the centered $2b$th moment involves at most $2b$ distinct names, so it equals the fully independent moment.
In that case, for $|s|\le1$, $\E e^{s(X-\mu)}\le e^{\mu s^2}$.
For $u\le2\mu$, applying Markov's inequality with $s=\pm u/(2\mu)$ gives the two-sided tail $2e^{-u^2/(4\mu)}$.
For $u>2\mu$, the choices $s=\pm1$ give $2e^{\mu-u}\le2e^{-u/2}$.
Thus the tail is bounded for all $u\ge0$ by $2e^{-u^2/(4\mu)}+2e^{-u/2}$.
Integrating $2b u^{2b-1}$ against this bound gives $2(4\mu)^b b!+2\,2^{2b}(2b)!$, which is bounded by \eqref{eq:sm1}.

For $\mu\ge64b$, the ratio $(4b\mu+16b^2)/(\mu-b)^2$ decreases with $\mu$.
Markov's inequality therefore bounds $\Pr(X<b)$ by $4(272/3969)^b\le2^{-b}$.
For $\mu\le128b$, the upper tail is at most $4(528/384^2)^b\le2^{-b}$.
For an integer $r$ not exceeding the independence order, $\E(X)_r\le\mu^r$, where $(X)_r$ is the falling factorial.
Taking $r=b+1$ proves the absent-name bound using $r!\ge(r/e)^r$.
Taking $r=2b$, each factor in $(C+1)_{2b}$ is at least $16N_E p$, which proves the universe cap when $C<N_E$.
When $C=N_E$, overflow is impossible.
All seeds and scratch have the stated bounds.
Different banks use independent seeds; the estimates within a bank apply to each fixed set of names.
\end{proof}

For a nonempty cut of size at most $512b$, level zero is decoded exactly.
Otherwise, either the deepest retained sample has mean at least $64b$, or a shallower dyadic level has mean in $(64b,128b]$.
Equations \eqref{eq:sm1}--\eqref{eq:sm2}, followed by the appropriate bound for the top-level data structure or bitmap, are precisely the per-cut guarantees used in Sections~\ref{sec:unrestricted-proof} and~\ref{sec:deletion-budget-proof}.
This reasoning also supplies $b$ distinct sampled neighbors in a singleton bootstrap.
It remains valid conditional on earlier banks, because the input stream is fixed and the current bank is fresh.

\subsection{Pointwise source encoding}
\label{app:pointwise-source-proof}

This pointwise-error adaptation of the lossless source encoding~\cite{KNPWWY2017,NelsonYu2019} uses only the adaptive-filling density calculation \eqref{eq:sparse-adaptive-density}--\eqref{eq:sparse-density-loss} in Appendix~\ref{sec:graph-frontier}.
It does not invoke the perturbed input-law conclusion of Lemma~\ref{lem:sparse-perturbed-ur}.

\begin{proof}[Proof of Lemma~\ref{lem:pointwise-source}]
Set $\alpha=100/L$, $q=1-\alpha$, $z=\alpha m$, and
\[
 R=\left\lfloor\frac{\ln z}{100\alpha}\right\rfloor,
 \qquad r_i=\lfloor m(1-q^i)\rfloor\quad(0\le i<R),
 \qquad r_R=m.
\]
Choose $\beta_*$ small enough that $L\ge20000/\ln100$.
These choices ensure $\alpha\le1/2$, $R\ge2$, and $R=\Theta(L\log(2+m/L))$ uniformly, including $m=\Theta(L)$.
For $i<R-1$, $r_{i+1}-r_i\ge z^{49/50}-1>1$, because $q^R\ge e^{-2\alpha R}\ge z^{-1/50}$.
In particular, all queried sets have size less than $m$.

Expose a fresh independent decoder tape and an independent uniform permutation of $[U]$ at each of the $R$ stages; denote their entire collection by $\Pi$.
These tapes are independent of $(S,M)$.
Start at $T_0=\varnothing$ and initialize the recovered set $S_{\mathrm{rec}}=\varnothing$.
At stage $i$, run the decoder on $(M,T_i)$.
If its output belongs to $S\setminus T_i$, record a success flag, collect the output in $S_{\mathrm{rec}}$, and add it to $T_i$.
Otherwise, record a failure flag.
Fill to size $r_{i+1}$ with the first available members of $S$ in the stage's permutation.
This operation always exists, and the recovered members are distinct.

For fixed $S,M$, the distribution of $T_i$ satisfies
\begin{equation}
\Pr(T_i=B\mid S,M)
 \le \binom{m}{r_i}^{-1}
       \prod_{j=0}^{i-1}\frac{m-r_j}{r_i-r_j}
 \le \beta^{-1/4}\binom{m}{r_i}^{-1}
 \label{eq:pe2}
\end{equation}
for every $r_i$-subset $B$ of $S$; the empty product at $i=0$ is one.
The combinatorial adaptive-filling and rounding calculations \eqref{eq:sparse-adaptive-density}--\eqref{eq:sparse-density-loss} of Lemma~\ref{lem:sparse-perturbed-ur}, with starting index zero, prove \eqref{eq:pe2}, uniformly down to $m=L$.
They allow every adaptive forced choice; averaging over the fresh decoder tapes therefore also permits randomized choices.

At stage $i$ the current decoder tape is fresh, independently of the history that selected $T_i$.
For an independent uniform $r_i$-subset of $S$, pointwise error and averaging over $S,M$ give error at most $\beta$.
The density domination \eqref{eq:pe2} consequently gives actual stage error at most $\beta^{3/4}$.
Consequently,
\begin{equation}
\E|S_{\mathrm{rec}}|\ge R(1-\beta^{3/4})\ge R/2.
 \label{eq:pe3}
\end{equation}
Thus density domination controls the adaptive query sequence while only the decoder tapes need to be fresh.

Given $M$, $\Pi$, the $R$ flags, and the residual set $B=S\setminus S_{\mathrm{rec}}$, a lossless decoder reconstructs $S$.
It reruns each successful query, and fills from the unused residual members in the stage permutation.
An element recovered at a later successful stage was not a filler at an earlier stage; excluding such future elements therefore leaves every earlier list of first fillers unchanged.
Induction reconstructs the same queried sets and outputs.
Fresh decoder tapes are shared in this lossless reconstruction, so no determinization of the original decoder is needed.

The flags reveal $a=|S_{\mathrm{rec}}|$, so the conditional residual entropy is at most $\E\log\binom{U}{m-a}$.
Since $\Pi$ is independent of $(S,M)$, losslessness and the entropy chain rule imply
\begin{equation}
I(S;M)\ge\log\binom{U}{m}-R-
                   \E\log\binom{U}{m-|S_{\mathrm{rec}}|}
 \ge\E|S_{\mathrm{rec}}|\log\frac{U-m}{m}-R.
 \label{eq:pe4}
\end{equation}
The last inequality follows by cancelling factorials.
Choose $C$ large enough that $\log((U-m)/m)$ absorbs the flag cost in \eqref{eq:pe4}.
Equations \eqref{eq:pe3}--\eqref{eq:pe4} and the estimate for $R$ prove \eqref{eq:pe1}.
\end{proof}

\subsection{External-information accounting for the coordinate reduction}
\label{app:coordinate-reduction-proof}

\begin{proof}[Proof of Lemma~\ref{lem:external-coordinate}]
Let $\pi$ be the joint law of $(X^k,Y^k,M,O)$ and let $W$ be the event that all $k$ answers are correct.
The proof uses the coordinate construction of Nelson and Yu~\cite{NelsonYu2019}, as follows.
Choose a uniform permutation $\kappa$ of $[k]$, independently of $\pi$, and independent uniform positions
\[
 s\in\{k/2+1,\ldots,k\},\qquad t\in\{1,\ldots,k/2\}.
\]
The first view sets $i=\kappa(s)$, $g=\kappa([1,s-1])$, and $h=\kappa([t,k])\setminus\{i\}$.
Equivalently, the second view sets $i=\kappa(t)$, $g=\kappa([1,s])\setminus\{i\}$, and $h=\kappa([t+1,k])$.
These induce the same distribution of $(i,g,h)$: in both views, the disjoint sets $g\setminus h$, $g\cap h$, $h\setminus g$, and $\{i\}$ have sizes $t-1$, $s-t$, $k-s$, and $1$, respectively, and their labels are uniform under the random permutation $\kappa$.
In both views $i\notin g\cup h$.
Draw $r$ from $\pi[X_g,Y_h\mid W]$.

For each $\lambda=(i,g,h,r)$, let $\mathcal D'_\lambda$ be the law of $(X_i,Y_i)$ conditional on $(X_g,Y_h)=r,W$.
On input $(x,y)$ the generalized protocol samples its message from
\[
 \pi[M\mid X_i=x,Y_i=y,(X_g,Y_h)=r,W],
\]
and, given $(m,y)$, samples its output from
\[
 \pi[O_i\mid Y_i=y,M=m,(X_g,Y_h)=r,W].
\]
The message may depend on Bob's input; the argument below bounds this dependence.
Private output coins can be removed by the Bayes-optimal determinization in Lemma~\ref{lem:sparse-perturbed-ur}.

\emph{External-information accounting.} Let $\mathsf{A}=(X^k,Y^k)$.
Under every original one-way protocol,
\begin{equation}
I_\pi(\mathsf{A};M)=I_\pi(X^k;M)=\mathcal J.
 \label{eq:bi5}
\end{equation}
For fixed $\kappa,t$, write $X_{\rm head}=X_{\kappa([1,k/2])}$, $X_{\rm tail}=X_{\kappa([k/2+1,k])}$, and $Y_{\rm suffix}=Y_{\kappa([t,k])}$.
Average over $r$ and telescope over $s$ in the first view.
The resulting mean coordinate information is exactly
\begin{equation}
\begin{aligned}
\E_\lambda I_{\mathcal D'_\lambda}(X;\widehat M\mid Y)
 &=\frac2k\E_{\kappa,t}
   I_\pi(X_{\rm tail};M\mid X_{\rm head},Y_{\rm suffix},W)\\
 &\le\frac2k I_\pi(\mathsf{A};M\mid W)\\
 &\le\frac{2(\mathcal J+1)}{kp}
 \le\frac{4(\mathcal J+1)}k.
\end{aligned}
\label{eq:bi6}
\end{equation}
The first inequality is just the information chain rule: the conditioning tuple $(X_{\rm head},Y_{\rm suffix})$ is a subtuple of $\mathsf{A}$, so it and $X_{\rm tail}$ reveal no more than $\mathsf{A}$.
The second uses \eqref{eq:bi3} and \eqref{eq:bi5}.
There is no additional conditioned message, output, future source row, or decoder coin in this step.
The permutation only chooses coordinates and is independent before the success conditioning.

\emph{Approximate one-way property.} Using the second view and telescoping over $t$, the mean value of $I(\widehat M;Y\mid X)$ is
\begin{equation}
\frac2k\E_{\kappa,s}
 I_\pi\bigl(Y_{\kappa([1,k/2])};M
   \mid X_{\kappa([1,s])},Y_{\kappa([k/2+1,k])},W\bigr).
 \label{eq:bi7}
\end{equation}
Without conditioning on $W$, the two arguments are conditionally independent: all source $X$ coordinates for the first group of $Y$ coordinates are already supplied, and the pairs are independent.
The message is generated from $X^k$ alone.
Equation \eqref{eq:bi2} bounds \eqref{eq:bi7} by $2\log(1/p)/k\le2/k$.

\emph{Decoder error.} The original $O_i$ is correct under the law conditioned on $W$.
The constructed decoder resamples it without knowing $X_i$.
Pinsker's inequality bounds the resulting error by a constant times
\[
 \sqrt{I_\pi(X_i;O_i\mid Y_i,M,(X_g,Y_h)=r,W)}.
\]
Average, apply Jensen, enlarge $O_i$ to $O$, and telescope in the first view.
The quantity inside the averaged square root is at most
\begin{equation}
\frac2k\E_{\kappa,t}
 I_\pi(X_{\rm tail};O\mid X_{\rm head},Y_{\rm suffix},M,W)
 \le\frac{2\log(1/p)}k.
 \label{eq:bi8}
\end{equation}
For the final inequality use \eqref{eq:bi2}.
To check the required unconditioned independence, the omitted $Y$ coordinates have positions below $t\le k/2$, and their corresponding $X$ coordinates are all in $X_{\rm head}$.
Given these head coordinates, the omitted $Y$ coordinates are independent of the tail, the message, and the supplied $Y$ coordinates.
The decoder depends only on $Y^k,M$ and its fresh private coins.
This proves mean error $O(k^{-1/2})$ even when $W$ involves those decoder coins.

\emph{Distribution perturbations.} For each $(i,g,h)$, the original $X_i$ and $Y_i$ are individually independent of $(X_g,Y_h)$.
Apply \eqref{eq:bi1} and the relative-entropy chain rule to obtain, after averaging over $r$, both marginal KL bounds at most $\log(1/p)\le1$.
For the forward conditional KL bound, use the second view and the reverse-order chain rule for the first half of the $Y$ coordinates.
Its mean is
\begin{equation}
\frac2k\E_{\kappa,s}
 \E_{\mathsf{C}\sim\pi[\mathsf{C}\mid W]}
 D_{\rm KL}\bigl(
   \pi[Y_{\kappa([1,k/2])}\mid \mathsf{C},W]
   \Vert \pi[Y_{\kappa([1,k/2])}\mid \mathsf{C}]\bigr),
 \quad
 \mathsf{C}=(X_{\kappa([1,s])},Y_{\kappa([k/2+1,k])}).
 \label{eq:bi9}
\end{equation}
The reference conditional factors into the relevant $\mathcal D[Y_j\mid X_j]$, since $s>k/2$.
The conditional relative entropy in \eqref{eq:bi9} is at most $\log(1/p)$ by \eqref{eq:bi1} and the chain rule.
This gives $2/k$.
The reverse conditional KL bound follows in the first view by the forward-order chain rule for $X_{\rm tail}$, conditioned on $X_{\rm head},Y_{\rm suffix}$; its reference conditional factors into $\mathcal D[X_j\mid Y_j]$.
It has the same $2/k$ bound.

There are seven nonnegative quantities: two marginal KL terms, two conditional KL terms, approximate-one-way information, decoder error, and coordinate information.
Applying Markov's inequality to each with a sufficiently large absolute factor and taking a union bound selects a single $\lambda$ satisfying all the displayed bounds simultaneously.
This proves \eqref{eq:bi4}.
\end{proof}

\subsection{Source-cardinality accounting in the information lower bound}
\label{sec:graph-frontier}

Constants are absolute unless otherwise indicated.
The $N$-fold relation asks for one element of $S_i\setminus T_i$ for every $i\in[N]$, where $T_i\subsetneq S_i\subseteq[U]$.

Lemma~\ref{lem:sparse-perturbed-ur} refines the source-size accounting in Nelson and Yu~\cite[Appendix~B]{NelsonYu2019}, whose encoding builds on Kapralov et al.~\cite[Section~3.1]{KNPWWY2017}.
The encoding below retains the source cardinality $m=|S|$ as a free parameter: the universe-only theorem chooses $m=\sqrt{U\log(1/\beta)}$, whereas here $m$ controls the total number of deletions in the lower-bound subclass.
The following proof records the rounding and perturbed-distribution estimates needed for this parameter extension.
The indexed-block application in Section~\ref{sec:indexed-block} then charges external information.

\begin{proof}[Proof of Lemma~\ref{lem:sparse-perturbed-ur}]
If the generalized protocol's output uses additional private coins, replace it, for each pair $(M,T)$, by a deterministic Bayes-optimal output under $\mathcal D'$.
This cannot increase its error and does not change the joint law of $(S,T,M)$, any KL term, or either information cost.
Communication protocols here have no computational restriction.
Write $O(M,T)$ for this deterministic output function; it is the same function in every reuse and in the lossless decoder.

\emph{Hard distribution and number of stages.} Take
\[
 \alpha=100/L,\quad q=1-\alpha,\quad z=\alpha m,\quad
 R=\left\lfloor\frac{\ln z}{100\alpha}\right\rfloor,
 \qquad r_j=\lfloor m(1-q^j)\rfloor\quad(0\le j<R).
\]
Since $m\ge L$, it follows that $z\ge100$ and $R=\lfloor L\ln(100m/L)/10000\rfloor$.
Taking $\beta_0$ small enough that $L\ge20000/\ln100$ ensures both $\alpha\le1/2$ and $R\ge2$, including at $m=L$.
Define $r_R=m$ only for the last filling step of the encoding.
Let $S$ be uniform among the $m$-subsets of $[U]$.
Choose $i_0$ uniformly from $\{0,\ldots,R-1\}$, then choose $T$ uniformly among the $r_{i_0}$-subsets of $S$.

For $j<R-1$,
\[
 r_{j+1}-r_j\ge\alpha m q^j-1
 \ge z^{49/50}-1>1,
\]
because $q^R\ge e^{-2\alpha R}\ge z^{-1/50}$.
Thus all cardinalities are distinct, all $T$ are proper subsets, and every encoding step can add a newly found element.
Also, by reducing $\beta_0$ if necessary,
\begin{equation}
 R=\Theta\!\left(L\log(2+m/L)\right).
 \label{eq:sparse-stages}
\end{equation}

\emph{An exact density bound for adaptive filling.} Fix $S$, a message $M$, and a starting index $i_0$.
Start from a uniform $r_{i_0}$-subset $T_0$ of $S$.
In stage $j$, a deterministic rule may add one selected member of $S\setminus T$, then fill uniformly at random to size $r_{j+1}$.
The rule can depend on everything exposed so far.
At every stage $i<R$, for every fixed $r_i$-subset $A\subseteq S$,
\begin{equation}
 \Pr(T_i=A\mid S,M,i_0)
 \le\binom{m}{r_i}^{-1}
       \prod_{j=i_0}^{i-1}\frac{m-r_j}{r_i-r_j}.
 \label{eq:sparse-adaptive-density}
\end{equation}
To see this, $T_0$ must lie in $A$, which has probability $\binom{r_i}{r_{i_0}}/\binom{m}{r_{i_0}}$.
Conditional on every earlier set lying in $A$, the largest probability that the next set still lies in $A$ occurs when the optional forced element is in $A$.
Its value is at most
\[
 \frac{\binom{r_i-r_j-1}{r_{j+1}-r_j-1}}
      {\binom{m-r_j-1}{r_{j+1}-r_j-1}}.
\]
If no element is forced, the corresponding probability is no larger.
Multiplying these conditional bounds and cancelling factorials gives \eqref{eq:sparse-adaptive-density}.
If a forced element lies outside $A$, the desired event has probability zero.
This proof permits adaptive choices, without assuming independence of the selected elements.

The next estimate bounds the product, including the integer rounding.
For $j<i$, writing $h=i-j$, its factor is at most
\[
 \frac{mq^j+1}{mq^j(1-q^h)-1}.
\]
Writing $b=mq^j(1-q^h)=mq^i(q^{-h}-1)$ gives $b\ge mq^i\alpha h\ge z^{49/50}h>2$.
The logarithm of this factor is therefore at most
\[
 -\ln(1-q^h)+\ln\left(1+\frac1{mq^j}\right)
                 -\ln(1-1/b)
 \le-\ln(1-e^{-\alpha h})+\frac3{z^{49/50}h}.
\]
This uses $mq^j\ge b$, $\ln(1+x)\le x$, and $-\ln(1-x)\le2x$ for $0\le x\le1/2$.
Summing, and weakening $1/h$ to $1$, gives
\[
 \ln\prod_{j=i_0}^{i-1}\frac{m-r_j}{r_i-r_j}
 \le\sum_{h\ge1}-\ln(1-e^{-\alpha h})
       +\frac{3R}{z^{49/50}}.
\]
For the decreasing nonnegative function $f(x)=-\ln(1-e^{-x})$, the sum is at most $\alpha^{-1}\int_0^\infty f(x)\,dx=\pi^2/(6\alpha)$.
Moreover,
\[
 \frac{3R}{z^{49/50}}
 \le\frac{3}{100\alpha}\frac{\ln z}{z^{49/50}}
 <\frac{1}{100\alpha}\qquad(z\ge100).
\]
Consequently, the logarithm is less than $2/\alpha=L/50$.
The following weaker bound is convenient:
\begin{equation}
 \Pr(T_i=A\mid S,M,i_0)
 \le\beta^{-1/4}\binom{m}{r_i}^{-1}.
 \label{eq:sparse-density-loss}
\end{equation}
The estimate is uniform down to $m=L$.

\emph{Lossless encoding under the actual perturbed law.} Draw $(S,T_0,M)$ according to $\mathcal D'$ and the generalized protocol.
Public randomness consists of an independent uniform permutation of $[U]$ for each stage; it is independent of this triple.
Finite marginal and forward conditional KL ensure that $|S|=m$, $T_0\subsetneq S$, and $|T_0|=r_{i_0}$ for a unique $i_0$.
Write $T_0$ and $M$, initialize the recovered set $S_{\mathrm{rec}}=\varnothing$, and reuse the output function on $(M,T)$ in stages $i_0,\ldots,R-1$.
If the output belongs to $S\setminus T$, record a one-bit success flag, collect that output in a set $S_{\mathrm{rec}}$, and add it to $T$.
Otherwise, record a zero flag.
Fill $T$ to $r_{j+1}$ using the smallest remaining elements of $S$ in that stage's public permutation.
Finally, write $S\setminus(T_0\cup S_{\mathrm{rec}})$.
Pad the flags to $R$ bits.

The decoder knows $T_0$, $M$, the flags, the final residual set, and the public permutations.
It regenerates successful outputs and fills each stage from the still unconsumed residual elements.
This reproduces the encoder: an element that is to be recovered in a later successful stage was not selected as a filler in an earlier stage, so excluding all such elements cannot change an earlier selection of the smallest fillers.
Induction on stages proves exact reconstruction.
Fresh permutations per stage make the uniform filling in \eqref{eq:sparse-adaptive-density} literal; their use does not change this losslessness argument.

Write $\mathcal I=I(S;M\mid T)$.
Uniformity of the base law's $S$ marginal and the first KL bound give $H(S)\ge\log\binom{U}{m}-K_0$.
Uniformity of the base conditional $S\mid T=t$ and the reverse conditional KL bound give
\[
 I(S;T_0)\le\log\binom{U}{m}
 -\E\log\binom{U-|T_0|}{m-|T_0|}+\eta.
\]
Conditioning on the public permutations, the message contributes exactly $\mathcal I$ information; the flags contribute at most $R$; the residual set contributes at most
\[
 \log(R+1)+
 \E\log\binom{U-|T_0|}{m-|T_0|-|S_{\mathrm{rec}}|}.
\]
The factorial ratio of the two binomial terms is at least $((U-m)/m)^{|S_{\mathrm{rec}}|}$.
Lossless decoding therefore implies
\begin{equation}
 \mathcal I\ge
  \E|S_{\mathrm{rec}}|\log\frac{U-m}{m}
  -R-\log(R+1)-K_0-\eta.
 \label{eq:sparse-encoding}
\end{equation}

\emph{Decoupling for the success analysis.} Let $P$ be the actual law of $(S,T_0,M)$.
Let $\widetilde P$ first draw $S\sim\mathcal D'[S]$, then draw $T_0\sim\mathcal D[T\mid S]$ and $M\sim P[M\mid S]$ independently conditional on $S$.
Pinsker, conditional Jensen, the bound on $I(M;T_0\mid S)$, and the forward conditional KL bound imply
\begin{equation}
 \|P-\widetilde P\|_{\rm TV}=O(\sqrt\eta)=O(\beta).
 \label{eq:sparse-decoupling}
\end{equation}
One can obtain this in two steps: first replace $M\mid S,T_0$ by $M\mid S$ under the original $(S,T_0)$ marginal, then replace $T_0\mid S$ by the base conditional.
This avoids conditioning a uniform starting subset on a message that depended on that subset.

First apply \eqref{eq:sparse-decoupling} to a single decoder test.
Its failure indicator, as a function of $(S,T,M)$, lies in $[0,1]$ and has expectation at most $\beta$ under $P$.
Thus, under $\widetilde P$, a fresh base conditional test $T$ with the same $S,M$ has expected error
\[
 p_{\mathrm{err}}\le\beta+\|P-\widetilde P\|_{\rm TV}=O(\beta).
\]
For fixed $S,M$, let $p_{\mathrm{err},i}(S,M)$ be its error on a uniform $r_i$-subset.
Since $i_0$ is uniform and independent of $M$ conditional on $S$, \eqref{eq:sparse-density-loss} gives
\begin{align*}
 \E_{\widetilde P}|S_{\mathrm{rec}}|
 &\ge\frac{R+1}{2}
  -\beta^{-1/4}\,
    \E_{S,M}\frac1R
       \sum_{i_0=0}^{R-1}\sum_{i=i_0}^{R-1}p_{\mathrm{err},i}(S,M)\\
 &\ge\frac R2-R\beta^{-1/4}p_{\mathrm{err}}
  =\frac R2-O(R\beta^{3/4}).
\end{align*}
Apply \eqref{eq:sparse-decoupling} a second time to transfer the expected number of recovered elements back to the actual law.
The two initial laws are passed through the same randomized encoding channel, using independent public permutations.
Total variation cannot increase under this channel, and $0\le|S_{\mathrm{rec}}|\le R$, so
\[
 \bigl|\E_P|S_{\mathrm{rec}}|-\E_{\widetilde P}|S_{\mathrm{rec}}|\bigr|
 \le R\|P-\widetilde P\|_{\rm TV}
 =O(R\sqrt\eta)=O(R\beta).
\]
Thus $\E_P|S_{\mathrm{rec}}|\ge R/3$ for small enough $\beta_0$.
Choose $C_0$ large enough that $\log((U-m)/m)$ absorbs the absolute $O(R+K_0)$ terms in \eqref{eq:sparse-encoding}.
Equations \eqref{eq:sparse-stages} and \eqref{eq:sparse-encoding} give the claimed lower bound.
\end{proof}

\subsection{The insertion-only baseline}
\label{app:insertion-baseline}

\begin{lemma}[Insertion-only forest baseline]
\label{lem:randomized-forest-baseline}
For $v\ge2$, a spanning-forest algorithm with error at most $1/4$ needs $\Omega(v\log(v+1))$ bits on insertion-only simple streams with one report, for both peak-space measures of Appendix~\ref{app:machine-conventions}.
\end{lemma}
\begin{proof}
For $v\ge3$, draw a uniformly random labeled tree $\mathsf T$ on $[v]$; there are $M=v^{v-2}$ such trees.
Insert its $v-1$ edges in canonical order and request one report.
The only spanning forest of a tree is that tree itself.
All insertion prefixes have the same length.

Apply Lemma~\ref{lem:finite-simulation} with one common cutoff to this finite family of streams, with extra failure probability at most $1/8$.
Let $M_0$ be the tagged original complete prefix state, or a constant-size timeout symbol.
Under every fixed tree's original randomness its expected encoded length is at most $s+O(1)$, where $s$ is the worst-input expected peak space.
Averaging preserves this bound, and consequently
\[
 H(M_0)\le\E|M_0|+
 \log\bigl(e(\E|M_0|+1)\bigr)=O(s+1).
\]
The decoder, using $M_0$ and fresh coins, estimates $\mathsf T$ with error at most $3/8$.
Map every output that is not a labeled tree on $[v]$, including timeout, to a fixed tree.
Fano's inequality and the Markov chain $\mathsf T\leftrightarrow M_0\leftrightarrow\widehat{\mathsf T}$ give
\[
 H(M_0)\ge I(\mathsf T;M_0)\ge I(\mathsf T;\widehat{\mathsf T})
 \ge \log M-1-(3/8)\log M
 = (5/8)(v-2)\log v-1.
\]
Thus $s+1=\Omega(v\log(v+1))$, including bounded $v$ by the space-accounting convention.
The hard-peak case follows from the same argument or from counting bounded complete-state encodings.
\end{proof}

\subsection{Positive error at the zero-error space order}
\label{app:positive-error-exact}

\begin{theorem}[The zero-error space order at sufficiently high confidence]
\label{thm:small-budget-confidence}
There is an absolute $c_*>0$ such that, for every fixed $K_*>0$, the following holds.
If $0<\delta\le1/4$ and the integer historical-deficit budget $Z\ge0$ satisfies
\begin{equation}
Z+1\le c_*v,\qquad Z+1\le K_*\ln(1/\delta),
 \label{eq:ic2}
\end{equation}
then the optimal space complexity for one spanning-forest report under this historical-deficit promise is
\begin{equation}
\Theta_{K_*}\!\left(
 v(Z+1)\log\left(1+\frac v{Z+1}\right)\right)
 \label{eq:ic3}
\end{equation}
for both peak-space measures.
The lower bound already holds for the subclass with at most $Z$ total deletion operations and requires only one report on a simple graph.
The deterministic zero-error algorithm of Theorem~\ref{thm:zero-error-implementation} uses the matching amount of memory and permits any finite number of reports.
The bounded-error assertion concerns fixed legal simple-graph streams.
\end{theorem}
\begin{proof}
Write $m=Z+1$ and choose $K=\max\{2,2K_*\}$.
Take $N=\lfloor v/2\rfloor$ and $U=\lceil v/2\rceil$.
Choose $c_*$ small enough that $U\ge C_0m$, where $C_0$ is the exact-law universal-relation source-size constant.
Set
\[
 \beta=e^{-m/K}.
\]
The confidence assumption and $\delta\le1/4$ imply
\begin{equation}
\frac mK\le\tfrac12\ln(1/\delta)
 \le\ln(1/(2\delta)),\qquad \beta\ge2\delta.
 \label{eq:ic4}
\end{equation}
If $m$ exceeds a sufficiently large constant depending only on $K_*$, then $\beta<\beta_0$.
Relax the algorithm's error to $\delta'=\beta/2\ge\delta$ and apply Lemma~\ref{lem:exact-row-interface} with $L'=\ln(1/(2\delta'))=m/K$.
The query uses at most $m-1=Z$ deletions, and $m\ge L'$.
The resulting lower bound is
\[
 \Omega\!\left(N\frac mK\log(2+K)\log(U/m)\right)
 =\Omega_{K_*}\!\left(
 vm\log(1+v/m)\right).
\]
Here $U/m\ge C_0$ makes the two logarithmic forms comparable.
The finite cutoff uses the relaxed positive error budget and charges original-law entropy.
If $m$ is below the stated threshold, the insertion-only Cayley/Fano lower bound $\Omega(v\log(v+1))$ implies the same result with a constant depending on $K_*$.

Theorem~\ref{thm:zero-error-implementation} gives the matching $O(vm\log(1+v/m))$ hard cap by deterministically storing historical edge names and their live bits.
This deterministic algorithm attains the stated memory use under both peak-space measures, performs polynomial work, and does not depend on $\delta$ or report count.
\end{proof}

\subsection{Uniform source regimes for the confidence lower bound}
\label{app:source-regimes-proof}

\begin{proof}[Proof of Proposition~\ref{prop:nearlinear-source-bound}]
Let $c_*$ be the constant in Theorem~\ref{thm:small-budget-confidence}, and let $C_0,\beta_0$ be those in Lemma~\ref{lem:exact-row-interface}.
Choose
\[
 c\le\min\{c_*/2,1/(8C_0),1/16\},\qquad
 \eta_0=\min\{1/8,\beta_0/4\},
\]
and an absolute $v_0$ ensuring $\lfloor cv\rfloor\ge1$.
Fix an admissible $m$ and write $\Phi_m=t\log(1+m/t)\log(1+v/m)$.

If $m\le64t$, use the lower-bound proof of Theorem~\ref{thm:small-budget-confidence} with source size $m$ and the fixed constant $K_*=96/\ln2$.
Indeed, $m\le c_*v$, and $\log(1/\delta)\ge2$ gives
\[
 m\le64t\le K_*\ln(1/\delta).
\]
The source needs only $m-1\le Z$ total deletion operations.
The proof therefore gives
\[
 s+1=\Omega\bigl(vm\log(1+v/m)\bigr)
     =\Omega(v\Phi_m),
\]
using $t\log(1+m/t)\le m/\ln2$.
These streams are legal even for the total-deletion subclass at the monitor's original advertised budget $Z$.

If $m>64t$ and $\delta\le\eta_0$, apply \eqref{eq:ic1} with $N=\lfloor v/2\rfloor$, $U=\lceil v/2\rceil$, and source size $m$.
Its hypotheses hold:
\[
 L=\ln(1/(2\delta))=\Theta(t),\qquad
 m\ge L,\qquad U\ge C_0m,\qquad m-1\le Z.
\]
The restriction $m\le cv\le v/16$ yields, with absolute constants,
\[
 \log(2+m/L)=\Theta\bigl(\log(1+m/t)\bigr),\qquad
 \log(U/m)=\Theta\bigl(\log(1+v/m)\bigr).
\]
Thus \eqref{eq:ic1} gives $s+1=\Omega(v\Phi_m)$ again.

In the remaining case $\delta>\eta_0$, $t=O(1)$ and
\[
 \Phi_m=O\bigl(A\log(m+1)\bigr)
       =O\bigl(A\min\{A,\log(Z+2)\}\bigr)
       =O(AB_Z).
\]
Corollary~\ref{cor:all-budget-block-lower}, namely \eqref{eq:ab1}, therefore absorbs this term.
Its use of $\log(v+1)$ rather than $A$ changes only an absolute constant.

The constants work uniformly for every admissible $m$, so take the maximum and combine it with \eqref{eq:ab1}.
The maximum of two lower bounds is at least half their sum; a common hard distribution is unnecessary.
Every imported family has synchronized finite source prefixes.
Its common finite-family simulation cutoff, tagged original prefix state $M_0$, and bound $H(M_0)=O(s+1)$ remain unchanged.
In particular, the expected space charge is neither truncated nor conditioned on successful responses, and choosing $m$ does not alter the original law.
\end{proof}

\subsection{The finite distributional query law}
\label{app:finite-query-law-proof}

\begin{proof}[Proof of Lemma~\ref{lem:finite-query-law}]
Use the construction in the proof of Lemma~\ref{lem:pointwise-source}.
Namely, put $\alpha=100/L$, $q_0=1-\alpha$, $z=\alpha m$,
\[
 R=\left\lfloor\frac{\ln z}{100\alpha}\right\rfloor,
 \qquad r_i=\lfloor m(1-q_0^i)\rfloor\quad(0\le i<R),
 \qquad r_R=m.
\]
The same absolute choice of $\beta_*$ ensures that these are valid proper query sizes and $R=\Theta(L\log(2+m/L))$.
Given $S$, choose $i$ uniformly from $\{0,\ldots,R-1\}$ and then a uniform $r_i$-subset $T$ of $S$.
This defines $\mathcal D_\beta$.

Let $e_i$ be the decoder's average error for a uniform source and a uniform $r_i$-subset of that source, drawn independently of the message conditional on the source.
The displayed distributional premise says $\sum_{i=0}^{R-1}e_i\le R\beta$.
In the lossless source encoding from that earlier proof, the distribution of the adaptively generated query at stage $i$, conditional on each fixed $S,M$, is dominated by $\beta^{-1/4}$ times the uniform law on $r_i$-subsets of $S$.
This is exactly the combinatorial bound \eqref{eq:pe2}, which does not use an error assumption.
Fresh decoder coins therefore give total expected stage failures at most
\begin{equation}
\beta^{-1/4}\sum_{i=0}^{R-1}e_i\le R\beta^{3/4}.
                                                               \label{eq:sj2}
\end{equation}
The recovered set $S_{\mathrm{rec}}$ has expected size at least $R(1-\beta^{3/4})\ge R/2$.
The same flags, residual set, and independent tapes reconstruct $S$ losslessly.
The entropy calculation \eqref{eq:pe4} consequently yields
\[
 I(S;M)\ge \E|S_{\mathrm{rec}}|\log\frac{U-m}{m}-R
          =\Omega(R\log(U/m)),
\]
after increasing the absolute constant $C$.
Thus average uniform-query error suffices for \eqref{eq:sj1}.
\end{proof}

\section{Construction and Memory Guarantees for Binary Codes and Verification Fields}
\label{sec:binary-codec-upgrade}
\subsection{An entropy-sized binary codec with bounded total workspace}

Entropy-sized Hamming checksums from algebraic-geometric codes and error-locating-pair decoding are classical~\cite{ChengLi2021Asymmetric,Pellikaan92,DuursmaKotter94}.
The implementation below gives the uniform construction, checked decoding on arbitrary syndromes, and total-workspace bounds needed by the graph algorithm.
The construction uses the characteristic-two Garcia--Stichtenoth tower~\cite[Section~3]{GarciaStichtenoth1996Reciprocal} and two global Riemann--Roch evaluation tables.
\S\ref{sec:binary-rr-constructor} charges field descriptions, tables, and streamed construction scratch; no full constraint matrix or coefficient basis is retained.

\begin{lemma}[Characteristic-two sparse syndromes]
\label{lem:binary-paid-field-syndrome}
For every $n\ge2$ and $1\le k\le n$ there is a uniform deterministic constructor of a characteristic-two field $\mathbb K$ and a matrix $H\in \mathbb K^{h\times n}$ such that
\[
 h=O(k),\qquad \ell=\log|\mathbb K|=O(\log(en/k)).
\]
Every $k$-sparse vector over $\mathbb K$, with arbitrary nonzero amplitudes, is recoverable from its syndrome.
Setup, syndrome updates, and decoding have polynomial bit work with absolute exponent.
Their total workspace, including full evaluation tables and output, is $O(nk\log(en/k))$ bits.
On an arbitrary syndrome the decoder returns a checked vector of support at most $k$ with that syndrome, or failure.
\end{lemma}

\begin{proof}
Fix $\varepsilon=1/8$ and a sufficiently large absolute $C_0\ge128$.
All finite parameter exceptions below use the identity matrix; their threshold and the resulting constant factor are absolute.
If $k\ge n/32$, use the identity over $\mathbb F_2$.
If $n/k\ge n^\varepsilon$, choose $\mathbb K=\mathbb F_{2^a}$ with $n<2^a\le2(n+1)$, distinct nonzero names $\alpha_1,\ldots,\alpha_n\in \mathbb K$, and $H_{j,i}=\alpha_i^j$ for $0\le j<2k$.
On any $y=(y_0,\ldots,y_{2k-1})$, return zero if $y=0$; otherwise try $t=1,\ldots,k$, skipping singular Hankel matrices $M_t=(y_{a+b})_{0\le a,b<t}$.
For each nonsingular trial solve
\[
 M_t(c_0,\ldots,c_{t-1})^{\mathsf T}
       =-(y_t,\ldots,y_{2t-1})^{\mathsf T},\qquad
 p(T)=T^t+\sum_{j=0}^{t-1}c_jT^j.
\]
Scan all $n$ names for roots of $p$; unless exactly $t$ occur, continue.
Solve their first $t$ Vandermonde moment equations, extend by zero, and return only after checking support at most $k$ and all $2k$ equations.
Return failure if no trial succeeds.

When the loop reaches the true support size $t=|S|$ of a nonzero promised vector, its Hankel matrix has the factorization
\[
 M_t=V_t\operatorname{diag}(x_S)V_t^{\mathsf T},\qquad
 (V_t)_{a,i}=\alpha_i^a\quad(0\le a<t,\ i\in S).
\]
The square Vandermonde matrix $V_t$ and the diagonal matrix $\operatorname{diag}(x_S)$ are invertible, so $M_t$ is nonsingular.
The solved recurrence therefore implies $p(\alpha_i)=0$ for every $i\in S$, so its $t$ roots are precisely those names and the amplitude solve recovers $x$.
This argument uses arbitrary nonzero amplitudes and works in characteristic two.
Every set of at most $2k$ columns of $H$ is independent by its initial Vandermonde rows, so two vectors of support at most $k$ cannot satisfy the same $2k$ syndrome equations.
Thus a smaller trial may have a nonsingular Hankel matrix, but its candidate, supported on at most $t<|S|$ names, cannot pass the final syndrome check.
The same uniqueness argument shows that $y=0$ has only the zero promised preimage.
The fixed trials use $O(k^4+nk^2)$ field operations and $O(k^2+n)$ auxiliary field words.
Enumerating monic degree-$a$ binary polynomials and testing irreducibility constructs $\mathbb K$ in polynomial work and $O(a)$ scratch.
Including the full table and output, the space is $O(nk\log n)$ bits, which is $O(nk\log(en/k))$ in this branch.

It remains to treat $k<n/32$ and $k>n^{1-\varepsilon}$.
Choose the least power of two
\begin{equation}
q=2^r\ge C_0(n/k+1),\qquad q<2C_0(n/k+1),\qquad \mathbb K=\mathbb F_{q^2}.
 \label{eq:b1}
\end{equation}
Bounded irreducible-polynomial enumeration constructs $\mathbb K$.
Enumeration and binary elimination give the subfield $\mathbb F_q$ and a fixed binary basis of it.
Even a full $q^2$-entry enumeration buffer is absorbed by the later total space bound.

Use the reciprocal tower
\begin{equation}
F_0=\mathbb K(x_0),\qquad
 x_{i+1}^q+x_{i+1}=\frac{x_i^q}{x_i^{q-1}+1}.
\label{eq:b2}
\end{equation}
The classical full-level statements \cite[Lemmas~5.2 and~5.4]{GarciaStichtenoth2007ExplicitTowers} hold for every prime power $q$: the steps have degree $q$, base infinity is unique and totally ramified, and each of the $q^2-q$ values with nonzero $\mathbb K/\mathbb F_q$ trace splits completely at every level.
The exact genus formula \cite[Proposition~1.1(iii)]{PellikaanStichtenothTorres1998Semigroups} gives $g(F_i)\le q^{i+1}$.
In characteristic two these good values are exactly $\mathbb K\setminus\mathbb F_q$.
If $\alpha$ is good, then $\alpha^{q+1}/(\alpha^q+\alpha)\in\mathbb F_q^\times$; the next full equation has $q$ distinct roots in $\mathbb K$, all again good.

The remark following Proposition~3.1 of~\cite{BP14} describes the existence of intermediate fields for general prime powers $q$.
We give an explicit full binary flag for $q=2^r$ and account for its description and arithmetic below.
For $0\le s\le r$, let $H_s$ be the span of the first $r-s$ vectors of the selected basis of $\mathbb F_q$.
Define
\begin{equation}
L_s(T)=\prod_{a\in H_s}(T-a),\qquad
 P_s(T)=\prod_{b\in L_s(\mathbb F_q)}(T-b).
\label{eq:b3}
\end{equation}
Subspace polynomials are linearized: inductively, adjoining one basis element $a$ replaces $L(T)$ by $L(T)^2+L(a)L(T)$, with $L(a)\ne0$.
This also proves that their derivatives are nonzero constants.
The composite $P_s(L_s(T))$ is monic of degree $q$, has all elements of $\mathbb F_q$ as its distinct roots, and hence equals $T^q+T$.
All coefficients and products in \eqref{eq:b3} can be computed by serial polynomial arithmetic over $\mathbb K$ in polynomial work and $O(q^2)$ field words of space.
Retaining the full binary flag uses no larger bound.

Put $z_s=L_s(x_{i+1})$.
It satisfies the monic separable equation
\begin{equation}
P_s(z_s)=\frac{x_i^q}{x_i^{q-1}+1}
\label{eq:b4}
\end{equation}
of degree $e=2^s$.
In the full degree-$q$ translation extension, the stabilizer of $z_s$ is exactly $H_s$: additivity gives $L_s(x_{i+1}+a)-L_s(x_{i+1})=L_s(a)$.
Thus $[F_i(z_s):F_i]=e$ and $[F_{i+1}:F_i(z_s)]=q/e$.
These subfields form a nested chain, and consecutive degrees are doubled.
Their good places are the $H_s$-cosets of the full split fibers, so there are exactly
\begin{equation}
N=(q^2-q)M,\qquad M=e q^i
\label{eq:b5}
\end{equation}
designated good places.
The full infinity descends to a unique rational infinity.
In the full cover $z_s$ has pole order $q/e$, because its leading monomial has that degree and the full top coordinate has pole order one.
Dividing by the cover ramification index $q/e$ gives pole order one in the intermediate field.
Hurwitz gives
\begin{equation}
g(F_i(z_s))\le\frac{g(F_{i+1})-1}{q/e}+1
 \le qM+1.
\label{eq:b6}
\end{equation}
Consequently, the first full or partial good-point count at least $n$ has $n\le N<2n$ and
\begin{equation}
qM=\frac{N}{q-1}<\frac{2n}{q-1}=O(k/C_0).
\label{eq:b7}
\end{equation}
The rational-base case is absorbed in the fixed cutoff: in the dense branch $q=O(n^{1/8})$, so $n\le q^2-q=O(n^{1/4})$ implies bounded $n$.

Let $E$ denote the chosen intermediate level and set $g_0=qM+1$.
By \eqref{eq:b7}, increasing $C_0$ and the fixed finite cutoff ensures $g(E)\le g_0\le k/4$.
Use the first $n$ designated good places $P_1,\ldots,P_n$, and put
\begin{equation}
s=2g_0+k,\qquad \mathcal U=L_E(sP_\infty),\qquad
 \mathcal V=L_E(2sP_\infty).                            \label{eq:b8}
\end{equation}
Since $s\ge2g(E)$, the Riemann--Roch dimension formula gives $u=s+1-g(E)>k$ and $h=2s+1-g(E)=O(k)$.
Retain their two evaluation tables
\begin{equation}
A=(f_i(P_j))_{i\le u,j\le n},\qquad
 H=(h_i(P_j))_{i\le h,j\le n}                           \label{eq:b9}
\end{equation}
for bases of $\mathcal U$ and $\mathcal V$.

Next bound the space and work needed to construct these bases.
The full cover $F_m/E$ has degree $h_0=q/e\le q$, with $B=h_0M=q^m$ and $m=O(\log n)$.
Its pole bound for either table is $D\le2h_0s$.
Thus
\begin{equation}
qB=h_0qM=O(qk)=O(n),\qquad D=O(qk)=O(n).              \label{eq:b10}
\end{equation}
Proposition~\ref{prop:binary-rr-entry-oracle} and Corollary~\ref{cor:binary-rr-invariant-kernel} give matrices with $n\operatorname{polylog}n$ rows and columns and entries generated in $n\operatorname{polylog}n$ field words of scratch.
The good-point count follows from $g_0\le k/4$ and the branch condition $k<n/32$: these give $2s=4g_0+2k\le3k<n\le N$, and hence $D\le2h_0s<h_0N=N_{F_m}$.
Lemma~\ref{lem:entry-oracle-kernel} generates each kernel coefficient and basis evaluation on demand; no full coefficient-basis matrix is stored.
Only the $O(nk)$ entries in \eqref{eq:b9} remain.
The dense-ratio assumption $k>n^{1-\varepsilon}$ absorbs all $n\operatorname{polylog}n$ construction scratch in $O(nk)$ field words.
Enumeration of good tuples and top-translation orbits has polynomial work: at most $q^{2m+2}=q^2B^2$ candidate tuples need be scanned, and an orbit representative can be recognized by at most $q$ translations.
The number of nested entry-oracle layers is fixed, so the work exponent is absolute.

To decode $y\in \mathbb K^h$, set $a_0=2s+1<n$ and $I=\{1,\ldots,a_0\}$.
Zero counting makes $H_I$ full row rank: a nonzero element of $\mathcal V$ has at most $2s$ zeros.
Solve $H_Iw=y$ and form
\begin{equation}
G=A_I\operatorname{diag}(w)A_I^{\mathsf T}.
                                                               \label{eq:b11}
\end{equation}
Since every $f_if_j$ belongs to $\mathcal V$, a valid syndrome $y=Hx$ satisfies
\begin{equation}
G=A_S\operatorname{diag}(x_S)A_S^{\mathsf T},
            \qquad S=\supp(x).            \label{eq:b12}
\end{equation}
In the basis $(f_i)$, the evaluation map of $\mathcal U$ at $S$ is $z\mapsto A_S^{\mathsf T}z$.
It is surjective whenever $|S|\le k$: $s-|S|\ge2g(E)-1$, so the Riemann--Roch dimension formula gives a kernel of codimension $|S|$.
Its transpose $A_S$ therefore has full column rank.
Since the coordinates of $x_S$ are nonzero, the left factor $A_S\operatorname{diag}(x_S)$ is injective and hence
\begin{equation}
\ker G=\ker A_S^{\mathsf T},        \label{eq:b13}
\end{equation}
also in characteristic two.
Because $u>k$, choose a nonzero $z\in\ker G$.
The function $f=\sum_i z_if_i$ is nonzero; its zero set $\mathcal Z$ among the selected places contains $S$ and has size at most $s$.
Evaluation of $\mathcal V$ onto $\mathcal Z$ is likewise surjective, since $2s-|\mathcal Z|\ge s\ge2g(E)-1$.
Thus $H_{\mathcal Z}$ has full column rank.
Solve $H_{\mathcal Z}b=y$, extend by zero outside $\mathcal Z$, and check its support bound and all syndrome equations.
This recovers every promised sparse field vector.
For any other input, a missing kernel, inconsistent solve, or failed final check returns failure; all loops and matrices remain bounded.

The $O(k)$-dimensional decoder matrices and $O(nk)$ retained table entries give polynomial work for field arithmetic, elimination, serial evaluation, and checks, with $O(nk\log q)$ bits; $\log|\mathbb K|=2\log q=O(\log(en/k))$.
\end{proof}

\begin{corollary}[Binary entropy-sized sparse syndrome]
\label{cor:binary-entropy-codec}
There is a deterministic binary matrix $C$ with $r\le n$ and $r=O(k\log(en/k))$ rows whose syndrome uniquely determines every binary vector of weight at most $k$.
It has a polynomial-work, all-input checked decoder, and its setup, matrix, decoding scratch and output use $O(nk\log(en/k))$ bits.
\end{corollary}
\begin{proof}
Expand each entry of $H$ from the preceding lemma in a fixed binary basis of $\mathbb K$.
This gives $C_0$ with $h\ell$ binary rows and, for binary $x$, exactly the binary expansion of $Hx$.
Decode by regrouping the syndrome, running the field decoder, then requiring all recovered coordinates to belong to $\{0,1\}$ and verifying all binary equations.
If $h\ell>n$, use the binary identity instead; this branch still has $n=O(k\log(en/k))$.
The previously constructed tables and all temporary representations fit the same total space bound.
Otherwise, take $C=C_0$.
\end{proof}

\subsection{Streamed linear algebra and Riemann--Roch construction}
\label{sec:entry-oracle-linear-algebra}

\begin{lemma}[Rank and kernel computation using linearly many registers]
\label{lem:entry-oracle-kernel}
Let $A$ be an $R$-by-$C$ matrix over an explicitly represented field $\mathbb K$.
Suppose an oracle evaluates each entry deterministically, using $W$ bits of scratch and at most $T$ field/bit operations.
Write $d=\min\{R,C\}$.
Assume every intermediate field element has an $\ell$-bit representation and field primitives, including inversion and zero testing, use $O(\ell)$ scratch bits; any retained field description is charged separately.
Count each field primitive and each bit operation as one operation.
There is a deterministic algorithm that selects a nonsingular rank-sized minor, represents a basis of $\ker A$ by an entry oracle, and solves every consistent linear system with coefficient matrix $A$.
Its auxiliary space is
\[
 O\bigl(W+(R+C)(\ell+\log(2+R+C))\bigr),
\]
apart from any explicitly retained output table.
Its work and number of entry-oracle calls are bounded by an absolute polynomial in $R,C,T$.
If each field primitive takes at most $\tau_{\mathbb K}$ bit operations, total bit work is $\poly(R,C,T)(1+\tau_{\mathbb K})$.
The algorithm works over arbitrary fields and characteristics, using exact determinant tests to select its rank-sized minor.
\end{lemma}

\begin{proof}
For a $b$-square matrix $B$ accessed by its entries, we implement the Samuelson--Berkowitz recurrence~\cite{Berkowitz84} sequentially to compute its characteristic polynomial and determinant using $O(b)$ field registers and $O(b^4)$ arithmetic operations and entry calls.
To obtain this space bound, process the trailing principal matrices from the bottom right.
Given the characteristic coefficient vector of the trailing $(b-1)$-square matrix $D$, write the next matrix as
\[
 \begin{pmatrix}a&u^{\mathsf T}\\w&D\end{pmatrix}.
\]
Generate the Toeplitz coefficient vector $1,-a,-u^{\mathsf T}w,-u^{\mathsf T}Dw,\ldots, -u^{\mathsf T}D^{b-2}w$ serially.
Two length-$(b-1)$ registers suffice for the current vector $D^jw$ and its next iterate.
The old characteristic vector, this Toeplitz vector, and their convolution fit a constant number of length-$(b+1)$ registers.
Discard the old registers after the convolution.
Neither $D$ nor a matrix power is stored.
This division-free recurrence works in every characteristic.

Maintain index lists $I,J$ of equal length $r$ such that $A_{I,J}$ is nonsingular, initially the empty minor with determinant one.
Enumerate each unused row $i$ and unused column $j$ and test
\[
              \det A_{I\cup\{i\},J\cup\{j\}}.
\]
If it is nonzero, append $i,j$ and restart the scan.
If every bordered determinant is zero, the current minor has maximum rank.
Indeed, after arranging its rows and columns first and writing $A=\left(\begin{smallmatrix}P&B\\C&D\end{smallmatrix}\right)$, all tested determinants are $\det(P)$ times entries of $D-CP^{-1}B$.
If all vanish then this Schur complement is zero, so $\operatorname{rank}A=r$.
Conversely, a nonzero Schur-complement entry supplies a valid extension.
At most $d$ successful extensions occur and at most $(d+1)RC$ determinants are tested.
Index tests and scans require only the lists, not an $R$-by-$C$ mask.

Let $J^c$ be the free columns and $P=A_{I,J}$.
For each $j\in J^c$, the associated kernel vector has coordinate one at $j$, zero at the other free columns, and pivot coordinates
\[
                 z=-P^{-1}A_{I,j}.
\]
Cramer's rule generates each coordinate as a ratio of two $r$-square determinants in the same workspace.
The chosen rows span the row space, so the vectors annihilate $A$; their $C-r$ distinct free coordinates prove independence and completeness.
Retaining $I,J,J^c,\det P$ gives a linear-size basis oracle supporting arbitrary entry order; a materialized kernel table is charged separately.

For a right-hand side $b$, set the free coordinates to zero, solve $Pz=b_I$ by the same determinant ratios, and verify all original equations.
This either returns a solution or detects inconsistency.
The rank-zero and empty-dimensional cases use the evident empty lists and zero-vector checks.

All determinant calls are sequential; an entry call begins only after the caller has stored its linear-size registers.
The entry oracle's scratch thus adds $W$, rather than multiplying it by the number of calls.
For a fixed number of nested entry-oracle layers, work remains polynomial with an absolute exponent and workspace is the sum of the simultaneously live layer bounds.
This guarantee assumes an absolute bound on nesting depth, independent of tower height and input size.
\end{proof}

For the tower, entry generation uses Proposition~\ref{prop:binary-rr-entry-oracle}.

\subsubsection{A streamed Riemann--Roch constructor}
\label{sec:binary-rr-constructor}

Let $q=2^r\geq8$, let $\mathbb K=\mathbb F_{q^2}$ be explicitly represented, and consider the reciprocal Garcia--Stichtenoth tower
\[
 F_m=\mathbb K(x_0,\ldots,x_m),\qquad
 x_j^q+x_j=R(x_{j-1}),\qquad
 R(X)=\frac{X^q}{X^{q-1}+1}.
\]
Write $B=q^m$ and let $P_\infty$ be its unique place above infinity.
These degree and infinity facts have a local verification.
If $x_{j-1}$ has a simple pole at its level's infinity, so does $R(x_{j-1})$.
A place above it, of ramification index $e$, satisfies $qv(x_j)=-e$.
Hence $q\mid e$; since the defining polynomial has degree $q$, the step has degree $q$, infinity is unique and rational, and $x_j$ has a simple pole there.
Induction gives $-v_\infty(x_j)=q^{m-j}$ and $[F_m:\mathbb K(x_0)]=B$, also after any constant-field extension.
For use in the explicit construction, put
\[
 g_j=x_j^{q-1}+1,\qquad \pi_{-1}=1,\qquad
 \pi_j=\prod_{i=0}^j g_i,\qquad
 \Pi=\prod_{j=0}^{m-1}\pi_j.
\]

\begin{lemma}[An explicit rational-coordinate span]
\label{lem:binary-rr-trace-span}
For $m\geq1$, every function of $F_m$ regular away from $P_\infty$ belongs to
\begin{equation}
 \frac1\Pi\sum_{0\leq e_1,\ldots,e_m<q}
       \mathbb K[x_0]x_1^{e_1}\cdots x_m^{e_m}.
 \label{eq:binary-rr-trace-span}
\end{equation}
The denominator $\Pi$ is regular away from infinity, with pole order
\begin{equation}
 W=mqB-\frac{q(B-1)}{q-1}\leq mqB.
 \label{eq:binary-rr-denominator-weight}
\end{equation}
\end{lemma}

\begin{proof}
First, verify the classical regularity statement~\cite[Lemma~3.4]{PellikaanStichtenothTorres1998Semigroups} by valuations: $\pi_{j-1}x_j^e$ has no finite pole for $1\leq j\leq m$ and $0\leq e<q$.
Fix a finite place $P$ and put $a_j=v_P(x_j)$ and $u_j=v_P(\pi_j)$, with $u_{-1}=0$.
Since $a_0\geq0$, also $u_0\geq0$.
Inductively, $u_j\geq0$, and if $a_j<0$ then $u_j\geq-a_j$.
Indeed, when $a_j<0$ the tower equation gives $qa_j=v_P(R(x_{j-1}))$.
If $a_{j-1}<0$, this is $a_{j-1}$, and the induction hypothesis gives $u_{j-1}\geq-qa_j$.
If $a_{j-1}=0$, it is $-v_P(g_{j-1})$, and $u_{j-1}=u_{j-2}+v_P(g_{j-1})\geq-qa_j$.
The case $a_{j-1}>0$ cannot yield $a_j<0$.
Thus $u_j=u_{j-1}+(q-1)a_j\geq-a_j$ when $a_j<0$; when $a_j\geq0$, both $u_{j-1}$ and $v_P(g_j)$ are nonnegative.
Consequently, $u_{j-1}+ea_j\geq0$ for $0\leq e<q$, proving the claimed regularity, and every $\pi_j$ is finite-regular.

For the degree-$q$ extension $F_j/F_{j-1}$, the basis $1,x_j,\ldots,x_j^{q-1}$ has trace-dual basis
\[
 d_e(x_j)=x_j^{q-1-e}+\boldsymbol1_{e=0},\qquad0\leq e<q.
\]
The difference quotient of $T^q+T-R(x_{j-1})$ has coefficients $d_e(x_j)$, and the derivative is $1$, so no derivative normalization is needed in the trace-dual formula.
Equivalently, $\operatorname{Tr}_{F_j/F_{j-1}}(x_j^e d_f(x_j))= \boldsymbol1_{e=f}$ follows by summing over $x_j+a$, $a\in\mathbb F_q$.
Trace transitivity now shows that
\[
 b_{\boldsymbol e}=\prod_{j=1}^m\pi_{j-1}x_j^{e_j},\qquad
 b^*_{\boldsymbol e}=\prod_{j=1}^m
              \frac{d_{e_j}(x_j)}{\pi_{j-1}}
\]
are dual bases of $F_m/\mathbb K(x_0)$: take the trace from the top step down, cancel the two top scaling factors, and obtain one Kronecker delta at each step.
The elements $b_{\boldsymbol e}$ are integral over $\mathbb K[x_0]$ by the regularity just proved.
For a finite-regular $f$, trace preserves integrality and $\mathbb K[x_0]$ is integrally closed, so
\[
 c_{\boldsymbol e}=
 \operatorname{Tr}_{F_m/\mathbb K(x_0)}(f b_{\boldsymbol e})\in \mathbb K[x_0].
\]
The dual expansion $f=\sum_{\boldsymbol e}c_{\boldsymbol e} b^*_{\boldsymbol e}$ proves~\eqref{eq:binary-rr-trace-span}, since each numerator has degree less than $q$ in every $x_j$.
Finally, $-v_\infty(\pi_j)=qB-q^{m-j}$, and summing these $m$ weights proves~\eqref{eq:binary-rr-denominator-weight}.
\end{proof}

\begin{proposition}[Principal-part constraints with an explicit entry oracle]
\label{prop:binary-rr-entry-oracle}
For integers $m\geq1$ and $D\geq0$, there is an explicitly indexed matrix over $\mathbb K$ whose kernel, interpreted as rational-function coefficients, is $L_{F_m}(DP_\infty)$.
Its column and row counts are
\[
 C=D+W+1=O(D+mqB),\qquad R_0=O(m^2qB).
\]
A requested matrix entry or good-place evaluation of a column is generated deterministically in absolute polynomial time in $q,B,D$, using
\[
 O\bigl(m^2(D+mqB)\bigr)\quad \mathbb K\text{-words}
 \quad\text{and}\quad
 O\bigl((m+\log q+\log(D+B+1))^4\bigr)\quad\text{additional bits}.
\]
Only one place and its coordinate series are processed at a time.
\end{proposition}

\begin{proof}
The raw columns are the functions
\begin{equation}
 \psi_{a,\boldsymbol e}=
 \frac{x_0^a x_1^{e_1}\cdots x_m^{e_m}}{\Pi},
 \qquad a\geq0,\quad0\leq e_j<q,
 \label{eq:binary-rr-raw-columns}
\end{equation}
of infinity weight at most $D$.
Their weights $aB+\sum_j e_jq^{m-j}-W$ run without repetition through all integers $-W,-W+1,\ldots$.
Thus there are exactly $D+W+1$ columns, and they are linearly independent.
By Lemma~\ref{lem:binary-rr-trace-span}, they span every member of $L(DP_\infty)$: a term of weight greater than $D$ cannot cancel against any other term at infinity.

Every coordinate has total pole degree at most $B$.
In fact the reciprocal coordinates $y_i=1/x_{m-i}$ satisfy the same tower equations.
Extending from $\mathbb K(x_j)$ in both directions therefore gives $[F_m:\mathbb K(x_j)]\leq q^m=B$.
Since $aB\leq D+W$, a selected column has total pole degree at most
\begin{equation}
 \Lambda=D+2W+m(q-1)B=O(D+mqB).
 \label{eq:binary-rr-column-height}
\end{equation}
All finite poles lie above $x_0\in\mathbb F_q$.
Indeed, outside these base places all $x_j$ and $g_j$ are units, inductively by the tower equation.
At a finite geometric place $P$ above this set, a common pole bound for all columns is
\begin{equation}
 N_P=(q-1)\sum_{j=1}^m\max\{0,-v_P(x_j)\}+v_P(\Pi).
 \label{eq:binary-rr-local-pole-bound}
\end{equation}
The last term is nonnegative by Lemma~\ref{lem:binary-rr-trace-span}.
Divisor degrees, after a splitting constant-field extension, give
\begin{equation}
 \sum_P N_P\leq m(q-1)B+W=O(mqB).
 \label{eq:binary-rr-total-principal-parts}
\end{equation}

The following argument proves the needed part of the classical local description~\cite[Theorem~1]{AleshnikovKumarShumStichtenoth2001Splitting}.
Choose $d$ to be the least power of two at least $\max(1,m)$ and construct $\mathbb K'/\mathbb K$ of degree $d$, so $d<2m$.
Enumerating monic degree-$d$ polynomials over $\mathbb K$ and applying the finite-field irreducibility test takes at most $q^{2d}\leq B^4$ tests, using modular polynomial powering and gcd and retaining one candidate and its arithmetic scratch.
On the $2d$-dimensional $\mathbb F_q$-space $\mathbb K'$, the operator $N(a)=a^q+a$ is nilpotent, since $2d$ is a power of two and $N^{2d}=0$.
Its kernel is $\mathbb F_q$, of dimension one, so it has one nilpotent Jordan block.
In particular, $\ker N^j\subseteq\operatorname{im}N$ for $j<2d$.
Therefore, for every $\ell\leq\lfloor m/2\rfloor$, all tuples
\begin{equation}
 \alpha_0\in\mathbb F_q^*,\qquad
 \alpha_j^q+\alpha_j=\alpha_{j-1}\quad(1\leq j\leq\ell)
 \label{eq:binary-rr-residue-tuple}
\end{equation}
exist in $\mathbb K'$, with exactly $q$ choices at each successive step.
The residue equations are linear over $\mathbb F_2$, so their affine solution spaces can be enumerated serially by binary elimination in the chosen field representation.

For $v(z)>0$ put $\mathcal A(z)=\sum_{u\geq0}z^{q^u}$, and put $\mathcal A(0)=0$.
This is the unique positive-valuation solution of $Y^q+Y=z$.
Fix $1\leq t\leq m$ with $\ell=m-t\leq t$, set $\tau=x_0$, choose a tuple \eqref{eq:binary-rr-residue-tuple}, and compute
\begin{equation}
 x_i=\mathcal A(R(x_{i-1}))\ (1\leq i<t),\qquad
 x_t=\alpha_0+\mathcal A(R(x_{t-1})).
 \label{eq:binary-rr-positive-recursion}
\end{equation}
For $1\leq j\leq\ell$, compute
\begin{equation}
 b_j=\frac{\alpha_0^2}{x_{t-j}},\qquad
 z_j=R(x_{t+j-1})+b_j^q+b_j+\alpha_{j-1},\qquad
 x_{t+j}=b_j+\alpha_j+\mathcal A(z_j).
 \label{eq:binary-rr-polar-recursion}
\end{equation}
Since $\alpha_0\in\mathbb F_q^*$ and $1/R(u)=u^{-q}+u^{-1}$, the tower equation gives, with $y=x_{t-j+1}$,
\[
 b_j^q+b_j=\alpha_0^2(x_{t-j}^{-q}+x_{t-j}^{-1})
          =\frac{\alpha_0^2}{y^q+y}.
\]
To verify that these formulas define Laurent series with $v(z_j)>0$, first take $j=1$ and write $y=x_t$ and $\alpha=\alpha_0$.
Then
\[
 R(y)+\frac{\alpha^2}{y^q+y}+\alpha=R(y+\alpha),
\]
which has positive valuation.
For $j\geq2$, write $y=x_{t-j+1}$ and $X=x_{t+j-1}=\alpha_0^2/y+\alpha_{j-1}+w$, where $v(w)>0$.
Using $R(X)=X+X/(X^{q-1}+1)$ and the formula for $b_j^q+b_j$ gives
\[
 z_j=w+\frac{X}{X^{q-1}+1}
       +\alpha_0^2\frac{y^{q-2}}{1+y^{q-1}}.
\]
Here $v(y)>0$ and $v(X)<0$, so the last two terms have positive valuation because $q>2$.
The zero function is allowed with valuation $+\infty$.
Thus the series solve all tower equations, and they satisfy
\begin{equation}
 v(x_i)=q^i\ (i<t),\qquad v(x_t-\alpha_0)=q^t,\qquad
 x_{t+j}=\frac{\alpha_0^2}{x_{t-j}}+\alpha_j+w_j,
 \quad v(w_j)>0.
 \label{eq:binary-rr-local-expansion}
\end{equation}
In particular, $v(x_{t+j})=-q^{t-j}$.
Each tuple defines a rational place of $\mathbb K'F_m$, unramified above zero since $x_0=\tau$.
Distinct tuples give distinct places: their residues recover the tuple, and an unramified rational completion over the fixed base parameter has only one embedding into $\mathbb K'((\tau))$.

For $t\geq\lceil m/2\rceil$ use these direct branches.
For $0\leq t<\lceil m/2\rceil$, use the same construction on $y_i=1/x_{m-i}$ with first nonzero residue at $m-t$, and invert back.
When $t\geq1$, the original first nonzero coordinate is $x_t$, and $v(x_0)=q^{m-2t}$, which is the ramification index over base zero.
The number of branches and their indices are
\begin{center}
\begin{tabular}{@{}ccc@{}}
\toprule
Range & Number of branches & Ramification index\\
\midrule
$t\geq\lceil m/2\rceil$ & $(q-1)q^{m-t}$ & $1$\\
$1\leq t<\lceil m/2\rceil$ & $(q-1)q^t$ & $q^{m-2t}$\\
\bottomrule
\end{tabular}
\end{center}
The reciprocal branches are distinct for the same completion reason, using $y_0$ as parameter.
The all-zero direct branch is unramified and regular for every column.
Above base zero the sum of ramification indices of these rational places is $1+\sum_{t=1}^m(q-1)q^{m-t}=B$, proving completeness by the fundamental degree inequality.
In the separate case $t=0$, $y_m=\alpha_0+O(\tau^{q^m})$ with exact leading order $q^m$; thus $x_0$ has residue $\alpha_0^{-1}$ and $v(x_0-\alpha_0^{-1})=B$.
This gives the unique index-$B$ place above each nonzero element of $\mathbb F_q$.
All needed places consequently split in the constructed field $\mathbb K'$.
There are at most $(m+1)(q-1)q^{\lfloor m/2\rfloor}$ nonzero tuples in total.
These are generated one at a time, and the series give the valuations in~\eqref{eq:binary-rr-local-pole-bound}.

It remains to bound the series precision and workspace.
Represent a nonzero Laurent series by its valuation and its next $p$ coefficients, called relative precision $p$.
Multiplication, division, and positive integer powers preserve relative precision.
Addition loses only the increase of valuation above the smaller input valuation.
Under the constructed embeddings of $\mathbb K'F_m$ into $\mathbb K'((\tau))$, the series $\mathcal A(R(x_{i-1}))$ in \eqref{eq:binary-rr-positive-recursion} equal $x_i$ for $i<t$ and $x_t-\alpha_0$ for $i=t$, while $\mathcal A(z_j)=x_{t+j}-b_j-\alpha_j$.
Their total pole degrees are at most $B$ and $2B$, respectively, also for the reciprocal branches.
Each ordinary-function intermediate sum in \eqref{eq:binary-rr-positive-recursion}--\eqref{eq:binary-rr-polar-recursion} has total pole degree at most $3qB$: $R(x_i)$ and $b_j^q$ have degree at most $qB$, and $b_j$ has degree at most $B$.
Every summand has valuation at least $-qB$.
A nonzero sum has positive valuation at most $3qB$, so one addition loses at most $4qB$ coefficients.
There are at most six such additions in a tower step, and an allowance of $32qB$ per step suffices.
For nonzero $z$ with $\nu=v(z)>0$, the terms $z^{q^u}$ in $\mathcal A(z)$ have strictly increasing valuations $q^u\nu$.
If $z$ is known to relative precision $p$, the Frobenius power $z^{q^u}$ has error of valuation at least $q^u(\nu+p)$, which is greater than $\nu+p$ for $u\ge1$.
The term $u=0$ therefore determines both the leading valuation and the limiting absolute precision $\nu+p$, so summation preserves relative precision $p$.
If a truncated intermediate is zero beyond absolute precision $3qB$, the same degree bound certifies that the function is zero; its $\mathcal A$ value is then zero.
Forming $g_i=x_i^{q-1}+1$ loses at most another $2qB$ coefficients.
Starting with
\begin{equation}
 p=\Lambda+(64m+8)qB+1=O(D+mqB)
 \label{eq:binary-rr-working-precision}
\end{equation}
therefore determines every required negative coefficient of every column exactly.
Indeed, after at most $m$ tower steps and the formation of the $g_i$, the remaining relative precision of each nonzero series is at least $p-(32m+2)qB>\Lambda+1$.
Throughout these computations, every intermediate addition has absolute precision at least $p-(32m+3)qB>3qB$, so the zero certification above is valid before the next operation.
Products and powers are truncated directly to the target relative precision: Frobenius uses coefficient powers and index shifts, and binary powering retains a constant number of target-length arrays.
Only $m+1$ coordinate series and a constant number of temporary series are live.
Since each coefficient uses $d<2m$ $\mathbb K$-words, their workspace is $O(m^2(D+mqB))$ $\mathbb K$-words.
The field-construction, place, coefficient, and exponent counters, together with binary elimination on $O(m\log q)$ bits, fit the stated additional bit bound.
Ordinary truncated polynomial arithmetic gives absolute polynomial work in $q,B,D$.

For each $P$, impose that the coefficients of $\tau^{-N_P},\ldots,\tau^{-1}$ vanish, expanding each equation from $\mathbb K'$ to $\mathbb K$ in a fixed basis.
Equation~\eqref{eq:binary-rr-total-principal-parts} gives at most $d\sum_P N_P=O(m^2qB)$ rows.
A row is indexed by a place tuple, negative exponent, and $\mathbb K'$-coordinate.
Its kernel consists exactly of the functions with no finite pole and infinity pole at most $D$.

At a good point, ordinary substitution evaluates a column.
For $\alpha\in \mathbb K\setminus\mathbb F_q$, $R(\alpha)=\alpha^{q+1}/(\alpha^q+\alpha)\in\mathbb F_q^*$.
The equation $Y^q+Y=R(\alpha)$ has exactly $q$ roots in $\mathbb K$, all outside $\mathbb F_q$.
Thus every good tuple has $q$ good extensions at each step.
Enumerate such full points by starting with $x_0\in \mathbb K\setminus\mathbb F_q$ and testing the $q^2$ possible field values at each subsequent coordinate.
There are $(q^2-q)B$ full tuples.
Regenerating a requested tuple takes polynomial work in $q,B$ and retains one tuple.
Intermediate good points are enumerated in the same way using the displayed subspace-polynomial equation; testing the $q^2$ final coordinate values selects a canonical full lift.
This also uses only one tuple and polynomial work.
\end{proof}

\begin{corollary}[One kernel for an intermediate binary level]
\label{cor:binary-rr-invariant-kernel}
Let $E=F_m^H$, where $H\leq\mathbb F_q$ acts by $x_m\mapsto x_m+a$, $a\in H$, and write $h=|H|$.
For an integer $s\geq0$, put $D=hs$ and suppose that the full tower has at least $D+1$ good points.
There is an explicitly indexed matrix with
\[
 C=O(D+mqB),\qquad R=O\bigl(m^2qB+(D+1)\log q\bigr)
\]
columns and rows whose kernel, evaluated at one full lift of each chosen intermediate point, is exactly $L_E(sP_\infty)$.
It has the entry workspace and absolute polynomial work of Proposition~\ref{prop:binary-rr-entry-oracle}.
\end{corollary}

\begin{proof}
Append to the principal-part matrix, for each element $a$ of a binary basis of $H$ and each of $D+1$ distinct good full points $P$, the equation
\begin{equation}
 \sum_j c_j\bigl(\psi_j(\sigma_aP)-\psi_j(P)\bigr)=0.
 \label{eq:binary-rr-invariance-rows}
\end{equation}
Generate these entries by good-point substitution.
The principal-part rows give $f\in L_{F_m}(DP_\infty)$; since $\sigma_a$ fixes infinity, $\sigma_af-f$ has the same pole bound and its $D+1$ zeros force it to vanish.
Thus~\eqref{eq:binary-rr-invariance-rows} is equivalent to $H$-invariance.
The kernel is $L_{F_m}(hsP_\infty)\cap E=L_E(sP_\infty)$, because infinity has ramification index $h$ over $E$.
Evaluations are independent of the chosen lift.
These difference equations avoid division by the group order $h$, which is not invertible in characteristic two when $H$ is nontrivial.
\end{proof}

Combining Corollary~\ref{cor:binary-rr-invariant-kernel} with Lemma~\ref{lem:entry-oracle-kernel} constructs a basis evaluation table on $N$ selected good intermediate points.
Each basis coordinate is generated by a determinant ratio, and each evaluation is a sum of $C$ raw-column evaluations.
These are a fixed number of nested oracle layers.
Apart from the retained table of at most $N(s+1)$ field entries, the total space in bits is
\[
 O\!\left(
 m^2(D+mqB)\log q+
 (R+C)(\log q+\log(2+R+C))+
 (m+\log q+\log(D+B+1))^4\right).
\]
The total work is an absolute polynomial in $q,B,D,N$.
At full height zero, the same interface uses the polynomial basis $1,x_0,\ldots,x_0^D$ directly.

\subsection{Constructing the binary verification field with polynomial work on every tape}
\label{sec:bounded-binary-field}

The deterministic exhaustive search for a degree-$w$ binary irreducible polynomial can be replaced by a capped random search.
This concerns the independent verification field only; it does not change the sparse-code matrix or the sampling law.

\begin{lemma}[Binary-field construction on every tape]
\label{lem:bounded-binary-field}
For integers $w\ge2$ and $a\ge1$, there is a randomized procedure which uses $O(w+\log(a+1))$ bits and polynomial bit work in $w+a$ on every tape.
It returns either an explicitly represented field $\mathbb F_{2^w}$ or a failure flag.
Its failure probability is at most $2^{-a}$.
Conditional on any returned representation, a fresh $w$-bit string is an exactly uniform field element.
\end{lemma}
\begin{proof}
Let $I_w$ be the number of monic irreducible degree-$w$ polynomials over $\mathbb F_2$.
The factorization of $X^{2^w}-X$ gives
\[
 2^w=\sum_{d\mid w} dI_d.
\]
For every $d$ one has $dI_d\le2^d$.
Every proper divisor of $w$ is at most $w/2$, and for $w\ge2$,
\[
 \sum_{d=1}^{\lfloor w/2\rfloor}2^d
 =2^{\lfloor w/2\rfloor+1}-2\le2^{w-1}.
\]
Consequently, $I_w\ge2^{w-1}/w$.
A uniform monic degree-$w$ polynomial, sampled by its $w$ free coefficient bits, is therefore irreducible with probability at least $1/(2w)$.

Use at most $2wa$ independent trials.
For each sampled polynomial $f$, successively compute $X^{2^i}\bmod f$ for $1\le i\le\lfloor w/2\rfloor$, and require
\[
 \gcd(f,X^{2^i}-X)=1
\]
at every step.
An irreducible degree-$w$ polynomial passes every test, since no factor degree divides a smaller $i$; a reducible one has a factor of degree $d\le w/2$ and fails at $i=d$.
Return the first passing representation (with no algebraic failure event), or failure at the cap.

The probability of reaching the failure flag is at most
\[
 (1-1/(2w))^{2wa}\le e^{-a}\le2^{-a}.
\]
Repeated squaring, polynomial division and gcd use a fixed number of $O(w)$-bit polynomial buffers and polynomial work.
The current trial, power and degree indices use $O(\log(w+a))$ bits.
Failed trial data are discarded.
The output representation is a single degree-$w$ polynomial and its arithmetic is polynomial in $w$.
The stated space bound follows, including the trial counter.
Finally, the polynomial-basis representation is a bijection from $w$-bit strings to field elements, regardless of which irreducible polynomial was selected.
\end{proof}

\begin{corollary}[One shared field with independent verification points]
\label{cor:bounded-binary-cut-verification}
Consider the simple-graph recovery banks from Section~\ref{sec:unrestricted-cut-interface}, with at most $64v^3$ potential candidate tests in the ideal execution of a report.
Put
\[
 A=\lceil\log(v+1)\rceil,\quad
 \tau_0=\lceil\log(64/\delta)\rceil,\quad
 w=\tau_0+10A+20.
\]
Construct one binary field using the preceding lemma with failure exponent $a=\tau_0+10A+20$, independently of the code and sampling hashes.
If construction fails, retain a failure flag and return the empty forest at reports.
Otherwise, choose independent uniform verification points in this field for the banks.
This replacement has polynomial work and adds at most $\delta/64$ field-construction error, while preserving the $O(A+\tau_0)$-bit fingerprint precision and workspace per vertex row.
The total probability of a false candidate being accepted is at most $\delta/64$.
\end{corollary}
\begin{proof}
The setup failure is at most $2^{-a}\le\delta/64$.
For a successful field representation, the difference between two distinct crossing-edge sets is a nonzero binary polynomial of degree at most $N_E=\binom{v}{2}$.
Its evaluation at a fresh uniform point vanishes with probability at most $N_E/2^w$.

In the ideal execution, verification comparisons are replaced by exact equality tests.
The ideal cuts and candidate lists depend only on the input, sampling hashes and deterministic sparse decoder; they do not depend on the verification field representation or its points.
Condition on that ideal data and on any successfully constructed field.
Applying the polynomial root bound to each potential test and taking a union bound over all such tests, including those the real execution might skip, gives the bound
\[
 64v^3N_E/2^w\le\delta/64.
\]
Outside this event, actual and ideal reports agree by induction.
The space bound includes a constant number of field buffers, the retained representation, the failure flag, trial indices and points; the work bound in the lemma holds on every tape.
On the exact graph-storage branch the entire field procedure is omitted.
\end{proof}

\section{Implementation of Historical Edge Storage and Parameter Headers}
\label{app:supplementary-implementations}

\subsection{Compressed storage of historical edge names and live bits}
\label{app:certificate-representation}

\begin{proof}[Implementation in Lemma~\ref{lem:dense-compressed-certificate}]
Here $N_E=\binom{v}{2}$ and $A=\lceil\log(v+1)\rceil$.
To implement the compressed representation, set $C_q=\min\{N_E,q(v-1)\}$.
If $C_q\ge N_E/2$, a fixed two-bit state per name, together with $O(vA)$ scratch, has the order in \eqref{eq:forest-df7}.
Otherwise, store $j=|S|$, the combinatorial rank of its $j$-subset of $[N_E]$, and its $j$ live bits in name order.
Its cap is
\[
 O\left(\log\binom{N_E}{C_q}+C_q+vA\right)
 =O\bigl(vq\log(1+v/q)\bigr).
\]
A name lookup scans the ordinary serial combinatorial unranking procedure, keeping only a residual rank, a remaining cardinality, and the current name.
To insert a new name, scan the old subset in this fashion and accumulate the rank of the merged subset, while shifting the live-bit word at the insertion position.
All intermediate binomial values and ranks are at most $\binom{N_E}{C_q}$; a multiplication before an exact division may use another $O(A)$ bits.
A constant number of registers of these widths suffices.
Counters and indices use $O(A)$ bits.
Thus even these temporary copies fit the displayed cap.

Forest traversal of $W$ uses an ordinary $O(vA)$ component structure and a serial name scan.
All these representation operations take polynomially many bit operations.
The endpoint acceptance test is implemented in Appendix~\ref{app:certificate-polynomial}.
\end{proof}

\subsection{Polynomial operations on compressed historical edge storage}
\label{app:certificate-polynomial}

\begin{proof}[Proof of Lemma~\ref{lem:poly-historical-certificate}]
Keep exactly the rank-plus-live-bits representation of \eqref{eq:forest-df7}, or its fixed two-bit-per-name branch when $q(v-1)\ge\binom{v}{2}/2$.
All lookup, insertion, deletion and traversal operations there already have polynomial work: their scans have at most $N_E=\binom{v}{2}$ names, and binomial coefficients can be formed by at most $N_E$ exact multiply/divide steps, with their intermediate bit lengths charged.
It remains to implement the historical endpoint cut test.

During a test the historical set is fixed.
Replace each undirected stored edge by two labelled original directed arcs, one in each direction, both of capacity one.
Retain the two original-arc flow values as two bits in the same name order as $S$.
Initially all are zero.
An original arc with flow zero has a forward residual arc; one with flow one has its intrinsic reverse residual arc.
Original antiparallel arcs and intrinsic reverse arcs retain different labels.
Thus no ambiguity arises when two residual arcs have the same ordered endpoints.

A breadth-first search finds an augmenting endpoint path by scanning the compressed names to enumerate residual arcs.
Its vertex queue, visited bits and predecessor data use $O(v\log(v+1))$ bits.
A predecessor records the original stored-name position, original orientation and whether this traversal is intrinsic reverse; all indices have $O(\log(v+1))$ bits.
A path augmentation flips the corresponding original flow bits.
Its bottleneck is one.
Conservation and the interval $[0,1]$ are preserved by the usual residual-path update, so every augmentation increases the integer flow value by one.
The maximum endpoint value is at most the source degree, hence at most $v-1$.
Stop early and reject the insertion once the flow value reaches $q$.
If there is no augmenting path first, its residual reachable set is an endpoint cut with capacity equal to the flow value, so accept exactly when that value is below $q$.
The two directed arcs assigned to an undirected name count precisely one unit on any directed cut; therefore this is the same undirected cut test as in \eqref{eq:forest-df6}.

Each search, augmentation and compressed-name access is polynomial in $v$, and there are at most $v-1$ augmentations, independently of the preceding stream length.
The set $S$ is changed only after the flow scratch has been discarded.
The transient flow uses $2|S|$ bits.
These bits and $O(v\log(v+1))$ BFS scratch fit \eqref{eq:forest-pu1}: in the sparse representation both are absorbed by $\log\binom{N_E}{C_q}+C_q+v\log(v+1)$, where $C_q=q(v-1)<N_E/2$; in the dense branch they cost $O(v^2)$.
Old/new set ranks and shifted live-bit words need only a constant number of such registers.
The proof of all invariants of the maintained edge sets is unaffected because the acceptance predicate is identical.
\end{proof}

\subsection{Parameter selection and header capping}
\label{app:budget-finite-work}
The algorithm may select the least of the explicit exact, budget-sensitive (when $Z\le v$), and budget-independent space estimates.
Their dimensions and constant-factor estimates are computed by integer bit lengths, products, and comparisons; one need not know an optimal lower-bound source or solve an analytic optimization problem.
On the sketching branches, these estimates are represented by integers of $O(A)$ bits and computed before any array is allocated.

Cap $Z$ at $v+1$, where the budget-independent or exact branch suffices, and cap the confidence exponent at $v$, where $t>v$ selects the exact option.
Otherwise $\tau_0\le v+5$, bounding all matrix, bank, field, and counter dimensions.
Parameter reading follows Appendix~\ref{app:machine-conventions}.

The main implementation uses the polynomial code and verification-field constructors together with the operations on compressed edge sets and live bits above.
These parameter caps make all dimensions polynomial in $v$, so the implementation in Section~\ref{sec:budget-upper-bound} runs in polynomial time on every tape.
Reports leave persistent live bits, syndromes, hashes, and fields unchanged.
The all-tape hard cap also bounds expected peak space under the original coins.

\subsection{Finite headers and rounded report confidence}
\label{app:joint-header-implementation}

The following argument completes the implementation of Theorem~\ref{thm:joint-report-implementation}.
The algorithm runs the one-report procedure at error $\delta/Q$.
The following parameter representation realizes the same space order with finite header buffers and polynomial time per event.
Use the notation of that theorem and write $B_Z=\min\{A^2,\log^2(Z+2)\}$.

While reading the headers, saturate $Z$ at $v+1$, $Q$ at $2^v$, and the confidence exponent at $v$.
Saturation of $Q$ or confidence makes $\tau\ge v\ge\mu$, selecting the deterministic algorithm.
Otherwise, retain the exact value of $Q$ and put $a=\lceil\log(1/\delta)\rceil$, $c_Q=\lceil\log Q\rceil$, $\rho=2^{-(a+c_Q)}$, and $z'=1+a+c_Q$.
Then $Q\rho\le\delta$ and $\tau\le z'\le\tau+2\le5\tau/3$.
For the entropy envelope \eqref{eq:graph-envelope-main}, decreasing the first logarithm at every fixed source size gives $\Gamma_{v,\mu}(cz)\le c\Gamma_{v,\mu}(z)$ for $c\ge1$.
Thus the single-report algorithm at error $\rho$ has space $O(vAB_Z+v\Gamma_{v,\mu}(\tau))$ despite rounding.
The exact option, whenever selected by the parameter rules, also fits this bound by Theorem~\ref{thm:budget-space-implementation}; on each sketching branch, every field degree and loop count is polynomial in $v$ by \eqref{eq:forest-pu4} in Section~\ref{sec:budget-upper-bound}.
Very long headers are read through their delimiters without retaining a position counter, and the reading time is charged separately.
The $O(v)$-bit header buffers are absorbed by the $vA$ baseline.
The same hard cap bounds the expected peak space under the original coin law.

\subsection{Finite-search alternatives for recovery, verification, and cut tests}
\label{sec:finite-search-alternatives}

For readers concerned only with the space bounds, this subsection gives exhaustive-search implementations of the algorithms in Sections~\ref{sec:unrestricted-proof} and~\ref{sec:deletion-budget-proof}, retaining the same space order at the cost of exponential running time.
For the historical-deficit task this is $O(\mathcal L'(v,Z,t))$ bits, with setup, update, and report time bounded by $2^{O(v^2+t)}$ times a polynomial in $v+t$, in addition to the time spent reading the headers.
Reports remain read-only.
This subsection gives the search routines and their workspace bounds; the graph invariants and parameter choices are unchanged.

\paragraph{Finite search for the recovery code.}
\label{sec:finite-core-alternatives}
The following construction uses the matrix-existence guarantee of Corollary~\ref{cor:binary-entropy-codec}, but does not invoke its polynomial-time constructor.
With exhaustive syndrome decoding, it supports the two-pass reconstruction and space analysis in the proof of Lemma~\ref{lem:dense-bilinear-cut}; the running-time bound is the finite one stated above.

\begin{lemma}[Finite-search binary recovery matrix]
\label{lem:dense-binary-code}
For integers $v\ge2$ and $K\ge1$ there is a deterministic binary matrix $H\in\mathbb F_2^{r\times v}$, where
\begin{equation}
r\le v,\qquad
 r=O\bigl(\min\{v,K\log(1+v/K)\}\bigr),             \label{eq:forest-df3}
\end{equation}
such that every vector of Hamming weight at most $\min\{2K,v\}$ is uniquely determined by its syndrome $Hz$.
The matrix can be found by a uniform exhaustive search using $O(vr+v)$ bits.
Given $H$ and a syndrome, exhaustive decoding uses $O(v+r)$ additional bits and either returns that unique sparse vector or declares that none exists.
\end{lemma}
\begin{proof}
Read $K$ with saturation at $v$, discarding subsequent header digits and their count.
If $2K>v$, use $H=I_v$ and decode without a sparsity restriction.
This includes every saturated header and satisfies \eqref{eq:forest-df3}, since $K\log(1+v/K)=\Theta(v)$ whenever $K>v/2$.
Otherwise, the retained $K$ is exact.
Corollary~\ref{cor:binary-entropy-codec}, with $(n,k)=(v,2K)$, supplies a binary matrix with at most $v$ rows and $O(K\log(1+v/K))$ rows that distinguishes all vectors of weight at most $2K$.
Equivalently, its kernel contains no nonzero vector of weight at most $q=\min\{4K,v\}$: any such vector splits into two vectors of weight at most $2K$.
Only this existence guarantee is used here, not the corollary's constructor or workspace.

For $r=1,\ldots,v$, enumerate binary $r\times v$ matrices in bitwise order, using each matrix as its own enumeration state.
For each matrix enumerate every $v$-bit vector, skipping zero and weights above $q$, and reject the matrix if a tested syndrome is zero.
Stop at the first matrix passing every test.
Existence ensures that its row count satisfies \eqref{eq:forest-df3}; preceding stages have no larger row count.
The matrix, current vector, syndrome accumulator and indices use $O(vr+v)$ bits.
The search tests at most $v2^{v^2+v}$ matrix--vector pairs.

To decode, enumerate vectors of weight at most $2K$, returning the first whose syndrome matches, or failure after exhausting them.
The kernel test ensures uniqueness.
This takes $O(v+r)$ additional bits and at most $2^v$ vector tests, so all work is finite and parameter-bounded.

\end{proof}

\paragraph{Finite search for the verification field.}
Let $w$ be the degree in \eqref{eq:forest-df5}.
Enumerate all monic binary polynomials $f$ of degree $w$, testing $\gcd(f,X^{2^i}-X)=1$ for $1\le i\le\lfloor w/2\rfloor$.
A reducible polynomial has an irreducible factor of degree at most $w/2$, so exactly the irreducible candidates pass.
Repeated squaring, polynomial multiplication, and gcd use a constant number of $O(w)$-bit registers.
An irreducible polynomial exists, so at most $2^w$ candidates are tested.
Retain the successful polynomial; a fresh $w$-bit string then gives an exactly uniform field element.

\paragraph{Finite-search cut tests and running-time bounds.}
\label{app:budget-finite-search-details}
Use the same compressed historical state and parameter caps as in Appendix~\ref{app:budget-finite-work}.
To test an insertion with threshold $q$, enumerate the $v$-bit cut indicators separating its endpoints.
For each, count crossing stored names by serial access, saturating the count at $q$.
This requires $O(v+A)$ bits beyond the compressed set and at most $2^v$ cut iterations.
It implements the same acceptance predicate as the polynomial-time test in Lemma~\ref{lem:poly-historical-certificate}.

The matrix search in \eqref{eq:forest-df3} stores at most $vr\le v^2$ bits and tests at most $v2^{vr}$ matrices, each against at most $2^v$ vectors.
Its dimension counters use $O(A)$ bits.
Syndrome decoding uses at most $2^v$ vector tests, and field construction tests at most $2^w$ candidates with $w=O(A+t)$.
All remaining loops and arithmetic have polynomial length in $v+t$.
The stated exponential bound thus holds on every tape; each search retains only its current candidate, test state, and bounded indices.
These states, the decoding workspace, and the temporary report forest fit the same space estimates used in Section~\ref{sec:deletion-budget-proof}.
Reports leave the persistent edge sets and live bits, syndromes, hashes, and field representations unchanged.
The hard space cap also bounds expected peak space under the original coins.

\section{Detailed Comparison with Prior Work}
\label{sec:extended-related-work}

\paragraph{Extending the fully dynamic confidence frontier.}
Ahn et al. provide cut cancellation and independent recovery rounds~\cite[Section~3.1]{AGM2012}.
For fixed $0<\epsilon<1$ and $2^{-v^{1-\epsilon}}<\delta\le1/4$, Nelson and Yu prove a space lower bound for every one-pass streaming algorithm for this task that matches the memory guarantee of the reparameterized AGM algorithm; they also give its good/bad-round high-confidence analysis~\cite[Theorem~3 and Appendix~A]{NelsonYu2019}.
The constant-error and inverse-polynomial-error space orders are therefore classical.
Sawettamalya and Yu~\cite[Theorem~1.3]{SawettamalyaYu2026} improve the space bound for $k$-edge-connectivity certificates in dynamic streams, with success probability at least $1-v^{-5}$.
The first main result concerns the full confidence curve, including the regime in which $\log(1/\delta)$ approaches $v$ without a fixed polynomial gap.
Our algorithm reduces both the width of a recovery row and the sampling and repair depth.
An entropy-optimal replacement for an edge-domain sparse-recovery primitive alone would leave the latter costs unchanged.
Neighborhood recovery followed by contraction already appears in dynamic-stream spanner constructions~\cite[Sections~2 and~4]{KapralovWoodruff2014}; here its role is to reduce the initial number of unfinished components within the claimed confidence and space bounds.

\paragraph{Vertex-domain recovery and total working memory.}
Our cut interface stores short vertex-neighbor syndromes and constructs the two-sided cut measurement at query time, recovering current original edge pairs.
Its sparse-recovery domain is the $v$ vertices, rather than the $\binom{v}{2}$ possible edge names.
At threshold $b\le v$, an entropy-optimal edge-domain representation still pays $\log(\binom{v}{2}/b)=\Theta(\log v)$ per sparse entry, whereas the vertex-domain codec permits $\log(1+v/b)$.
The algorithm combines the sampled-universe bitmap with a shortened repair schedule to turn this smaller coordinate cost into the claimed space saving; the shared code tables, verification data and decoding scratch are charged as well.

Entropy-optimal Hamming checksums based on algebraic-geometric codes are classical; Cheng and Li explicitly recall this fact~\cite[Introduction and Section~3]{ChengLi2021Asymmetric}.
The decoder used here employs the classical error-locating-pair method \cite{Pellikaan92,DuursmaKotter94}.
The explicit tower functions and local branches used in the constructor have classical antecedents \cite{PellikaanStichtenothTorres1998Semigroups,AleshnikovKumarShumStichtenoth2001Splitting}.
This implementation supplies the required identities and a streamed matrix schedule, accounting for construction and decoder workspace.
These details are needed because a short syndrome or a communication-length guarantee does not by itself bound total peak memory.

Two-sided matrix sketches are established~\cite{DasarathyEtAl2015Tensor}, and staged recovery for tensor-product measurements appears in~\cite[Section~II-B]{HeJoseph2025Hierarchical}.
Bilinear sketches also occur in graph-stream cost estimation~\cite[Section~2.1]{AhnCormodeGuhaMcGregorWirth2015}.
Dynamic support sampling has a standard recovery--selection formulation \cite{CF14}, while invertible Bloom lookup tables and Newton-identity decoders give related set-reconciliation primitives \cite{GoodrichMitzenmacher2011IBLT,EppsteinGoodrich11}.
Code-based recovery already supports arbitrary small cuts through edge-domain labels~\cite[Sections~4.2 and~7.4]{IzumiEmekWadayamaMasuzawa2025}.
Sparse-neighborhood recovery with a source set fixed before the stream and side information supplied afterward appears in Assadi and Shah~\cite[Section~3.2 of the full version]{AssadiShah2022Matching}; this primitive alone does not identify the endpoint inside the source set~\cite[Section~5.2, footnote~10 of the full version]{AssadiShah2022Matching}.
\paragraph{Optimal spanning-forest space under a report-time deletion promise.}
Prior work on maximal matching~\cite{KKD2025} and hypergraph sparsification~\cite{KPS2025} bounds the total number of deletion operations.
For graphs, the reweighted subgraph produced by Khanna et al.~\cite{KPS2025} with approximation error below one preserves connected components; extracting a spanning forest therefore supplies actual original edges.
We require only that at most $Z$ previously inserted edge names are absent at a report, while allowing arbitrarily many deletions and restorations between reports.
For one report, our algorithms and matching space lower bounds determine the optimal space complexity, including the vertex and budget logarithms and the full confidence dependence.
The lower bounds already hold on streams with at most $Z$ deletion operations in total, so the same characterization also applies to that subclass.
Relative deletion promises~\cite{JayaramWoodruff2018} parameterize a different restriction on the stream.
The density-sensitive hybrid algorithm of~\cite{DeManGillBenderDhulipalaTench2026} uses $O(\min\{v+e,v\log v\log(2+e/v)\})$ edge/bucket records, with high probability for polynomial-length streams and current edge count $e=|E|$~\cite[Theorem~1.2]{DeManGillBenderDhulipalaTench2026}.
Their bucket fields and explicit-neighbor representation use $\Theta(\log v)$ bits per record.
This density-sensitive tradeoff complements our parameterization by the advance bound $Z$ and arbitrary confidence.
We use entropy-compressed historical edge sets with live bits and vertex-domain syndromes to resolve the bit cost.

Khanna et al. study bounded-deletion streams for distance oracles and spanners~\cite{KhannaKonradPutterman2026}, extending deletion-sensitive streaming to distance preservation.
Bounded-independence edge sampling for combinatorial graph properties is studied by Putterman et al. \cite{PuttermanVadhanZaripov2026}.
Our spanning-forest algorithm returns current original edges and accounts for sampling data and recovery workspace in its memory usage.
Classical work of Nagamochi and Ibaraki gives sparse connectivity certificates~\cite{NagamochiIbaraki1992}; a streaming formulation is given in~\cite[Lemma~3.7]{JinKapralovMahabadiVakilian2024}.
Our algorithm stores its accepted historical edge names and their live bits in compressed form and keeps the minimum-cut workspace within the same entropy-scale space budget.
It permits an advance bound on historical edge names absent at reports, even when intermediate deletions and restorations are unrestricted.

\paragraph{Source encoding for space lower bounds.}
The all-error universal-relation complexity is known \cite{KNPWWY2017,NelsonYu2019}; optimizing the corresponding bound gives the source-size choice $m\asymp\sqrt{U\log(1/\delta)}$.
The forest lower-bound argument transfers this encoding to the graph family while retaining the source-size and deletion constraints.
The compact-membership and retrieval literature provides classical space baselines for membership and associated-value queries \cite{Carter78,DP08}.
External-information accounting allows a query to delete from one block while charging information across all blocks.
Exact channel simulation with shared randomness~\cite[Result~1]{HJMR2010} preserves each source's message law and bounds communication averaged over the source distribution and protocol randomness.
It therefore transfers corresponding expected-communication lower bounds to external-information lower bounds.
Our information interface makes the source-size dependence explicit for the deletion constraint and uses the information chain rule to charge independent graph neighborhoods.

\paragraph{Joint correctness and lower bounds for repeated reports.}
Simultaneous correctness across many queries is standard in streaming and dynamic data structures.
Hassidim et al. explicitly describe allocating failure probability $\delta/Q$ to each of $Q$ queries and applying a union bound when the stream is independent of the algorithm's randomness~\cite[Introduction]{HassidimEtAl2020Robust}.
Continuous monitoring studies the related requirement that all estimates along a stream be correct simultaneously; B{\l}asiok et al. show that the direct union-bound cost can be improved for norm estimation~\cite[Definition~1 and Introduction]{BlasiokDingNelson2017Monitoring}.
For dynamic graph connectivity, Gibb et al. give joint correctness over polynomially many updates~\cite[Theorem~3.1]{GibbKapronKingThorn2015}; forest caches also have precedents~\cite{TenchEtAl2025Landscape}.

Our repeated-report result determines the optimal dependence on $Q$ and $\delta$ for spanning-forest recovery under the report-time bound $Z$.
The algorithmic guarantee follows by applying the standard $\delta/Q$ allocation to our read-only single-report algorithm.
Our additional result is a matching space lower bound that allows information learned from earlier reports to remain in the algorithm's state.
Our update and report sequences are fixed independently of the algorithm's randomness; this guarantee does not assert robustness to adaptively chosen streams.
The lower bound requires an additional argument: one source is restored and queried repeatedly, and the algorithm may retain information learned from earlier answers.
Low-error-round extraction appears in tracking lower bounds \cite[Section~7 of the full version]{Blasiok2018}; same-source requests and learned retained states also occur in communication and quantum streaming~\cite{GilboaJainMcClean2025,Voronova2026Replenishment}.
The forest lower bound additionally requires preservation of the source law and control of the expected state length after earlier reports.
Conditioning jointly on earlier success can bias the restored source.
Instead, after selecting one fixed low-loss round by averaging under the original source law, the proof normalizes the surviving state channel separately at each source and draws that source from its original law.
The current query remains fresh conditional on the source.
The sourcewise survival bound controls both the error and expected message length by constant factors, so the original expected-peak charge is preserved.
This is the additional step connecting low-error-round extraction to the repeated-report forest lower bound; it allows earlier answers to change the retained state without changing the fixed-input model or replacing the hard source distribution by a success-conditioned one.

\bibliographystyle{alpha}
\begingroup
\interlinepenalty=10000
\bibliography{references}
\endgroup

\section*{AI Assistance}

The main proofs were developed by the human authors.
AI tools assisted with organizing and documenting the proofs and with parts of the proof development for selected theorems.
All AI-assisted mathematical material was reviewed by the human authors.
ChatGPT and Gemini were used extensively to assist with writing and revising the manuscript.
The human authors reviewed the final manuscript in full and take ultimate responsibility for its mathematical content and the manuscript as a whole.

\end{document}